# Software tools for quantum control: Improving quantum computer performance through noise and error suppression

Harrison Ball, Michael J. Biercuk,[*] Andre Carvalho, Jiayin Chen, Michael
Hush, Leonardo A. De Castro, Li Li, Per J. Liebermann, and Harry J. Slatyer

*Q-CTRL, Sydney, NSW Australia & Los Angeles, CA USA*

Claire Edmunds, Virginia Frey, Cornelius Hempel, and Alistair Milne

*ARC Centre for Engineered Quantum Systems, The University of Sydney, NSW Australia*

(Dated: July 4, 2020)

Effectively manipulating quantum computing hardware in the presence of imperfect devices and control systems is a central challenge in realizing useful quantum computers. Susceptibility to noise critically limits the performance and capabilities of today's so-called noisy intermediate-scale quantum (NISQ) devices, as well as any future quantum computing technologies. Fortunately, quantum control enables efficient execution of quantum logic operations and quantum algorithms with built-in robustness to errors, and without the need for complex logical encoding. In this manuscript we introduce software tools for the application and integration of quantum control in quantum computing research, serving the needs of hardware R&D teams, algorithm developers, and end users. We provide an overview of a set of python-based classical software tools for creating and deploying optimized quantum control solutions at various layers of the quantum computing software stack. We describe a software architecture leveraging both high-performance distributed cloud computation and local custom integration into hardware systems, and explain how key functionality is integrable with other software packages and quantum programming languages. Our presentation includes a detailed mathematical overview of central product features including a flexible optimization toolkit, engineering-inspired filter functions for analyzing noise susceptibility in high-dimensional Hilbert spaces, and new approaches to noise and hardware characterization. Pseudocode is presented in order to elucidate common programming workflows for these tasks, and performance benchmarking is reported for numerically intensive tasks, highlighting the benefits of the selected cloud-compute architecture. Finally, we present a series of case studies demonstrating the application of quantum control solutions derived from these tools in real experimental settings using both trapped-ion and superconducting quantum computer hardware.

## CONTENTS

* Also ARC Centre for Engineered Quantum Systems, The University of Sydney, NSW Australia

## I. INTRODUCTION

The emergence of commercially-available quantum computing (QC) hardware at the scale of a few tens of qubits has led to an explosion of interest in NISQ (noisy intermediate scalable quantum) devices [1]. There are even now several software stacks allowing end-users and software developers to explore quantum computing over the cloud [2–5]. Commensurate with this growth in programming frameworks has been a expansion of efforts focused on application mapping and algorithmic development to identify applications yielding any commercially relevant computational advantage [6]. However, as in conventional software engineering, functionality and computational advantage ultimately rests on lower-level abstractions deeper in the computational stack such as compilers, and more fundamentally, on hardware device performance [7].

One of the earliest software methodologies proposed for quantum computing described a 4-phase design flow [8], transforming high-level algorithms to mid-level representations such as QASM (quantum assembly language), ultimately to be compiled down to device-specific instructions sent to the quantum hardware. Later work outlined a more detailed layered architecture [9] including a pipelined control cycle from the application layer down to the physical hardware layer, with mid-tier processes such as QEC in between.

Since then various QC programming languages, simulators, and compilers have been devised. Starting at the highest levels of the stack, freely-available examples include Quipper [10], a quantum program compiler implemented in Haskell, the LIQ$Ui|\rangle$ simulator [11] written in F#, and the ScaffCC compiler [12] designed for a C-style language, which compiles gate sets in QASM and supports program analysis and low-level optimizations. Further down the stack, efficient scheduling architectures with reduced communication overheads and increased parallelism have been proposed to accommodate the relatively short lifetimes of quantum information in quantum hardware [13–15]. Other compiler performance gains have been identified by considering device-specific optimizations, gate set choices and communication topologies [16].

More recently a variety of Python-based languages have been developed providing greater integration and functionality across various abstraction layers. ProjectQ [17–20] is a toolflow to optimize, simulate and compile quantum programs for different hardware backends. Qiskit [2, 21] is a general-purpose compiler framework generating OpenQASM [22], the language used to create and compile quantum programs on IBM's Quantum Experience [23]. pyQuil [4, 24, 25] generates Quil [24], the compiler language used for the Rigetti Computing system. Cirq [5] is Google's software library for writing, optimizing and running quantum circuits on hardware backends or simulators.

Nonetheless, the central impediment to realizing practical, functional machines in the NISQ-era and beyond remains the influence of noise and error in quantum hardware itself, despite these various advances in quantum software development. Electromagnetic noise in its various forms diminishes coherent lifetimes through the process of decoherence, and reduces the fidelity of quantum logic operations when imperfect quantum devices are manipulated by faulty classical hardware. This critically limits the range of useful computations achievable on quantum hardware, measured *e.g.* by circuit depth or quantum volume [26]. Overwhelmingly, the tools and frameworks introduced above focus on the design, implementation, and optimization of algorithms near the top of the quantum computing software stack, and do not directly address this most fundamental challenge in the field.

Developing techniques that improve the robustness of quantum hardware against noise and error is critical for pursuing commercially-viable applications. One approach to this problem comes through the implementation of low-level error-suppression strategies derived from the field of quantum control [27–36]. This discipline draws insights from classical control engineering - frequently associated with the stabilization of unstable hardware - though successful translation to the quantum domain requires modification of fundamental concepts. For instance, quantum systems used for quantum computing are typically nonlinear (control over qubits is formally bilinear), noise-processes in real quantum hardware are generally colored, and measurement has strong back-action on the controlled system.

The existing literature on quantum control comprises a wide range of complex techniques and diverse approaches to achieve error robustness in quantum computers [32, 37–51]. These strategies have been widely identified as an important complement to algorithmic error-mitigation approaches such as QEC [9, 44, 52, 53], due to their potential to improve resource-efficiency by reducing physical-qubit error rates. Experimental demonstrations have validated the utility of quantum control in mitigating noise in quantum hardware [54–59], leveraging longstanding insights from fields such as NMR [60]. Similarly, optimal control has begun to emerge as a powerful technique to manipulate complex Hilbert spaces [61–63], or optimize experimental efficiency [64–66]. Early hints of progress moving beyond proof-of-principle demonstrations towards system integration have emerged as well, placing greater focus on real hardware limitations (*e.g.* timing constraints, power and bandwidth limitations, and availability of controls) [34, 49, 67–70], moving beyond single-qubit settings [71–74], and extending their applicability to *realistic* multi-qubit NISQ devices.

The diversity of quantum control techniques and changing levels of hardware-knowledge among quantum computing end users highlight a need for a unified software framework supporting the integration of quantum control techniques with both differing hardware systems and high-level software abstractions. Such an approach is strongly aligned with emerging community expectations; a prime indicator of this is the release of OpenPulse [21], the Qiskit language providing cloud access to IBM backends at the *analog layer*, motivated by the need "to explore noise in these systems, apply dynamical decoupling and perform optimal control theory". However, a historic reliance on customized local code for quantum control tasks is cost-inefficient, harms reproducibility, fails to deliver on the most up-to-date knowledge from the research community, and has substantial negative consequences as students and staff inevitably move on from current roles and support ceases.

In this manuscript we introduce an infrastructure software package aimed at addressing these challenges, focused on providing access to state-of-the-art quantum control techniques, and enabling integration into the quantum computing stack. These tools have been designed to meet the following central objectives:

1 To advance the performance of real quantum hardware by delivering optimized control strategies. More concretely, to enable efficient characterization of error sources, identify and exploit system controllability, and generate instructions for real hardware to suppress the influence of noise and imperfection at the device level.

2 To deliver greater functionality from fixed quantum computational resources (measured in qubits, gates, and compute runtime) for users with a broad-range of experience and expertise in quantum computing hardware or quantum control.

3 To provide maintained access to complex and rapidly evolving technology, and deliver state-of-the-art computational resources for numerically intensive tasks via a modern cloud-compute architecture. This includes access to numerical techniques that benefit from or require specialized computational hardware such as GPUs.

4 To build cross-compatibility with existing workflows, programming languages, QC architectures, and access methods. These tools may be integrated into conventional programming workflows via Python, linking them to research code, cloud-based quantum computers, and custom QC hardware.

The remainder of this paper is organized as follows. First, we provide an overview of infrastructure software products for the development and deployment of quantum control in quantum computers in Sec. II. We then move on to present a technical, mathematical treatment of a novel quantum control capabilities we have developed and deployed in these packages in Sec. III. Our presentation includes a detailed discussion of new algorithmic approaches to: flexible numeric optimization using an engine built in TensorFlow and linking to various quantum control tasks; performance evaluation and validation using both numeric simulation and multidimensional filter functions; and control-hardware characterization via noise spectroscopy and Hamiltonian parameter estimation. In Sec. IV these functionalities are demonstrated through a series of case studies tied to challenges in real QC hardware. We provide experimental validation of the benefits of low-level quantum control in quantum computing hardware, demonstrate novel numerically optimized gate solutions for multiqubit gates, extract previously inaccessible information about noise sources in cloud quantum computer hardware, and demonstrate the impact of optimization at the circuit level for increasing noise robustness. We conclude with a brief summary and future outlook of forthcoming feature developments.

## II. SOFTWARE ARCHITECTURE AND INTEGRATIONS

The Q-CTRL infrastructure software suite is designed to improve hardware performance in the quantum computing stack through access to quantum control. Q-CTRL tools incorporate the following general classes of quantum control capability relevant to stabilizing quantum systems against hardware errors:

- Error-robust control selection, creation, and integration into quantum hardware.

- Flexible optimization for quantum logic, circuits, algorithms, and high-dimensional quantum systems, incorporating various constraints, nonlinearities, etc.

- Predictive error-budgeting and simulation of hardware and circuit performance in realistic laboratory environments.

- Hardware tuneup, characterization, and calibration at the microscopic level to identify and offset sources of noise, imperfection, and performance variability.

In this section we provide a brief introduction to the cloud-compute architecture in use, key software packages, and integration with both other software tools and hardware systems. We then provide a technical discussion of the quantum control functionalities enabled by this software in Sec. III.

### A. Cloud-compute architecture

All software is delivered via a cloud-compute architecture built around an application programming interface (API) coded in Python. Once the client has entered relevant inputs, this information is sent to the back-end and processed through the API. The client's data is taken through to the Python module, which performs the relevant computations and outputs objects based on the system inputs. The interface with the API varies based on the specific software in use as described in Sec. II B, and allows for custom application development by the user.

A substantial proportion of the codebase comprises orchestration of cloud-compute resources, data management, memory management, and the like, though we will focus primarily on the technical functionality of the core python package here. The back-end software architecture employs established and lightweight web interfaces (OpenAPI specifications, REST APIs and JSON), as well as performant and scalable architectural designs such as a three-tiered application with dedicated worker pools and work queues. The use of open standards enables the entire application stack to be deployed on any cloud—public, private, hybrid or on-premises—allowing users to determine appropriate and necessary tradeoffs between price, performance and privacy. For instance, a remotely managed on-premises-cloud instance allows full control over all sensitive data, while still ensuring the advantages of a cloud-compute architecture.

Our choice of Python for the API incorporates speed of development and support for collaboration with external developers, scientists, and partners. Both quantum scientists and programmers are typically familiar with Python, having used its libraries for tasks ranging from instrument control and advanced numerics to web design. Building in Python also leverages compatibility with global resources of open-source code, and web-based frameworks like Django, bringing embedded security features to safeguard against attacks such as SQL injection, request forgery, or cross-site scripting.

In certain circumstances the Python API incorporates special-purpose programming frameworks such as TensorFlow and Cython in order to deliver performance enhancements via access to cloud-based hardware infrastructure. In all cases, processing resources used in the execution of a computation are scaled by the software, and specialized accelerators such as GPUs are automatically accessed for tasks exceeding predefined thresholds of computational complexity such as high-dimensional optimizations executed using our TensorFlow-based tools (see Sec. III C).

### B. Package overview

Here we survey the the central software packages designed for the deployment of quantum control techniques in quantum computer hardware. Each delivers a targeted set of quantum-control capabilities to users with different backgrounds, interests, and objectives. Our core focus is on solutions which integrate quantum control into the lowest level of the QC software stack, although quantum control also provides benefits at higher levels as well, as demonstrated in Sec. IV E.

*BOULDER OPAL* offers a Python-based toolkit allowing users to develop and deploy quantum control in their hardware or theoretical research. All technical features and core capabilities of Q-CTRL packages described in Sec. III are accessible via this package, making this the core toolkit in our offering. In order to facilitate integration into conventional programming environments, BOULDER OPAL includes a light Python package wrapper that is downloaded locally and orchestrates calls to the web API. All computationally intensive tasks remain the responsibility of the core computational engine in the cloud. Typical users - academic and industrial R&D experts and quantum hardware experts.

*BLACK OPAL* helps users learn about quantum computing and quantum control in the NISQ era by taking advantage of a graphical interface with interactive visualizations. It is designed to assist in building intuition for complex concepts such as the meaning of entanglement in quantum circuits, or the impact of noise on circuit func-tionality. Features are delivered as a web-based API service providing users with a graphical front-end interface incorporating guided tours, configuration wizards, and integrated help; examples of an interactive visualization for single and multi-qubit gates are shown in App. F. The front-end prepopulates common system configurations for superconducting and trapped-ion processors, or custom configurations may be input, and contains predefined libraries of known control solutions. Typical users - students, conventional developers, and newcomers in quantum computing and quantum control.

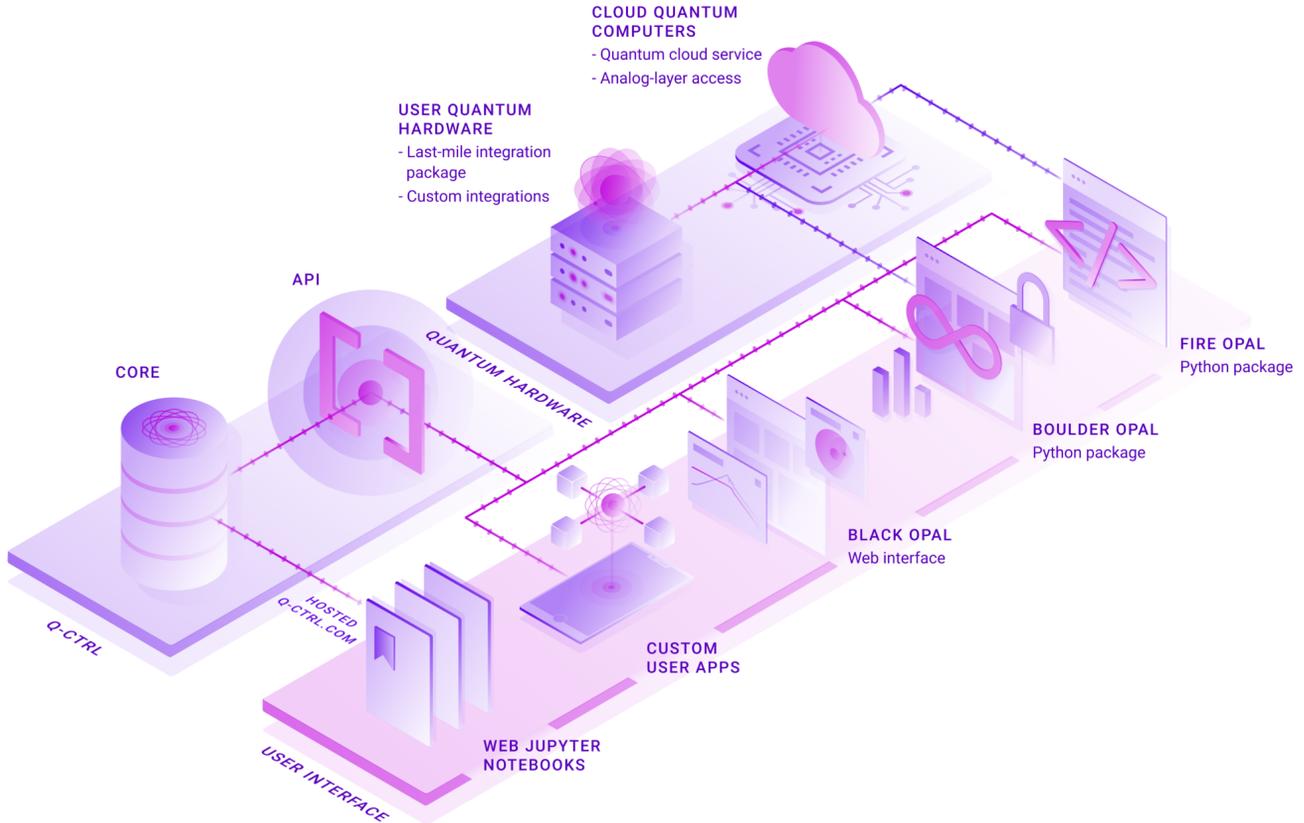

FIG. 1. Relationships between Q-CTRL software packages, demonstration of various means of user interaction, and links between cloud-compute resources, interfaces, and quantum computing hardware. For instance, BOULDER OPAL connects to the cloud engine via the API, is commonly accessed through a Python interface, or may be combined with the last-mile-integration package to enable direct integration into user quantum hardware. FIRE OPAL is also accessed via a python package and interfaces with cloud quantum computers. Meanwhile core functionality may be accessed outside of these products either via the Q-CTRL API by users building custom tools, or directly via hosted Jupyter notebooks by Q-CTRL research partners. As a standalone open-source python package, Open Controls is not included on this architectural diagram.

*FIRE OPAL* is a forthcoming package focused on embedding the benefits of quantum control into algorithmic design and execution. Key functionality includes analysis of algorithmic performance in the presence of realistic time-varying noise, embedding of error-robust quantum logic-operations into a compiled algorithm, and the integration of control theoretic concepts such as robustness through the structure of a compiled circuit. This toolkit is designed to be compatible with other compilers and can provide deterministic error robustness without the need for additional overhead such as repetition when adding engineered error in zero-noise-extrapolation schemes (see Sec. IV E for an example case study). Typical users - quantum algorithm developers and end-users focusing on application mapping without detailed knowledge of the underlying hardware.

*Open Controls* is an open-source Python package that includes established error-robust quantum control protocols from the open literature. The aim of the package is to be a comprehensive library of published and tested open-loop quantum control techniques developed by the community, with export functions allowing users to deploy these controls on custom quantum hardware, publicly available cloud quantum computers, or other parts of the Q-CTRL software suite. Typical users - quantum research teams contributing to or employing community-derived quantum control protocols and sequences.

### C. Compatibility with other programming languages

A typical mode of accessing the software packages described above comes from a lightweight Python wrapper or SDK which enables access to the API from within a standard Python interface. This approach brings the added advantage of compatibility with a wide variety of programming languages commonly employed in the quantum computing research community, as summarized in Sec. I. Q-CTRL provides Python adaptors for all open-source Python-based quantum computing languages, allowing integration of advanced control solutions into conventional programming workflows and execution on cloud hardware platforms.

The *qctrl-qiskit* convenience package provides export functions of Q-CTRL-derived control solutions or protocols to Qiskit [2]. For instance, dynamical decoupling sequences (used for implementing the identity operator in preservation of quantum memory) can be converted into Qiskit quantum circuits using these methods, accounting for approximations made in Qiskit circuit compilation and ensuring circuits are not compactified in such a way that undermines performance. Furthermore, control pulses can be exported from BOULDER OPAL in the OpenPulse format [21] which provides analog-level programming of microwave operations performed on hardware. As a complement, Q-CTRL has also developed calibration routines for IBM hardware employing the OpenPulse framework, again offered as convenience functions. Thus, appropriately formatted controls and calibration routines can then run on IBM's quantum computing hardware with an IBM Q account [23].

The *qctrl-pyquil* convenience package provides export functions to pyQuil [24, 25], and *qctrl-cirq* allows export Cirq [5]. At present, the absence of analog-level control access limits export functionality to timed sequences of standard control operations handled natively in these platforms. In both cases the translation layer ensures that the integrity of the sequence structure and timing is preserved within approximations made in sequencing in these two languages. pyQuil integration currently permits the execution of Q-CTRL-derived sequences on quantum hardware provided by Rigetti's Quantum Cloud Service.

### D. Integration with quantum hardware

The software tools described above integrate with a wide variety of hardware systems—from local laboratory-based quantum hardware to quantum-compute cloud engines ( Fig. 1). Both BOULDER OPAL and FIRE OPAL support integration into cloud-quantum computers using analog-layer access and/or appropriate convenience functions. However, a more powerful integration strategy may be pursued in the case of interfacing these software tools with custom quantum hardware.

At its most basic level, integration into custom user hardware involves converting waveforms described in software into physical outputs from hardware signal generators such as arbitrary waveform generators (AWGs), direct digital synthesizers (DDSs), and vector signal generators (VSGs). BOULDER OPAL permits exporting controls in a format tailored to hardware constraints such as sample rates, amplitude resolution, and data formats. Custom control pulses are exported into a format (*e.g.* CSV or JSON [21]) easily read by the experimental control stack. Q-CTRL provides a range of pre-built formatting scripts to translate control output into machine-compatible formats.

As an example, Q-CTRL has partnered with Quantum Machines, providing a direct interface between Q-CTRL protocols and their Quantum Orchestration Platform [75]. The Quantum Orchestration Platform is a software-hardware solution whose software interface is Quantum Machines' programming language called QUA, and an advanced hardware system allowing orchestration of QUA programs in real-time (*e.g.* waveform generation, waveform acquisition, classical data processing and real-time control-flow). This integration naturally permits control formatting matched to QUA, but also exploits parametric encoding to enable rapid hardware tuneup and calibration. Similarly, scripting in QUA allows exploitation of low-latency FPGA-based computation for the execution of real-time machine-learning routines or Bayesian updates via tools offered jointly by Q-CTRL and Quantum Machines. Using this framework we have recently demonstrated dephasing-robust single-qubit operations in a tuneable transmon device, indicating that the combination of Q-CTRL software with the Quantum Machines quantum orchestration platform stack permits faithful output of complex modulated control protocols. Beyond official Q-CTRL partners, custom scripts cover commonly encountered hardware solutions, leveraging the flexibility of Python programming.

Moving further down the experimental control hardware stack it also becomes possible to implement a variety of low-latency real-time processing tasks. Capabilities are based on core routines and techniques customized for a user's or vendor's hardware system (*e.g.* QUA development with Quantum Machines). An example is closed-loop optimization of control solutions in order to reoptimize error-robust controls as system noise sources drift or in the presence of unknown system responses [54, 76]. Numerically optimized controls may be used as a seed for optimization based on experimental measurements incorporating user-defined cost functions such as randomized benchmarking survival probabilities.

An essential aspect of this integration is efficient hardware calibration permitting determination of the analog voltages or digital commands required to achieve output signals with appropriate phase and amplitude (alternatively $I$ and $Q$) values. This helps account not only for residual amplitude modulation in the presence of other forms of modulation (*e.g.* phase modulation), but also cross-coupling and signal distortions encountered due to room-temperature hardware such as mixers. Hardware calibration can also leverage quantum control in order to gain access to information about signal distortions and transmission-line nonlinearities within experimental systems that are not easily characterized through conventional means. This approach is widely employed in most laboratories through basic protocols for qubit frequency determination such as Ramsey spectroscopy, drive-amplitude calibration through Rabi measurements, and more advanced protocols to estimate microwave phases [77] or identify quadrature cross-couplings [78, 79]. BOULDER OPAL currently offers a range of convenience tuneup functions for superconducting-circuits using tools described in Sec. III E:

1. Resonator-probe pulse optimization.

2. Square-wave-frequency-modulation [80, 81] qubit-resonance identification.

3. State-discriminators using ML-based classifiers (linear and gradient-boosting) [82].

4. Maximum-likelihood state estimation incorporat-

ing decay rates.

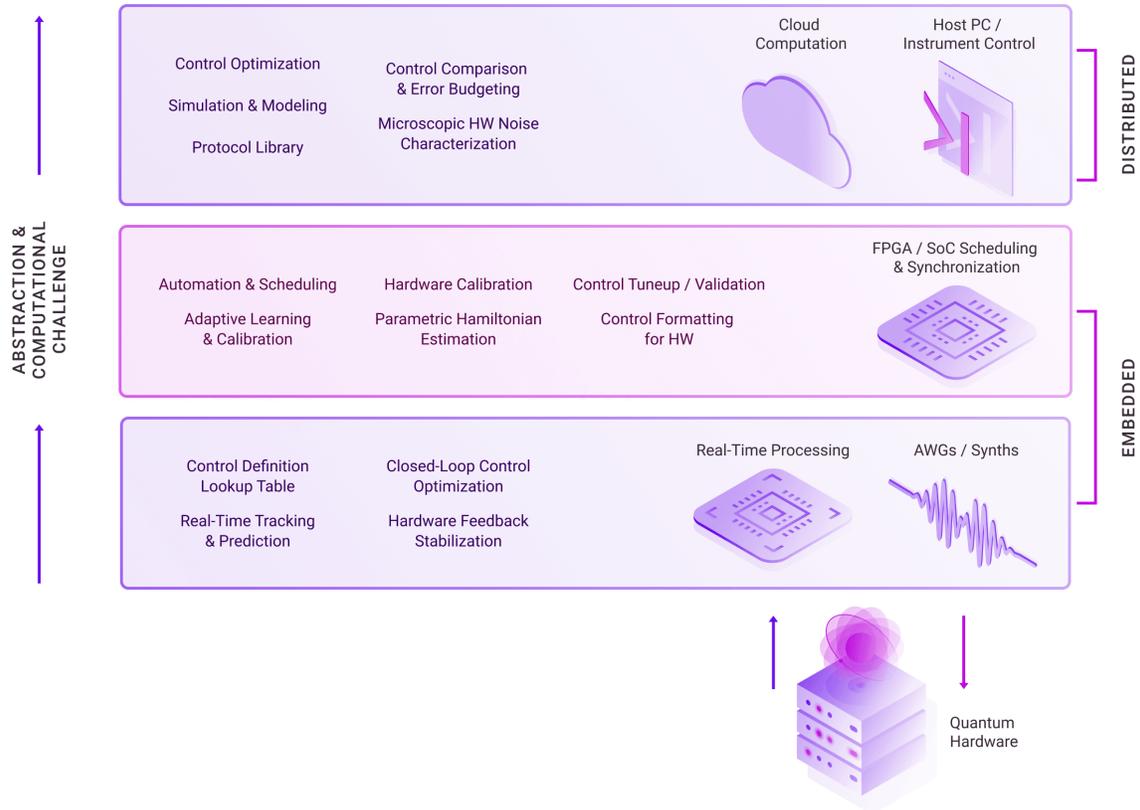

FIG. 2. Schematic overview of quantum control functions and their integration into a user's custom classical experimental control hardware. As tasks are abstracted further away from the operation of experimental hardware, the point of integration similarly rises in the experimental control stack. Functions are broadly characterized by their execution in either embedded or distributed computational hardware.

5. IQ-nonlinearity calibration [83].

6. Time [84] and frequency-domain impulse-response characterization.

7. Modulated spectroscopy for excited states.

8. Hamiltonian parameter estimation (*e.g.* control-rotation-axis identification).

These capabilities may be integrated into a user's existing experimental hardware control system and software stack either manually or via the encapsulated Last-Mile Integration (LMI) extension for BOULDER OPAL. This is delivered via a customized, local-instance Python package that runs in parallel with the user's experimental control stack to automate and schedule key tasks in control definition, calibration, and optimization. The LMI is responsible for the orchestration of critical tasks conduced across different hardware devices as shown in Fig. 2.

The LMI package maximizes automation via a suite of Python scripts for scheduling of essential tasks such as hardware calibration and characterization. As an example, the package permits scheduled noise sensing and reconstruction in order to detect changes in dominant noise power spectra. Again, all computationally intensive calculations are handled by the distributed cloud-compute engine (Fig. 1) with only scripting and experimental-control-software integration handled locally. Experimental calibration and noise characterization results are also logged on the cloud server and accessible to users through a developer's portal.

## III. TECHNICAL FUNCTIONALITY OVERVIEW

In this section we provide an overview of key technical capabilities afforded by quantum control as accessed through the packages introduced in Sec. II. Our presentation focuses on tasks relevant to driving performance enhancement in quantum computer hardware, though many other applications exist in quantum sensing, data fusion, and advanced medical imaging. Alongside our formal mathematical treatment we incorporate pseudocode to illustrate how this functionality is embodied in software features.

### A. General quantum-control setting

Here we establish the general control-theoretic setting for the creation of control solutions in multi-dimensional quantum systems. We write the total Hamiltonian as the sum of dynamical contributions from both control and noise interactions

$$H_{\mathrm{tot}}(t) = H_{\mathrm{ctrl}}(t) + H_{\mathrm{noise}}(t). \tag{1}$$

We assume that $H_{\mathrm{ctrl}}(t)$ is a deterministic component of $H_{\mathrm{tot}}(t)$ containing both an intrinsic *drift* term (*e.g.* the frequencies of the qubits) and *controllable* parts of the system (*e.g.* microwave *drives* or clock *shifts*). See App. B for generalized definitions of these terms. The error Hamiltonian $H_{\mathrm{noise}}(t)$ captures the influence of noise, is assumed to be small, and may also be a function of control Hamiltonian $H_{\mathrm{ctrl}}(t)$. The typical control problem may be split into two distinct tasks:

1. Design a control solution for a $D$-dimensional Hilbert space such that $H_{\mathrm{ctrl}}(t)$ implements a target unitary $U_{\mathrm{target}}$ at time $\tau$.

2. Design a control solution such that $H_{\mathrm{ctrl}}(t)$ is *robust* against noise interactions $H_{\mathrm{noise}}(t)$ over duration $[0, \tau]$.

The first task is typically referred to as an *optimal* control problem while the second is called *robust* control; Q-CTRL provides tools for both.

#### 1. Optimal quantum control

The *optimal* control setting reduces the following problem. Given the Schrödinger equation

$$i\dot{U}_{\mathrm{ctrl}}(t) = H_{\mathrm{ctrl}}(t)U_{\mathrm{ctrl}}(t) \tag{2}$$

the aim is to find $H_{\mathrm{ctrl}}(t)$ such that

$$U_{\mathrm{target}} = U_{\mathrm{ctrl}}(\tau) \tag{3}$$

where $U_{\mathrm{ctrl}}(\tau)$ is the evolved unitary at time $\tau$ and $U_{\mathrm{target}}$ is the target operation. This control problem, generally, can be considered a bilinear control problem as the controllable element of the equation of motion ($H_{\mathrm{ctrl}}(t)$) linearly *multiplies* the state. This is in contrast to the much more common linear control problems from classical control where the controllable element is linearly *added* to the state. Bilinear control problems typically do not have analytically tractable solutions, and instead must be solved numerically.

We define a measure of optimal control using a Fronenius inner product Eq. A1 to evaluate the operator-distance between $U_{\mathrm{ctrl}}(\tau)$ and $U_{\mathrm{target}}$. Specifically

$$\mathcal{F}_{\mathrm{optimal}}(\tau) = \left| \frac{1}{D} \left\langle U_{\mathrm{target}}, U_{\mathrm{ctrl}}(\tau) \right\rangle_F \right|^2. \tag{4}$$

This measure is bounded between $[0, 1]$, with perfect implementation of the target Eq. 3 corresponding to $\mathcal{F}_{\mathrm{optimal}}(\tau) = 1$. We define a corresponding infidelity measure as

$$\mathcal{I}_{\mathrm{optimal}}(\tau) = 1 - \mathcal{F}_{\mathrm{optimal}}(\tau). \tag{5}$$

A simple modification of this fidelity measure enables the optimal condition to be evaluated on a subspace of interest. Specifically

$$\mathcal{F}_{\mathrm{optimal}}^P(\tau) = \left| \frac{1}{\mathrm{Tr}\,(P)} \left\langle PU_{\mathrm{target}}, U_{\mathrm{ctrl}}(\tau) \right\rangle_F \right|^2 \tag{6}$$

where $P$ defines a projection matrix, enabling optimal control to be evaluated on a target subspace. Similarly, achieving high-fidelity state-transfer $|\psi_{\mathrm{initial}}\rangle \rightarrow |\psi_{\mathrm{final}}\rangle$ is equivalent to maximizing the state fidelity defined as

$$\mathcal{F}_{\mathrm{optimal}}^\psi(\tau) = |\langle \psi_{\mathrm{initial}}|U_{\mathrm{ctrl}}(\tau)|\psi_{\mathrm{final}}\rangle|. \tag{7}$$

As discussed in detail in Sec. III C, crafting numeric solutions for these control problems may be conveniently cast as cost-minimization; we provide specific numeric tools addressing this challenge. These measures strictly describe whether the ideal Hamiltonian $H_{\mathrm{ctrl}}$ implements the target evolution. They do not incorporate errors arising from stochastic noise processes. This is the subject of robust control, described below.

### 2. Robust quantum control

The *robust* control setting presents the multi-objective problem of achieving both tasks 1 and 2. In this case we consider the stochastic evolution of the total Hamiltonian $H_{\text{tot}}(t)$

$$i\dot{U}_{\text{tot}}(t) = H_{\text{tot}}(t)U_{\text{tot}}(t). \qquad (8)$$

Noisy dynamics contributed by $H_{\text{noise}}(t)$, therefore distort the final operation $U_{\text{tot}}(\tau)$ away from the ideal $U_{\text{ctrl}}(\tau)$. This effect may be isolated by expressing the total propagator as

$$\tilde{U}_{\text{noise}}(\tau) = U_{\text{tot}}(\tau)U_{\text{ctrl}}(\tau)^\dagger. \qquad (9)$$

The residual operator $\tilde{U}_{\text{noise}}(\tau)$ defined in Eq. 9 is referred to as the *error action operator*. This unitary satisfies the Schrödinger equation in an interaction picture co-rotating with the control, which we call the *control frame*. Specifically,

$$\tilde{U}_{\text{noise}}(\tau) = \mathcal{T}\exp\left[-i\int_0^\tau \tilde{H}_{\text{noise}}(t)dt\right] \qquad (10)$$

where $\mathcal{T}$ is the time-ordering operator, and

$$\tilde{H}_{\text{noise}}(t) \equiv U_{\text{ctrl}}(t)^\dagger H_{\text{noise}}(t)U_{\text{ctrl}}(t) \qquad (11)$$

defines the *control-frame Hamiltonian*. An analogous concept originally appeared in average Hamiltonian theory developed for NMR [85], but was called a *toggling* frame in that context because it considered only instantaneous operations.

Using the definition of $\tilde{U}_{\text{noise}}(\tau)$ in Eq. 9, the robust control problem may be formalized in terms of the dual conditions

$$U_{\text{target}} = U_{\text{ctrl}} \qquad (12)$$

$$\tilde{U}_{\text{noise}} = \mathbb{I} \qquad (13)$$

where $\mathbb{I}$ is the identity operation on the control system. The *robust* control problem therefore consists of augmenting the *optimal* control problem with the additional condition Eq. 13, describing how susceptible the system is to noise interactions under a given control Hamiltonian. We define the corresponding measure

$$\mathcal{F}_{\text{robust}}(\tau) = \left\langle \left| \frac{1}{D}\left\langle \tilde{U}_{\text{noise}}(\tau), \mathbb{I} \right\rangle_F \right|^2 \right\rangle, \qquad (14)$$

where the outer angle brackets $\langle \cdot \rangle$ denote an ensemble average over realizations of the noise processes, and the inner angle brackets $\langle \cdot, \cdot \rangle_F$ denotes a Frobenius inner product, defined by Eq. A1. Robustness is therefore evaluated as the noise-averaged operator distance between the error action operator Eq. 9 and the identity. This measure is bounded between $[0,1]$, with the robustness condition Eq. 13 perfect implemented when $\mathcal{F}_{\text{robust}}(\tau) = 1$. We define a corresponding infidelity measure as

$$\mathcal{I}_{\text{robust}}(\tau) = 1 - \mathcal{F}_{\text{robust}}(\tau) \qquad (15)$$

As above, a simple modification of this measure enables the robustness condition to be evaluated on a subspace of interest. Specifically

$$\mathcal{F}_{\text{robust}}^P(\tau) = \left\langle \left| \frac{1}{\text{Tr}\,(P)}\left\langle P\tilde{U}_{\text{noise}}(\tau), \mathbb{I} \right\rangle_F \right|^2 \right\rangle \qquad (16)$$

where $P$ defines a projection matrix, enabling robust control to be evaluated on a target subspace. And again, finding a robust control for a state transfer problem may be expressed

$$\mathcal{F}_{\text{robust}}^\psi(\tau) = \langle |\langle \psi_{\text{initial}}|\tilde{U}_{\text{noise}}|\psi_{\text{initial}}\rangle| \rangle. \qquad (17)$$

Note Eq. 14 does not include any information about whether $U_{\text{ctrl}}(\tau)$ implements a *particular* target gate $U_{\text{target}}$. In this sense, the robustness criterion is target-independent. Once again, solving these conditions requires a numerical approach subject to a multiobjective optimization routine. In practice, with sufficient control, it is always possible to satisfy both of these conditions and find a *robust* control that achieves the desired target operation with high fidelity. We tackle this problem using a number of novel approaches developed in Sec. III C, and demonstrated in real case studies in Sec. IV B.

### 3. Controllability

The number of controls required for complete control of a quantum system has previously been studied by Schirmer et. al [86]. When the controls are assumed to have unlimited bandwidth and power, the controllability of a quantum system can be determined by the Lie algebra generated by the control operators. If the Lie algebra generated by the controls spans the Hilbert space, the system is completely controllable. This result means, in some cases, far fewer controls are required to control a quantum system than the Hilbert space. Consider a quantum system of $m$ qubits in a Hilbert space of dimension $n = 2^m$. Let there be $2m$ single-qubit controls of the form $\sigma_i^x$, $\sigma_i^y$ for each qubit $i \in \{1, \ldots, m\}$. Additionally, let there be $m(m-1)/2$ two-qubit controls defined by coupling interactions of the form $\sigma_i^x\sigma_j^x$ for each qubit pair $(i,j)$. The total number of available controls therfore comes to $L = m(m+3)/2$. It can be shown that the Lie algebra generated by this set of control operators spans $\text{SU}(n)$. The number of controls required for complete controllability of this system therefore scales as $\mathcal{O}(\log_2(n)^2/2) < \mathcal{O}(n^2)$.

When the controls have limited bandwidth, time, and power the number of controls required for complete controllability can no longer be addressed analytically. An upper bound is set by $n^2 - 1$, the number of generators of the Lie algebra for $\text{SU}(n)$ (*e.g.* all Pauli operators for a single-qubit). In real quantum systems, however, the available controls comprise a subset of all possible controls, including only such interactions as are supported by the system architecture or control hardware.

With these device-dependent limitations in mind, it is not straightforward to make claims about how the number of required controls scale with the system dimension. The level of controllability must be made by accounting for the particular noise processes being targeted, the available controls supported by the system, and the limitations on evolution time imposed e.g. decoherence timescales. These are important considerations for producing realistic control solutions for physical devices.

### B. Performance evaluation for arbitrary controls

The evaluation of any measure for the fidelity of a robust-control operation as in Eq. 14 requires the computation of Eq. 10, which is generally challenging as control and noise Hamiltonians need not commute at different times. Characterizing control robustness and performance in realistic laboratory settings—especially for operations performed within large interacting systems—therefore requires simple, easily computed heuristics that aid a user in gaining intuition into control performance.

Here we introduce generalized multi-dimensional filter functions which serve as an engineering-inspired heuristic to determine noise susceptibility for arbitrary unitaries within high-dimensional Hilbert spaces. These objects express the noise-admittance of a control as a function of noise frequency, and reduce control selection to the examination of an easily visualized object similar to the Bode plot in classical engineering. Noise may be considered over a wide range of parameter regimes, from quasi-static (noise slow compared to $H_{\text{ctrl}}(t)$) to the limit in which the noise fluctuates on timescales comparable to or faster than $H_{\text{ctrl}}(t)$. We build on past single-qubit studies [33, 43, 46, 58, 87], assuming quasi-classical noise channels, to produce an explicit *basis independent* computational form incorporating all leading-order filter functions and extensible to higher-dimensional quantum systems [88] with enhanced computational efficiency.

#### 1. Modelling noise and error in D-dimensional systems

The error action operator $\tilde{U}_{\text{noise}}(\tau)$ for non-dissipative system-bath dynamics is treated as the unitary generated by an effective Hamiltonian $H^{(\text{eff})} = \Phi(\tau)/\tau$, such that

$$\tilde{U}_{\text{noise}}(\tau) \equiv e^{-i\Phi(\tau)}, \qquad \Phi(\tau) \equiv \sum_{\alpha=1}^{\infty} \Phi_\alpha(\tau). \qquad (18)$$

We obtain an arbitrarily accurate approximation for the unitary evolution using a Magnus series expansion [89, 90]. where the $\alpha$th Magnus term, $\Phi_\alpha(\tau)$, is computed as the sum of time-ordered integrals over permutations of $\alpha$th-order nested commutators of $\tilde{H}_{\text{noise}}(t_j)$, for $j \in \{1, ..., \alpha\}$ (see App. C 1).

Consider an arbitrary $D$-dimensional quantum system defined on the Hilbert space $\mathcal{H}$, and let the total control Hamiltonian in Eq. 1 have control and error terms of the form

$$H_{\text{ctrl}}(t) = \sum_{j=1}^{n} \alpha_j(t) C_j, \qquad (19)$$

$$H_{\text{noise}}(t) = \sum_{k=1}^{p} \beta_k(t) N_k(t). \qquad (20)$$

The control Hamiltonian, $H_{\text{ctrl}}(t)$, captures a target evolution generated by $n$ participating control operators, $C_j \in \mathcal{H}$. The noise Hamiltonian, $H_{\text{noise}}(t)$, captures interactions with $p$ independent noise channels. Distortions in the target evolution are generated by the noise operators $N_k(t) \in \mathcal{H}$, formally time-dependent such that $N_k(t) = 0$ for $t \notin [0, \tau]$. The noise fields $\beta_k(t)$ are assumed to be classical zero-mean wide-sense stationary processes with associated noise power spectral densities $S_k(\omega)$. The control-frame [33, 88] Hamiltonian takes the form

$$\tilde{H}_{\text{noise}}(t) = \sum_{k=1}^{p} \beta_k(t) \tilde{N}_k(t) \qquad (21)$$

where

$$\tilde{N}_k(t) \equiv U_{\text{ctrl}}^\dagger(t) N_k(t) U_{\text{ctrl}}(t) \qquad (22)$$

defines the noise operators in the control frame. Using a Magnus expansion as in Eq. 18, the noise action operator may then be approximated to the desired order.

For the purpose of calculating the error action operator Eq. 10 we are free to choose any gauge transformation of the form $\tilde{H}'_{\text{noise}} = \tilde{H}_{\text{noise}} + g\mathbb{I}$ which, up to a global phase, leaves the the dynamical evolution unchanged. With this freedom it is convenient to define the transformed Hamiltonian with the property that $\text{Tr}(P\tilde{H}'_{\text{noise}}) = 0$, namely tracelessness on the subspace associated with the projection matrix $P$, by choosing $g = -\text{Tr}(P\tilde{H}_{\text{noise}})/\text{Tr}(P)$. From Eq. 21, using the linearity of the trace and observing the noise variables $\beta_k(t)$ are scalar-valued for classical noise, we obtain

$$\tilde{H}'_{\text{noise}}(t) = \sum_{k=1}^{p} \beta_k(t) \tilde{N}'_k(t) \qquad (23)$$

where we define the *traceless* noise operators in the toggling frame as

$$\tilde{N}'_k(t) \equiv \tilde{N}_k(t) - \frac{\text{Tr}\left(P\tilde{N}_k(t)\right)}{\text{Tr}\left(P\right)}\mathbb{I}. \qquad (24)$$

Assuming the noise fields $\beta_k(t)$ are sufficiently weak, we truncate the Magnus expansion Eq. 18 at leading order and approximate the error action operator as

$$\tilde{U}_{\text{noise}}(\tau) \approx \exp[-i\Phi_1(\tau)]. \qquad (25)$$

Substituting into Eq. 16 the leading-order infidelity measure for robust control is approximated as

$$\mathcal{I}_{\text{robust}}(\tau) \approx \frac{1}{\text{Tr}(P)} \text{Tr}\left(P \left\langle \Phi_1(\tau)\Phi_1^\dagger(\tau) \right\rangle\right). \quad (26)$$

To obtain this expression we perform a Taylor expansion on Eq. 25, retain terms consistent with the leading-order approximation, and use the inherited property from Eq. 24 that $\text{Tr}(P\Phi_1) = 0$ (see App. C 2).

To compute the first order Magnus term we substitute Eq. 23 into Eq. C1, yielding

$$\Phi_1(\tau) = \sum_{k=1}^{p} \int_{-\infty}^{\infty} dt \beta_k(t) \tilde{N}'_k(t) \quad (27)$$

where we formally extend the limits of integration to $\pm\infty$, noting that $N'_k(t) = 0$ for $t \notin [0,\tau]$.

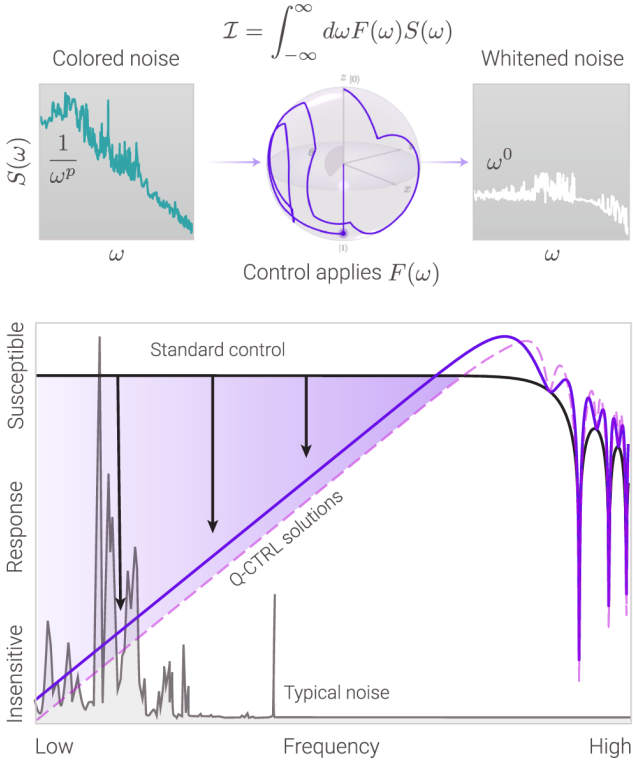

FIG. 3. Overview of the action of control as a noise filter at the operator level. Upper: An example of how colored noise enters expressions for the infidelity of a control operation as the overlap integral of the noise power spectrum and filter function for the control. A colored spectrum is thus whitened by the control through the physics of coherent averaging. Lower: Example filter function for an appropriately constructed noise-suppressing/filtering control. Such filters are low-frequency-noise suppressing; by reducing the filter function magnitude in a spectral range where the noise power spectrum is large, the fidelity of the operation is improved.

### 2. Multi-dimensional filter functions in the frequency domain

In order to efficiently compute Magnus contributions to the infidelity we move to the Fourier domain, and re-express contributions to error in a D-dimensional system

$$\Phi_1(\tau) = \sum_{k=1}^{p} \int_{-\infty}^{\infty} \frac{d\omega}{2\pi} G_k(\omega)\beta_k(\omega) \quad (28)$$

where the Fourier-domain functions

$$G_k(\omega) \equiv \mathscr{F}\{N'_k(t)\}(-\omega) \quad (29)$$
$$\beta_k(\omega) \equiv \mathscr{F}\{\beta_k(t)\}(\omega) \quad (30)$$

are defined according to the conventions set out in Eq. A5 and Eq. A5.

Substituting Eq. 28 into Eq. 26 then yields a compact expression for the leading-order robustness infidelity in the frequency-domain

$$\mathcal{I}_{\text{robust}}(\tau) \approx \frac{1}{2\pi} \sum_{k=1}^{p} \int_{-\infty}^{\infty} d\omega F_k(\omega) S_k(\omega) \quad (31)$$

Here, each noise channel $k$ contributes a term computed as an overlap integral between the noise power spectrum $S_k(\omega)$ and a corresponding filter function, $F_k(\omega)$. An approximation to the inclusion of higher-order Magnus terms for the infidelity may be obtained by exponentiating this expression, due to the similarity of the power-series expansion for an exponential function and the structure of the Magnus series [58].

The critical element for capturing the action of the control is the filter function, $F_k(\omega)$, relative to the $k$th noise channel. The explicit form of the filter function with respect to the projection matrix $P$ is defined (see App. C for details) as

$$F_k(\omega) = \frac{1}{\text{Tr}(P)} \text{Tr}\left(P G_k(\omega) G_k^\dagger(\omega)\right). \quad (32)$$

This expression may be simply recast in a form that is easily computed numerically, an essential task in software implementations. Let $p_l$ be the $l$th diagonal element of $P$, then the filter function may be expressed

$$F_k(\omega) = \frac{1}{\text{Tr}(P)} \sum_{l=1}^{D} p_l \sum_{q=1}^{D} \left|\left[G_k(\omega)\right]_{lq}\right|^2. \quad (33)$$

That is, take the Fourier transform of each matrix element of the time-dependent operator $\tilde{N}'_k(t)$, sum the complex modulus square of every element, weighted by the diagonal elements $p_l$, and divide through by $\text{Tr}(P)$, the dimension of the quantum system subspace. With this definition, we enable the efficient calculation of filter functions for single and multi-qubit gates, arbitrary high-dimensional systems, and complete circuits composed of multiple qubits and many operations. Thus we have a new computational device allowing the calculation of noise susceptibility for a wide range of elements relevant to quantum computation. An example application of this computational technique to the evaluation of noise susceptibility got a user-defined control, as realized in BOULDER OPAL, is presented in Algo. 1.

Given a noise power spectral density which represents realistic time-varying noise for a target noise operator (*e.g.* dephasing $\propto \sigma_z$), one may use the filter function to simply estimate operational fidelity; the net fidelity is given by the overlap integral of these two quantities as a function of frequency (Fig. 3). A high-fidelity control will minimize the filter function's spectral weight in frequency ranges where the noise power spectral density for a particular error channel is large. The predictive capabilities of this technique to control performance evaluation are experimentally validated for both single-qubit operations [58] and higher-dimensional systems (*e.g.* Mølmer-Sørensen gates [73]). An example of the predictive power of the filter function is presented in Fig. 8d,e for a variety of single-qubit controls.

---

**Algorithm 1** Filter function (FF)

---

$\{f\} \leftarrow \{f_1,...,f_n\}$ ▷ Arbitrary frequencies to evaluate FF
$H_{\text{ctrl}}(t) \leftarrow$ control Hamiltonian ▷ Eq. B1
$N_k(t) \leftarrow$ dynamical noise operator ▷ $k$th noise process
$P \leftarrow$ projection matrix
$\tau \leftarrow$ duration of control
$m \leftarrow$ samples
**function** FILTERFUNCTION($\{f\}$; $H_{\text{ctrl}}, N_k$ ) ▷ Abb. FF
   $\{H_{\text{ctrl},1},...,H_{\text{ctrl},m}\} \leftarrow$ SAMPLE($H_{\text{ctrl}}(t),\tau,m$)
   $\{N_{k,1},...,N_{k,m}\} \leftarrow$ SAMPLE($N_k(t),\tau,m$)
   $U_{\text{ctrl}} \leftarrow \mathbb{I}$
   **for** $i \in \{1,...,m\}$ **do**
      $\tilde{N}_{k,i} \leftarrow U_{\text{ctrl}}^{\dagger} N_{k,i} U_{\text{ctrl}}$ ▷ Eq. 22
      $\tilde{N}'_{k,i} \leftarrow \tilde{N}_{k,i} - \frac{\text{Tr}(P\tilde{N}_{k,i})}{\text{Tr}(P)}\mathbb{I}$ ▷ Eq. 24
      $U_{\text{ctrl}} \leftarrow \exp[-iH_{\text{ctrl},i}\Delta t] U_{\text{ctrl}}$ ▷ $\Delta t = \tau/(m-1)$
   **end for**
   $\{\tilde{N}'_k\} \leftarrow \{\tilde{N}'_{k,1},...,\tilde{N}'_{k,m}\}$ ▷ $t$-domain: Eq. 24
   $\{G_k\} \leftarrow$ DTFT($\{\tilde{N}'_k\},\Delta t,\{-f\}$) ▷ $f$-domain: Eq. 29
   **for** $G_{k,\nu} \in \{G_k\}$ **do**
      $F_{k,\nu} \leftarrow \frac{1}{\text{Tr}(P)} \sum_{l=1}^{D} p_l \sum_{q=1}^{D} \left| \left[ G_{k,\nu} \right]_{lq} \right|^2$ ▷ Eq. 33
   **end for**
   Returns $\{F_{k,\nu}\} \leftarrow \{F_{k,1},...,F_{k,n}\}$
**end function**

· · · · · · · · · · · · · · · · · · · · · · · · · · · · · · · · · · · · · · · · ·

**function** SAMPLE($A(t),\tau,s$)
   $A(t)$: array-valued function of time
   $t_i \leftarrow i\Delta t$ for $i \in \{0,...,s-1\}$, $\Delta t = \tau/(s-1)$.
   Returns $\{A(t_0),...,A(t_{s-1})\}$
**end function**

**function** DTFT($\{A\},\Delta t,\{f\}$)
   $\{A\}$: samples of array-valued function, $A(t)$ ▷ length $s$.
   $\Delta t$ : time between samples
   $\{f\}$: arbitrary frequencies ▷ length $n$
   Returns discrete-time Fourier transform of $\{A\}$ at $\{f\}$
**end function**

---

## C. Flexible optimization tools for quantum control

Precise manipulation and characterization of quantum systems has emerged as a key area of development for quantum physics and chemistry. In most settings - whether addressing questions of unitary-control in high-dimensional Hilbert spaces or implementing Hamiltonian parameter estimation - key tasks rapidly become analytically intractable and require the use of numeric optimization techniques. We have developed a versatile optimization engine (the optimizer) based on a GPU-compatible graph architecture coded in TensorFlow, compatible with (but not limited to) the efficient computational heuristics introduced in Sec. III B.

This toolkit enables rapid creation of high-fidelity unitary operations spanning both *optimal* and *robust control* in high-dimensional Hilbert spaces. Creation of an optimized control solution may be undertaken for individual gates, small interacting subcircuits, or complete algorithms. All such circumstances are efficiently incorporated using the system definition introduced here without the need for a change in the underlying toolkit.

Beyond broad applications in the optimization of unitary operations, the optimizer has emerged as a versatile tool used throughout the software packages described in Sec. II. For example, in Sec. III E we present an algorithm for noise spectral estimation based on convex optimization. The convex optimization procedure in this algorithm may be implemented using the flexible optimization engine, simply by expressing the convex objective function in terms of TensorFlow operations. The same tools are also used for Hamiltonian parameter estimation discussed in Sec. III E 2. Here the computational task is to identify system parameters given a set of input controls. This problem reduces to optimization of an objective function that maps candidate parameter values to the deviation between expected and measured system response. These simple examples demonstrate the flexibility and value of the optimizer engine across a wide range of tasks.

A technical description of the essential framework is presented below, covering the structure of the core optimization engine, parameterization of control variables, definition of cost functions, and efficient incorporation of a wide range of constraints. In Sec. IV we demonstrate these capabilities using higher-dimensional superconducting systems as important case studies.

### 1. Flexible optimizer framework

Mathematically we define the optimization problem as follows. Let $C(\boldsymbol{v})$ denote the cost function for optimization, where $\boldsymbol{v} = (v_1, v_2, ...)$ denotes an array of generalized control variables. The framework enforces no additional structure on the cost function. To benefit from gradient ascent methods it is necessary to calculate all partial derivatives of the gradient function

$$\vec{\boldsymbol{\nabla}}_{\boldsymbol{v}} C = \left( \frac{\partial C}{\partial v_1}, \ \frac{\partial C}{\partial v_2}, \ \dots \right). \tag{34}$$

In general calculating $\vec{\boldsymbol{\nabla}}_{\boldsymbol{v}} C$ is a complex computation requiring many applications of the chain rule, with strong dependence on the specific form of the cost function. These difficulties are naturally overcome, however, using TensorFlow as the optimizer framework. This benefits from an in-built gradient calculator based on the underlying tensor map, and machine-learning algorithms for minimizing the cost-function. Moreover, TensorFlow permits the calculation of *nonlinear* gradients. This is particularly relevant for systems where modulation of a given control variable does not simply modulate the associated Hamiltonian term linearly. For example, a parametrically-driven entangling gate for superconducting transmon qubits is implemented by modulating a flux drive in the lab frame, mapping to an interaction in the quantum system with *effective* coupling strengths functionally dependent on Bessel functions (see [Eq. 81](#) [91–93].

The optimizer naturally benefits from these advantages by programming in TensorFlow, however we do not employ in-built TensorFlow optimization routines; the optimizer is custom-built. In order to perform optimizations, controls must be parameterized and cost functions defined; the mappings $\boldsymbol{v} \mapsto C(\boldsymbol{v})$ and $\boldsymbol{v} \mapsto \vec{\boldsymbol{\nabla}}_{\boldsymbol{v}} C$ are then passed directly into standard gradient-based optimization algorithms, for example L-BFGS-B [94]. To support the general goal of multi-objective optimization, various cost metrics may be combined in a linear combination

$$C(\boldsymbol{v}) = \sum_{\mu} w_{\mu} C_{\mu}(\boldsymbol{v}), \tag{35}$$

where each component $C_{\mu}(\boldsymbol{v})$ measures a distinct aspect of the target performance as a function of the control variables $\boldsymbol{v}$, and the constants $w_{\mu}$ weight the relative importance of these contributions in the optimized result. This capability will be exploited in the implementation of the features described next.

### 2. Flexible optimizer features

The Q-CTRL optimizer, in addition to providing the infrastructure required for linking a user-defined TensorFlow cost function with a gradient-based optimization algorithm, provides a collection of convenience methods for automatically building the critical parts of the cost function. These methods abstract away the low-level details of common but non-trivial computations (for example efficient numerical integration of the Schrödinger equation), allowing the cost function to be composed from higher-level intuitive "building blocks". This design encapsulates the details of frequently employed yet complicated steps of the cost-function calculation within the internally-defined convenience methods. As a result these steps can be implemented efficiently and the code written by the user can be focused on describing the specific system under consideration, rather than implementing general-purpose algorithms. Importantly, these convenience methods do not constrain the types of cost function that can be represented; where a convenience method does not exist, arbitrary TensorFlow operations may be used instead. See [Table I](#) for examples of important functional descriptions of commonly used cost-function components.

These convenience methods, or building blocks, are designed around three primary data types: tensors (pure multi-dimensional arrays of numbers), piecewise-constant (PWC) scalar-valued functions of time, and PWC operator-valued functions of time. Starting from the raw control variables $\boldsymbol{v}$, which are tensors, a representation of the target system and cost function can be built by applying a chain of these methods.

A major consideration in the application of control is the generation of solutions that are practical to implement on real hardware, which motivates the inclusion of features which effectively constrain the optimization procedure. Limiting the search space via implementation of appropriately constructed constraints can assist in ensuring that controls meet hardware limitations, and also dramatically improve the general efficiency of the optimization problem. Such constraints may be naturally incorporated into the optimization framework via the definition of the cost function, as shown in [Table I](#). We have focused on providing a range of convenience methods to model constraints that meet the demands imposed by physical hardware limitations (see [Table II](#)). These methods allow any combination of constraints to be incorporated into the description of the system when creating the cost function.

One of the most important constraints to be considered is smoothing of control waveforms to accommodate bandwidth limits and finite response times from hard-ware. In general, smoothed solutions can be achieved through a number of supported techniques such as constraining the effective time-derivative of the control, or ensuring that all optimized waveforms incorporate linear time-invariant filters such as RC or sinc function. For example, as shown in Algo. 2, a band-limit constraint can be implemented simply by introducing a transformation on the signal prior as part of the cost function. Another example in which such a transformation incorpo-rates an RC-filter for the pulse waveform is highlighted in Fig. 11d. Importantly, as smoothed waveforms are eventually discretized in time for output on arbitrary waveform generators, the optimizer can include temporal discretization in order to ensure the optimal gate is produced by the sampled waveform.

| Optimizer cost type | Functional form: cost $C_{\mu}(\boldsymbol{v})$ | Description |
|---|---|---|
| Optimal | $\mathcal{I}_{\mathrm{optimal}}$ | Noise-free target unitary over $D$-dimensions |
| Robust | $\frac{1}{2\pi} F_k(0)$ | Quasi-static noise/ constant-offset |
| Robust | $\frac{1}{2\pi} F_k(\omega)$ | Fixed frequency noise suppression at $\omega$ |
| Robust | $\int_{\omega_1}^{\omega_2} S_k(\omega) F_k(\omega) \frac{d\omega}{2\pi}$ | Broadband noise suppression over $[\omega_1, \omega_2]$ |

TABLE I. Component cost functions to be included as desired in [Eq. 35](#) for an optimization task. Here $F_k(\omega)$ and $S_k(\omega)$ are the filter function and noise power spectral density respectively, associated with the $k$th noise channel.

| Optimizer convenience feature | Technical details |
|---|---|
| **Smooth controls:** The temporal variation in a control waveform is bounded to limit discontinuous transitions requiring high bandwidths. | A "smooth" waveform may be obtained via one of multiple methods: (i) Limiting the effective time-derivative for any signal, $$\|\alpha(\tau_i) - \alpha(\tau_{i-1})\| < (\delta\alpha)_{\max}.$$ (ii) Composing candidate control waveforms as superpositions in a basis (Fourier, Slepian [95], etc) before discretized sampling (CRAB). (iii) Passing non-smooth waveforms through filters prior to inclusion in the Hamiltonian (see below). |
| **Filtered controls:** Control pulses are transformed by a linear-time-invariant filter. | Discrete high-bandwidth pulses may be transformed into filtered waveforms using arbitrary linear time-invariant filters such as RC filters with specified high-frequency cutoff, $\omega_{\max}$, a sinc window function, or user-defined filters. The transformed waveform enacts the optimized control, and can include the time-discretization ultimately required for output on hardware, ensuring the sampled waveform remains optimized. Alternatively, the effect of known filters on control lines can be incorporated into the system definition, in order to find optimized controls that compensate for the effect of the filters. |
| **Symmetrized controls:** Control pulses are temporally symmetrized about the midpoint of the control. | Controls can be simplified via temporal symmetrization in order to produce waveforms which comprise half of the desired number of segments. In certain cases this may improve the overall efficacy of the desired solution. |
| **Bounded-strength controls:** The magnitude of a pulse waveform may be limited to ensure optimized solutions do not exceed physically motivated bounds. | Hard bounds may be enforced on any control variables. In particular, these bounds may be used to constrain the maximum value of signal waveforms, such that $$\|\alpha(t)\| \leq \alpha_{\max}$$ where $\alpha(t)$ is some signal of interest and $\alpha_{\max}$ defines its maximum (positive) permissible value. |
| **Fixed-control waveforms:** For any individual control, the pulse waveform may be held fixed and effectively frozen out of the variational search. | Pulse waveforms need not be functions of the control variables, and instead may be specified by fixed values. This functionality enables support for systems with time-dependent terms that should not be tuned by the optimizer (for example if they cannot be accessed by the control hardware). |
| **Concurrent vs interleaved controls:** Control pulses on different drives and shifts are executed sequentially or simultaneously. | In certain physical systems it is not possible to implement all controls simultaneously. This constraint involves transforming the optimization variables as $\boldsymbol{v} \rightarrow \boldsymbol{v} \cdot \boldsymbol{b}$, where $\boldsymbol{b}$ is a binary mask enforcing the required structure of interleaved operations. For $\boldsymbol{b}$ set to unity, controls may be applied concurrently. |

TABLE II. Optimization features captured through convenience functions available in the package. See Ref. [96] for example code and Algo. 2 for an example implementation.

Another challenge faced in almost any quantum control problem is numerical integration of the Schrödinger equation to calculate the time evolution of the system.

In our framework, this integration forms a step in the cost-function calculation like any other, and may thus be customized by the user in order to best meet the demands of their particular optimization problem. The framework offers several built-in GPU-optimized integration routines, based on matrix exponentiation (for piecewise-constant controls) and Runge-Kutta integration (for smooth controls or large systems for which full exponentiation is infeasible). For instance, one may consider a waveform distorted by a transmission line with a well-characterized response function. This response may be incorporated into the optimization using Runge-Kutta integration such that the transformed waveform still provides optimal dynamics at the quantum hardware. Such approaches and the associated convenience features are particularly valuable in the Hamiltonian parameter estimation routine employed in Sec. III E.

Finally, we also offer a CRAB-type [97, 98] optimization in which a waveform is selected from a superposition in a user defined basis and discretized in time. Such a representation fits naturally into our framework, where waveforms may be represented as arbitrary functions of control variables. This approach truncates the effective search space by limiting it to the associated Fourier coefficients, and is therefore independent of the granularity of the piecewise-constant discretization. The optimizer contains a flexible CRAB implementation that allows a variety of CRAB techniques (*e.g.* bases with randomized frequencies, fixed frequencies, optimizable frequencies, or user-defined bases [95]).

All of these features fit into the flexible optimization framework presented above, and may thus be arbitrarily combined to produce optimizable models of a wide variety of systems. Further description of the implementation of these features are provided in Table II, and detailed code-based demonstrations are available online [96].

As a concrete example we consider the creation of an optimized unitary operation manipulating a qubit. In a standard optimal control context, one typically seeks to minimize a single noise-free fidelity metric. Here, the control variables parameterize the control Hamiltonian $H_{\text{ctrl}}(\boldsymbol{v})$, such that the cost function obeys the functional dependency $C(\boldsymbol{v}) = C(U_{\text{ctrl}}(\boldsymbol{v}, \tau))$, where $U_{\text{ctrl}}(\boldsymbol{v}, \tau)$ is the evolved unitary after time $\tau$. To produce optimized controls that account for the impact of noise, however, one must introduce additional terms in the cost function to penalize controls that achieve a high-quality gate in a manner that is not robust to noise. Similarly, the definition of the cost function may include other constraints as articulated above.

---

**Algorithm 2** Sample optimization
---
**function** SampleCostFunction($\boldsymbol{v}$)
    $\tau \leftarrow$ total duration
    $\omega_{\text{cutoff}}$ band-limit for pulse
    $m \leftarrow$ number of optimizable pulse segments
    $U_{\text{target}} \leftarrow$ target gate
    $\alpha_0(t) \leftarrow$ PwcScalar($\tau, \boldsymbol{v}$)
    $\mathcal{K}(t) \leftarrow$ SincKernel($\omega_{\text{cutoff}}$)
    $\alpha(t) \leftarrow$ LtiFilter($\alpha_0(t), \mathcal{K}(t), m$)
    $\alpha(t)\sigma_x \leftarrow$ PwcOperator($\alpha(t), \sigma_x$)
    $H(t) \leftarrow \alpha(t)\sigma_x$
    $C \leftarrow$ OptimalCost($H(t), U_{\text{target}}$)
    Returns $C, \{\alpha(t)\}$
**end function**

**procedure** SampleOptimization
    $C(\boldsymbol{v}), \{\alpha(t)(\boldsymbol{v})\} \leftarrow$ SampleCostFunction
    $\boldsymbol{v}_{\text{optimized}} \leftarrow$ Optimize($C(\boldsymbol{v})$)
    $\alpha_{\text{optimized}}(t) \leftarrow \alpha(t)(\boldsymbol{v}_{\text{optimized}})$
    Returns $\{\alpha_{\text{optimized}}\}$ ▷ optimized, band-limited control
**end procedure**
························································

**Built-in methods for building and optimizing cost functions:**

**function** PwcScalar($\tau, \boldsymbol{\alpha}$)
    Returns PWC scalar $\alpha(t)$ taking value $\alpha_i$ on segment $i$
**end function**

**function** SincKernel($\omega_{\text{cutoff}}$)
    Returns kernel $\mathcal{K}(t)$ for a sinc filter with $\omega_{\text{cutoff}}$
**end function**

**function** LtiFilter($\alpha(t), \mathcal{K}(t), m$)
    Returns $m$-segment PWC discretization of $(\alpha * \mathcal{K})(t)$
**end function**

**function** PwcOperator($\alpha(t), A$)
    Returns PWC operator $A(t) = \alpha(t)A$
**end function**

**function** PwcOperatorSum($\{A_l(t)\}$)
    Returns PWC operator $A(t) = \sum_l A_l(t)$
**end function**

**function** OptimalCost($H(t), U_{\text{target}}$)
    Returns $\mathcal{I}_{\text{optimal}}$ for Hamiltonian $H(t)$ and target $U_{\text{target}}$
**end function**

**function** QuasiStaticRobustCost($H(t), \{N_k(t)\}$)
    Returns filter function values $\sum_k \frac{1}{2\pi} F_k(0)$
**end function**

**function** Optimize($C(\boldsymbol{v})$)
    Returns optimized values of $\boldsymbol{v}$
**end function**

$\cdots$

---

We provide an algorithmic example of such an optimization task for a single-qubit unitary in Algo. 2, in which we construct the cost function for realizing a band-limited optimized control pulse. First, the raw control variables can be converted to a PWC scalar, representing a (non-band-limited) control signal, by using the control variables as the per-segment scalar function values. Then, this raw signal can be convolved with a sinc filter kernel with a specific cutoff frequency, and re-discretized to a PWC scalar. This new signal is band-limited, and discretized in order to be implementable on real hardware. Next, the new signal is multiplied by a constant operator to represent a full Hamiltonian term. If necessary, multiple Hamiltonian terms can be similarly constructed, and summed to yield the overall Hamiltonian. Finally, the optimal cost is computed for the given Hamiltonian and target gate (see Table I for alternative costs). This cost function may then be passed to the optimizer, and the discretized band-limited signal extracted from the optimized system. Importantly, it is this band-limited signal that defines the optimized gate, and therefore the evaluation of the cost function—the initial non-band-limited signal is used merely as an intermediate step between the raw control variables and the signal of interest.

### 3. Optimizer performance benchmarking

In addition to application flexibility, the Q-CTRL optimizer provides advantages in time-to-solution. As shown in Fig. 4, in head-to-head performance benchmarking of local-instance implementations we find greater than two orders-of-magnitude performance improvements over an internal naïve optimizer based on NumPy, and $\sim 3-5\times$ typical advantage relative to optimization tools in the open-source QuTiP package [99, 100] for the representative problems treated here. The performance advantages of the Q-CTRL local instance implementation vary with the details of the selected system, but in all circumstances studied are seen to grow with Hilbert-space dimension, number of controls, and number of time-segments in a solution.

Additional benefits may be gained via implementation using customized cloud-compute resources for complex optimization tasks; support for these resources is a standard part of the BOULDER OPAL package introduced in Sec. II. We observe a fixed overhead of approximately three seconds associated with web-access and data-upload latencies, meaning that the Q-CTRL local instance outperforms cloud-based computations for simple optimization tasks (small Hilbert spaces with low segment counts). However, for Hilbert-space dimensions associated with problems spanning three to seven qubits, the benefits of the cloud compute engine are manifested as an approximately $10\times$ reduction in optimization runtime. Beyond seven qubits (equivalently Hilbert space dimension 128) the cloud-engine automatically routes calculations to a GPU, which changes the performance scaling with Hilbert-space dimension. In this regime extrapolated performance benefits relative to the local instance implementation approach two orders of magnitude; we have not performed optimizations using competing packages with durations beyond one hour. Full details of the Hamiltonian used in the optimization, software package versions, computational hardware employed, etc. are described in App. D.

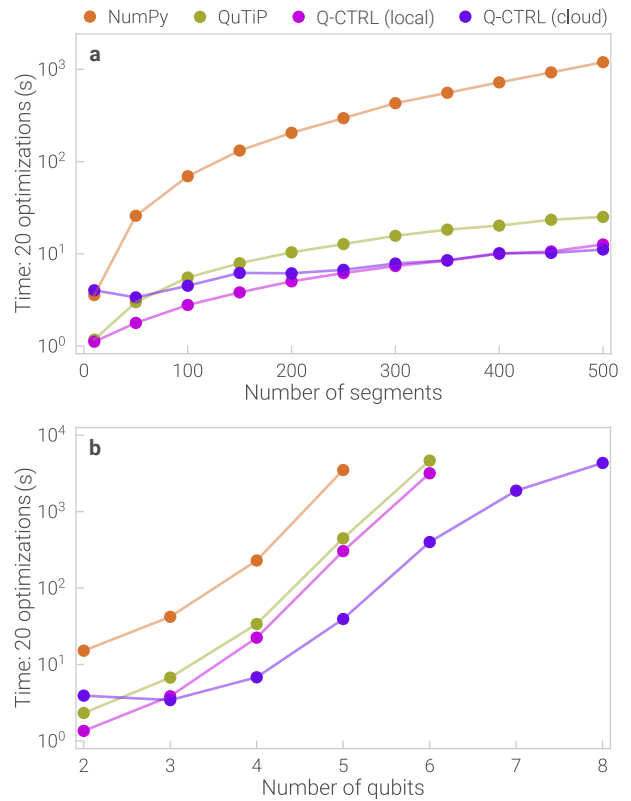

FIG. 4. Performance benchmarking of various optimization tools. Time-to-solution for 20 optimization runs is presented as a function of the optimization complexity, where the latter is measured by (a) number of control segments in fixed Hilbert space or (b) Hilbert-space size measured in qubits with a fixed control complexity. Panel (a) treats a four-qubit system with three-axis control applied to a single qubit within the larger space. Panel (b) considers a Rydberg atom array with two controls of 40 segments each. Cloud-based computation for the Q-CTRL package incurs a fixed overhead of approximately three seconds. See App. D for additional details.

### D. Time-domain simulation tools for realistic hardware error processes

A useful approach for analysing the dynamics of an algorithm or gate in the presence of noise is via time-domain simulation. If the noise-free evolution of the system is well-understood, simulation may be used to investigate system dynamics in the presence of different noise sources.

Simulation packages based on Schrödinger integration and matrix multiplication are a common feature of many existing software packages, including various open-source platforms. QuTiP [99, 100] supports numerical simulations of a wide variety of time-dependent open and closed quantum systems, enabling noise modelling through its `qutip.qip.noise` module. Krotov [101] implements gradient-based optimization algorithms based on Krotov's method, useful for exploring the limits of controllability in a quantum system. ProjectQ [17–20] provides a quantum computer simulator with emulation capabilities, equipped with various compiler plug-ins. pyGSTi [102] offers noise-modelling and characterization of single- or multi-qubit systems, with a noise model including stochastic gate errors and SPAM errors. The Ignes module within IBM's Python package Qiskit [2, 21] includes tools to simulate gate and small-circuit performance, as well as measure certain noise parameters. Similarly, pyQuil [4, 24, 25] is a Python library for executing and simulating programs via Quil [24], the compiler language developed by the Rigetti Computing.

For the most part, however, available simulation tools incorporate noise dynamics via quasi-static offsets, or fully stochastic depolarizing models. With the objective of supplementing existing numerical simulation packages, Q-CTRL provides efficient tools for simulating the dynamic evolution of arbitrary quantum systems subject to a broad class of noise processes. This includes common channels typically encountered in realistic laboratory environments such as correlated and colored semi-classical noise processes. User-defined noise PSDs may be used to characterize arbitrary noise generators in the Hamiltonian, enabling the user to simulate the impact on algorithmic performance.

The core Q-CTRL simulation module accepts a control Hamiltonian (expressed as drives, shifts and drifts, as described in App. B), together with any number of arbitrary piecewise-constant time-domain noise processes. These noise processes can multiplicatively perturb the moduli of the drive or shift controls, or contribute additively to the system Hamiltonian. From this information, the simulation module produces an overall piecewise-constant system Hamiltonian, and solves the Schrödinger equation via matrix exponentials to compute the unitary time evolution operator for the system at arbitrary times.

This package provides several key functions that enable efficient and useful simulation in noisy environments:

- Creation of a time-domain noise process from an input user-defined noise power spectral density.

- Incorporation of a user-defined time-series into a simulation, including data-series interpolation.

- Automated homogenization of time-segmentation of all input and software-defined time series in order to permit Schrödinger integration from data sets expressing different temporal discretization.

- Forward propagation of an initial input state subject to calculated time-evolution operators including noise.

- Calculation of ensemble-averaged density matrices over independent but statistically identical noise realizations.

In the remainder of this subsection we describe technical details of each of these functions.

### 1. Technical details of simulation functionality

A key function of the simulation package is to generate time-domain signals consistent with realistic noise processes observed in physical hardware, *e.g.* oscillator phase noise or magnetic field noise, typically characterized by a measurable noise PSD. Let the underlying noise spectral density of interest be denoted $S^{(1)}(\omega)$, defined as a *one-sided* PSD ($\omega \geq 0$), *e.g.* consistent with standard measurements from a spectrum analyzer. A physical measurement of this PSD takes the form of a discrete data series $\{S_k^{(1)}\}$ of $N$ samples, where $S_k^{(1)} \approx S^{(1)}(k\Delta\omega)$ for $k \in \{0, ..., N-1\}$, and $\Delta\omega$ defines the frequency resolution of the measurement. Samples from the corresponding two-sided spectrum are defined by symmetrizing and rescaling the one-sided spectrum as

$$S_k^{(2)} \equiv \begin{cases} S_0^{(1)} & \text{for } k = 0 \\ \frac{1}{2}S_k^{(1)} & \text{for } k = [1, \ldots, N-1] \\ \frac{1}{2}S_{2N-1-k}^{(1)} & \text{for } k = [N, \ldots, 2N-2] \end{cases} \quad (36)$$

such that $|\{S_k^{(2)}\}| = 2N - 1$. The corresponding discrete *amplitude spectral densities*, $\{X_k\}$, are defined such that $S_k^{(2)} = |X_k|^2$, permitting arbitrary choice of the complex phase of each $X_k$. Consequently,

$$X_k \equiv e^{i\phi_k}\sqrt{S_k^{(2)}}, \quad (37)$$

where

$$\phi_k = \begin{cases} 0 & \text{for } k = 0 \\ \sim \text{unif}(-\pi, \pi) & \text{for } k = [1, \ldots, N-1] \\ -\phi_{2N-1-k} & \text{for } k = [N, \ldots, 2N-2] \end{cases} \quad (38)$$

and the constraints imposed by the first and last cases ensure that $\{X_k\}$ has Hermitian symmetry, and thus corresponds to the spectrum of a *real* time domain process. The time series, $\{x_j\}$, generated by a given realization of $\{X_k\}$ is then obtained via a suitably-normalized inverse discrete Fourier transform, such that

$$x_j = \sqrt{\Delta\omega} \sum_{k=0}^{2N-2} X_k e^{2\pi i \frac{jk}{2N-1}}. \quad (39)$$

This yields a single random realization of a real-valued time-domain signal with a power spectrum matching the input spectrum $S^{(2)}(\omega)$.

---

**Algorithm 3** Simulator

---

$\{t\} \leftarrow$ times at which to simulate dynamics     ▷ Arbitrary
**procedure** CONTROLNOISE     ▷ Table IV
    drives $\leftarrow \{(\gamma_j(t), C_j) \mid$ for $j \in \{1,...d\}\}$
    shifts $\leftarrow \{(\alpha_l(t), A_l) \mid$ for $l \in \{1,...s\}\}$
    drift $\leftarrow D$
    **for** $(q(t), Q) \in$ drives, shifts **do**
        $\{q_s, \tau_s\} \leftarrow$ distinct segments for $q(t)$
        **if** Noise = True **then**
            $\{S^{(1,q)}\}, \Delta\omega^{(q)} \leftarrow$ sampled PSD for $q(t)$ noise
            $\{\delta q_t\} \leftarrow$ NOISESIGNAL($\{S^{(1,q)}\}, \Delta\omega^{(q)}, \{t\}$)
            $\{q_t\}, \{\delta q_t\} \leftarrow$ JOINTSEGMENTS($\{q_s\}, \{\delta q_t\}$)
            $\{q'_t\} \leftarrow \{q_t\} + \{\delta q_t\}$
        **end if**
    **end for**
    drives$' \leftarrow \{(\{\gamma'_{j,t}\}, C_j) \mid$ for $j \in \{1,...d\}\}$
    shifts$' \leftarrow \{(\{\alpha'_{l,t}\}, A_l) \mid$ for $l \in \{1,...s\}\}$
    drift$' \leftarrow$ drift
    $\{H_t^{\text{noisy-ctrl}}\} \leftarrow$ HAMILTONIAN(drives$'$, shifts$'$, drift$'$)
**end procedure**
**procedure** ADDITIVENOISE
    **for** $k \in \{1, ..., p\}$ **do**
        $N_k \in N_{\text{additive}}$     ▷ Algo. 1
        $\{S^{(1,k)}\}, \Delta\omega^{(k)} \leftarrow$ sampled PSD for additive noise
        $\{\beta_{k,t}\} \leftarrow$ NOISESIGNAL($\{S^{(1,k)}\}, \Delta\omega^{(k)}, \{t\}$)
    **end for**
    $\{H_t^{\text{add-noise}}\} \leftarrow \sum_{k=1}^{p} \{\beta_{k,t} N_k\}$     ▷ Eq. 20
**end procedure**
**procedure** SIMULATE
    $\{H_t^{\text{tot}}\} \leftarrow \{H_t^{\text{noisy-ctrl}}\} + \{H_t^{\text{add-noise}}\}$
    $\{U_t\} \leftarrow$ UNITARYEVOLUTION($\{H_t^{\text{tot}}\}, \{t\}$)
    $\{\psi_t\} \leftarrow \{U_t |\psi_0\rangle\}$     ▷ Eq. 46
**end procedure**

................................................................

**function** NOISESIGNAL( $\{S^{(1)}\}, \Delta\omega, \{t\}$ )
    Returns $\{x_t\}$     ▷ Eq. 40
**end function**

**function** JOINTSEGMENTS( $\{A_a, \tau_a\}_{a=1}^{N_a}, \{B_b, \tau_b\}_{b=1}^{N_b}, ...$ )
    series: $\{A_a\}$ and $\{B_b\}$
    segment durations: $\{\tau_a\}$ and $\{\tau_b\}$.
    joint segmentation: $\{A_j, B_j, \tau_j\}$     ▷ Eq. 41
    Returns $\{A_j, \tau_j\}_{j=1}^{N_j}, \{B_j, \tau_j\}_{j=1}^{N_j}, ...$
**end function**

**function** HAMILTONIAN(drives, shifts, drift)
    Returns $H_{\text{ctrl}}(t)$     ▷ Eq. B1
**end function**

**function** UNITARYEVOLUTION( $\{H_j, \tau_j\}, \{t\}$ )
    $\{H_j, \tau_j\}$ : segmented Hamiltonian
    $j$th segment: $\tau_j = t_j - t_{j-1}$     ▷ Eq. 42
    $\{t\}$: arbitrary times to U(t).
    Returns $\{U_t\}$     ▷ Eq. 43
**end function**

---

Given this form of a discrete, real, time series generated from a noise power spectral density (created as above or provided directly by the user), it may be desirable to perform simulation using a higher sampling rate than that native to the data (for example if only low-frequency noise is specified). The Q-CTRL simulation package enables this upsampling via Whittaker-Shannon interpolation. This produces a continuous-time function that interpolates the discrete time series, with a bandlimit set by the Nyquist frequency of the discrete data. This takes the functional form

$$x(t) = \sum_{k=-\infty}^{\infty} x_k \text{sinc}\left(\frac{t - k\Delta t}{\Delta t}\right), \qquad (40)$$

where $\Delta t$ is the time step between discrete samples in $\{x_k\}$. To approximate the infinite sum, the simulation package automatically performs periodic extension of the input series and truncation of the sum to accuracy within the domain of the original time series. Using Eq. 40, the discrete time series $\{x_k\}$ may then be resampled at arbitrary times $\{t\}$, yielding the upsampled (or otherwise) time-series $\{x_t\}$.

With discretized time-series data in hand it becomes possible to simulate the time evolution of a system via integration of the Schrödinger equation. However, in many cases the natural temporal discretizations will vary between different fields within the system. For example, rapidly-fluctuating noise sources may be defined on significantly shorter time scales than control fields, while quasi-static noise processes could be defined on longer time scales.

To enable simulation in such cases, all discretizations are automatically resampled on a shared grid prior to integration. This enables a user to simply input data series as-is and the package will handle all homogenization issues. For example, if a drive control pulse $\Omega(t)$ is defined on two segments of duration $\tau/2$ by $[\Omega_1, \Omega_2]$, but a noise process $\beta(t)$ is defined on three segments of duration $\tau/3$ by $[\beta_1, \beta_2, \beta_3]$, the joint discretization has six segments of duration $\tau/6$ defined by

$$
\tau/2 \overset{\Omega(t)}{\begin{bmatrix} \Omega_1 \\ \Omega_2 \end{bmatrix}} + \tau/3 \overset{\beta(t)}{\begin{bmatrix} \beta_1 \\ \beta_2 \\ \beta_3 \end{bmatrix}} \rightarrow \begin{matrix} \tau/6 \\ \tau/6 \\ \tau/6 \\ \tau/6 \\ \tau/6 \\ \tau/6 \end{matrix} \overset{\Omega(t) \quad \beta(t)}{\begin{bmatrix} \Omega_1 & \beta_1 \\ \Omega_1 & \beta_1 \\ \Omega_1 & \beta_2 \\ \Omega_2 & \beta_2 \\ \Omega_2 & \beta_3 \\ \Omega_2 & \beta_3 \end{bmatrix}}. \quad (41)
$$

Assuming the time-domain is jointly partitioned in this way, with respect to the various time-series of interest, the total Hamiltonian may be perfectly expressed in terms of $N$ piecewise-constant segments on the resampled time-domain, taking the form

$$
H(t) = \begin{cases} H_1 & \text{for } t \in [t_0, t_1] \\ \vdots \\ H_k & \text{for } t \in [t_{k-1}, t_k] \\ \vdots \\ H_N & \text{for } t \in [t_{N-1}, t_N] \end{cases} \quad (42)
$$

where $t \in [t_{k-1}, t_k]$ defines start and end times of the $k$th segment, for $k \in \{1, ..., N\}$, and where $t_0 \equiv 0$ and $t_N \equiv \tau$. Computing the unitary time-evolution operator $U(t, t_0)$ via Schrödinger integration is then equivalent to evaluating the matrix exponential product

$$U(t, t_0) = U(t, t_k)Q(t_k, t_0), \quad \text{for} \quad t \in [t_k, t_{k+1}] \quad (43)$$

$$U(t, t_k) \equiv e^{-iH_k(t - t_k)} \qquad \text{for} \quad k \in \{1, ..., N\} \quad (44)$$

$$Q(t_k, t_0) \equiv \prod_{i=1}^{k} U(t_i, t_{i-1}). \quad (45)$$

From Eq. 43 the unitary time-evolution operator may then be computed for arbitrary sample times yielding the time series $\{U_t\}$, where $U_t \equiv U(t, t_0)$. Using this functionality the Q-CTRL simulation package provides a function to propagate an given initial state $|\psi_0\rangle$ and evaluate the evolved state at arbitrary sample times within the desired evolution period, computed as

$$|\psi_t\rangle = U_t|\psi_0\rangle. \quad (46)$$

In general, however, calculating a single instance of the temporal evolution of the state is insufficient to understand the target dynamics, and an ensemble average over different noise realizations is required. The Q-CTRL simulation package provides a function to compute the mean density matrix associated with an ensemble of propagated state vectors. Given a set of state vectors $\{|\psi^m\rangle\}$ (for $1 \leq m \leq M$) produced from an ensemble of simulations corresponding to different noise realizations, the mean density matrix $\rho$ is given by

$$\rho = \frac{1}{M} \sum_{m=1}^{M} |\psi^m\rangle\langle\psi^m|. \quad (47)$$

### 2. Simulation example

These capabilities are demonstrated in Fig. 5. We model a superconducting qubit as an anharmonic three-level system incorporating leakage, and simulate the time-evolution under a control pulse implementing a NOT gate via Gaussian Half-DRAG (derivative removal by adiabatic gate) [103]. The simulation also incorporates multiple time-dependent noise processes each described by a distinct PSD. For ease of interpretation, in this example we implement a single quantum logic operation subject to high-frequency noise; however with this package it is easy to extend this simulation to complex multi-operation circuits experiencing noise on a variety of timescales.

This simulation includes time-varying phase noise on a microwave drive, a time-varying microwave detuning and an additive ambient dephasing field applied to the qubit. In each case an input power spectral density (left column) is converted to a time-series and combined with the relevant control channel (middle column), resulting in a noisy system representation (right column). The simulated performance in the ideal case is shown in Fig. 5d, indicating high fidelity state transfer from $|0\rangle \rightarrow |1\rangle$ with negligible population of the leakage level $|2\rangle$. However, in the presence of noise and leakage errors the fidelity of state transfer is reduced by approximately three orders of magnitude.

### E. Hardware characterization

Characterizing the noise profile of a quantum device is useful to identify opportunities for improving hardware, or implementing robust controls targeted at specific error sources. This includes Hamiltonian parameter estimation [104, 105] (e.g. determining phase offsets on control operations due to hardware imperfections), as well moving beyond generic averaged-error characterization routines [106] toward detailed microscopic characterization of time-dependent noise processes. In the filter function framework the latter properties are captured through the noise power spectral density (PSD) for various error channels in the system Hamiltonian. This information is also useful to evaluate control performance and pursue targeted pulse optimizations (both descried above).

In general detailed microscopic information about hardware noise processes and imperfections is not easily determined through conventional hardware calibration protocols. This limit may be overcome using a qubit as a measurement device to directly probe local dynamics or in-situ sources of signal distortion impacting system performance. In this subsection we describe software tools and techniques designed to employ the qubit as a transducer towards these tasks, covering both non-parametric noise spectral reconstruction and Hamiltonian parameter estimation.

### 1. Noise spectral estimation

Consider a noise Hamiltonian $H_{\text{noise}}(t)$ as in Eq. 20, comprised of multiple independent noise sources, each described by a corresponding PSD. Appropriately modulating qubit controls in the time domain can focus the measurement sensitivity to noise in a target spectral band, as well as selectively enhance sensitivity to a target noise operator. These objectives map to tuning a set of filter functions corresponding to the particular control-modulation scheme. The problem treated here is how to reconstruct these PSDs from the measurement record resulting from a given control-modulation scheme. We have developed spectral reconstruction packages allowing users to employ well conditioned control sequences. These are more flexible than existing spectrum reconstruction approaches [107] while in tests demonstrating superior reconstruction accuracy. Any relevant set of measurements may be employed to characterize a target noise channel, to produce an overall frequency-dependent sensitivity function employed in the reconstruction. The

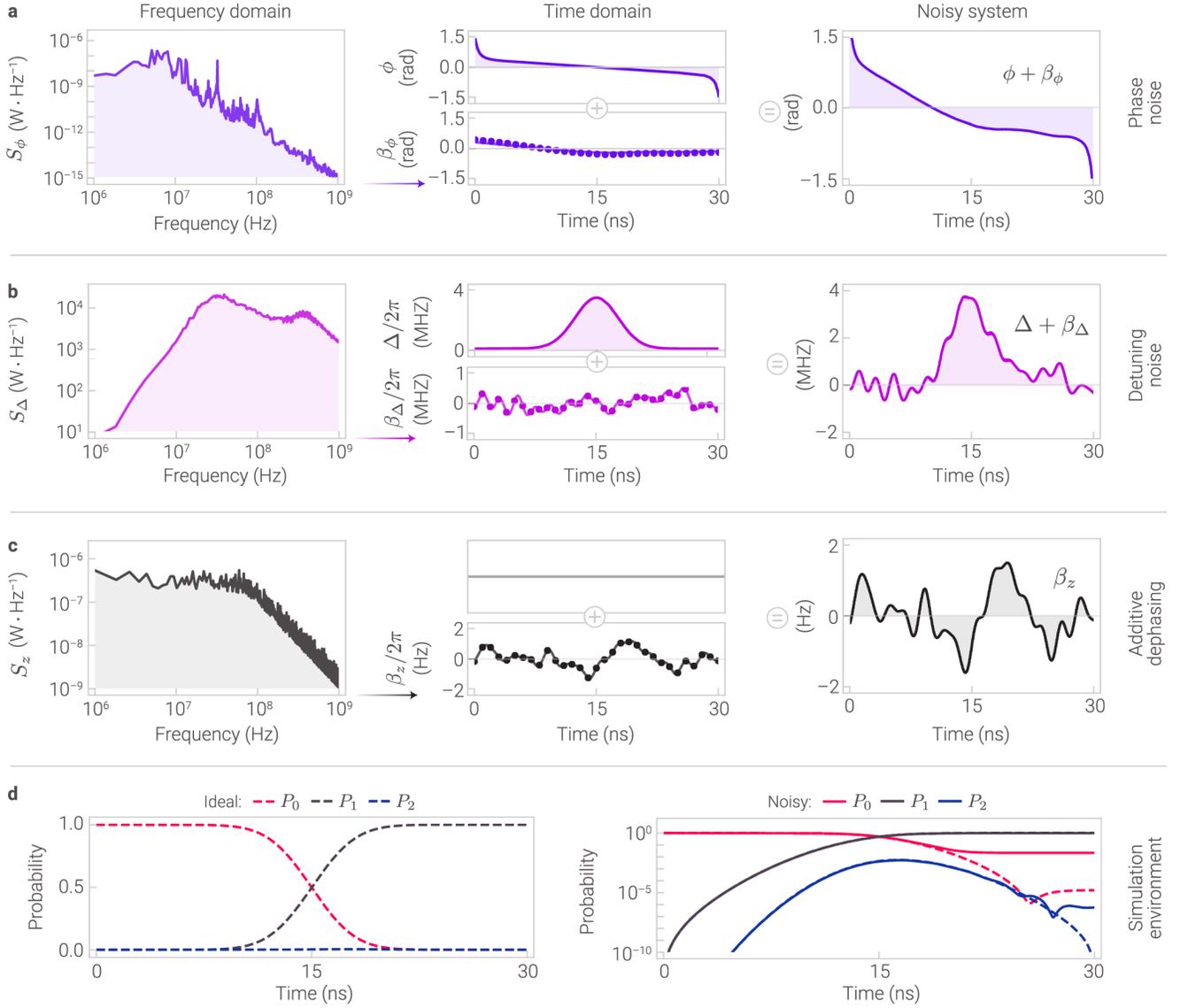

FIG. 5. Simulated time-evolution of a driven qutrit subject to leakage and various noise channels. Control comprises an off-resonant qubit drive $\gamma(t) = \Omega(t)e^{i\phi(t)}$ with detuning $\Delta(t)$, implementing a Gaussian DRAG pulse [103] with $I(t), Q(t)$ and $\Delta(t)$ channels (Eq. B17) proportional to a Gaussian envelope, its time derivative, and a Gaussian-square respectively. (a-c) Left column: noise PSDs associated to each noise channel input into the simulation tool. Middle column: time-domain waveforms for ideal controls (upper) and single-instance noise signals (lower). The latter include discrete samples (markers) from applying Eq. 39 to the corresponding PSDs, and continuous-time interpolation (solid lines) using Eq. 40 for high-precision simulation results. Right column: noisy waveforms summing ideal-control and noise contributions. Three different noise processes are treated: (a) phase noise $\phi(t) \to \phi(t) + \beta_\phi(t)$; (b) detuning noise $\Delta(t) \to \Delta(t) + \beta_\Delta(t)$; (c) ambient dephasing $\beta_z(t)\sigma_z$. (d) Evolution of state populations simulated during pulse application, including leakage and noise (Algo. 3). The DRAG pulse is designed to implement a $X_\pi$ gate, with ideal population transfer $[P_0, P_1, P_2] : [1, 0, 0] \to [0, 1, 0]$ Left: ideal DRAG pulse; Right: comparison of ideal (dashed) and noisy (solid) population evolution, plotted on log scale to resolve errors arising from the noise dynamics.

process of noise characterization follows a simple work-flow, highlighted schematically in Fig. 6:

1. Design control pulses with enhanced measurement sensitivity for probing the target noise process.

2. Implement controls on hardware device and obtain corresponding measurement data.

3. Perform data fusion on measurement results to re-construct the underlying noise power spectrum.

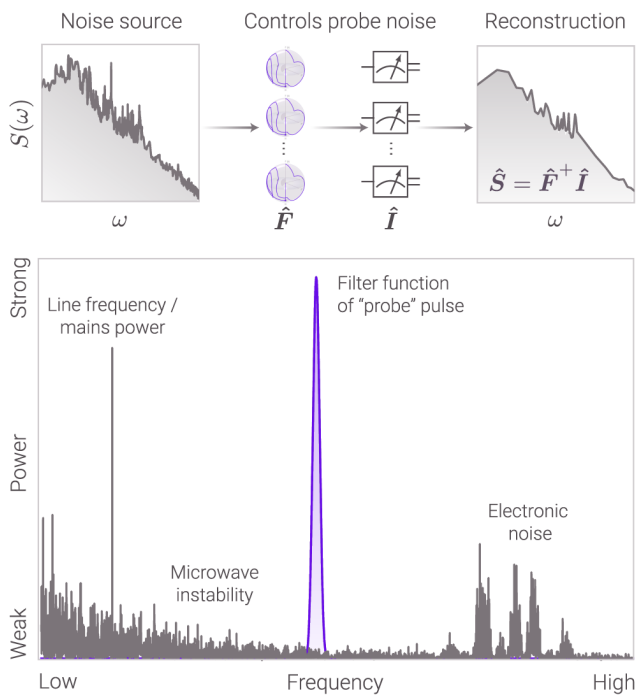

FIG. 6. Overview of the noise characterization process using multi-dimensional filter functions and SVD spectral inversion technique. Upper: A noise source is probed by a sequence of control solutions each providing sensitivity to a different spectral range, as determined by the multi-dimensional filter function. Measurement results are then used to produce a reconstruction of the actual spectrum experienced by the qubit, with degradation in fidelity determined by the available controls and numeric routine. Lower: Concept demonstrating how an appropriately constructed filter function can serve as a narrowband probe of underlying noise processes, giving access to different technical components of the noise spectrum. Selecting an alternate control can shift the peak in the filter function in order to allow broadband sampling of the noise spectrum.

In step (1), the design of appropriate control pulses depends on the type of noise being probed (described by the noise operator), the availability of controls supported by the device (described by the control operators), as well as any physical limitations set by the classical control hardware itself. A range of controls is available for characterizing familiar processes such as control-amplitude noise or ambient dephasing respectively. However the package supports generalizations to higher-order noise processes, e.g. Fig. 12 shows reconstruction results in a multi-qubit setting using a novel two-qubit control protocol. In cases where experimental simplicity is prioritized, dephasing-noise information can be obtained using timed sequences of simple driven rotations, often referred to as pulsed dynamical decoupling sequences [107]. Here quantum bit flips are sequenced in order to produce a filter function with a dominant peak at the frequency defined by the inverse interpulse delay. In contrast, shaped controls based on so-called Slepian waveforms [108, 109] are highly effective for the characterization of both control noise and dephasing [110]. These controls are provably optimal in terms of spectral concentration, i.e. how much spectral weight resides within a target band. Accordingly they mitigate issues of spectral leakage which cause unwanted out-of-band signals to contribute to the measurement as a form of interference. They can be thought of as mathematically optimal window functions applied directly to the qubit itself, restricting the qubit's sensitivity to noise.

In step (3), data fusion refers to algorithmic post-processing on sensing data [111, 112]. For our purposes, the sensors correspond to different controls/measurements used to infer the PSD. Choosing the data-fusion algorithm that produces the best inference on a given data set, however, can depend on a number of factors. For example, the type of noise being characterized, the available controls, or the quality of the measurement data. Depending on the required resolution, a trade-off exists in which the size of measurement data (and hence experimental complexity) may be reduced, at the cost increased numerical uncertainties under the data-fusion routine of choice. The noise reconstruction package supports two different inference methods: (i) a method based on singular-value decomposition Eq. 65, and (ii) a method based on convex optimization Eq. 66. Both support re-configurable fitting criteria for reconstructing PSDs from a given measurement record.

We now turn to a formal treatment linking the PSD to the actions of applied controls on the underlying quantum system, and the associated measurement outcomes. Consider a quantum system consisting of $p$ independent noise sources

$$H_{\text{noise}}(t) = \sum_{k=1}^{p} \beta_k(t) N_k(t) \tag{48}$$

where the $\beta_k(t)$ are stochastic scalar-valued noise fields with corresponding PSDs, $S_k(\omega)$. The structure of $H_{\text{noise}}(t)$ may be probed by defining a set of $c$ distinct control protocols

$$\{H_{\text{ctrl},j}(t)\}, \qquad j \in \{1, ..., c\} \tag{49}$$

and measuring the corresponding infidelities. From Sec. III B, and assuming the noise is sufficiently weak, the infidelities may be approximated as

$$\sum_{k=1}^{p} \int_{-\infty}^{\infty} \frac{d\omega}{2\pi} F_k^j(\omega) S_k(\omega) \approx \mathcal{I}^{(j)}, \qquad j \in \{1, ..., c\} \tag{50}$$

where $F_k^j(\omega)$ is the leading-order filter function associated with the $j$th control protocol and $k$th noise source. The filter functions may be computed using Eq. 33 given knowledge of the control Hamiltonians and dynamical noise generators, while the infidelities may be obtained from experiment.

Let $[\omega_{\min,k}, \omega_{\max,k}]$ denote the frequency domain spanned by the low- and high-frequency cutoffs, assuming they exist, for each PSDs in Eq. 50. We may then define the sample frequencies

$$\omega_{k,\ell} = \omega_{\min,k} + (\ell - 1)\Delta\omega_k, \qquad (51)$$

for samples $\ell \in \{1, ..., s(k)\}$, incremented by frequency steps

$$\Delta\omega_k = \frac{\omega_{\max,k} - \omega_{\min,k}}{s(k) - 1} \qquad (52)$$

on each domain $k \in \{1, ..., p\}$. Assuming sufficiently large sample numbers, $s(k)$, Eq. 50 may be recast as a discrete sum, with the integrals approximated using the trapezoidal rule. Specifically

$$\hat{I}^j = \frac{\Delta\omega_k}{2\pi} \hat{F}^j_{k,\ell} \hat{S}^{k,\ell} \left(1 - \frac{\delta_{\ell,1}}{2}\right) \left(1 - \frac{\delta_{\ell,s(k)}}{2}\right) \qquad (53)$$

where $\delta_{ij}$ is the Kronecker delta[113], and the sum runs implicitly over repeated tensor indices. Here we have introduced the following tensor notation for the various sampled quantities

$$F^j_k(\omega_{k,\ell}) \approx \hat{F}^j_{k,\ell} \pm \Delta\hat{F}^j_{k,\ell} \qquad (54)$$

$$S_k(\omega_{k,\ell}) \approx \hat{S}^{k,\ell} \pm \Delta\hat{S}^{k,\ell} \qquad (55)$$

$$\mathcal{I}^{(j)} \approx \hat{I}^j \pm \Delta\hat{I}^j \qquad (56)$$

where $\hat{Q}$ denotes the estimated value for the quantity $Q$, and $\Delta\hat{Q}$ denotes its uncertainty [114].

The challenge, then, is to obtain estimates, $\hat{S}^{k,\ell} \pm \Delta\hat{S}^{k,\ell}$ for the power spectral densities by inverting the relationship defined by Eq. 53, given knowledge of the measured quantites $\hat{I}^j \pm \Delta\hat{I}^j$ and computed values $F^j_{k,\ell} \pm \Delta F^j_{k,\ell}$.

As a first step we move to a discretized matrix representation. Namely,

$$\hat{F}\hat{S} = \hat{I}, \qquad (57)$$

where $\hat{F} = \begin{bmatrix} \hat{F}_1 & \hat{F}_2 & \cdots & \hat{F}_p \end{bmatrix}$ is a horizontal concatenation of matrices of the form

$$\hat{F}_k = \frac{\Delta\omega_k}{2\pi} \begin{bmatrix} \frac{1}{2}\hat{F}^1_{k,1} & \hat{F}^1_{k,2} & \cdots & \hat{F}^1_{k,s(k)-1} & \frac{1}{2}\hat{F}^1_{k,s(k)} \\ \frac{1}{2}\hat{F}^2_{k,1} & \hat{F}^2_{k,2} & \cdots & \hat{F}^2_{k,s(k)-1} & \frac{1}{2}\hat{F}^2_{k,s(k)} \\ \vdots & \vdots & \ddots & \vdots & \vdots \\ \frac{1}{2}\hat{F}^c_{k,1} & \hat{F}^c_{k,2} & \cdots & \hat{F}^c_{k,s(k)-1} & \frac{1}{2}\hat{F}^c_{k,s(k)} \end{bmatrix}, \qquad (58)$$

the estimated PSDs are concatenated vertically as

$$\hat{S} = \begin{bmatrix} \hat{S}_1 \\ \hat{S}_2 \\ \vdots \\ \hat{S}_p \end{bmatrix}, \qquad \hat{S}_k = \begin{bmatrix} \hat{S}^{k,1} \\ \hat{S}^{k,2} \\ \vdots \\ \hat{S}^{k,s(k)} \end{bmatrix}, \qquad (59)$$

and the estimated infidelities are arranged as

$$\hat{I} = \begin{bmatrix} \hat{I}^1 \\ \hat{I}^2 \\ \vdots \\ \hat{I}^c \end{bmatrix}. \qquad (60)$$

The matrix dimensions therefore satisfy

$$\dim(\hat{F}) = c \times n \qquad (61)$$

$$\dim(\hat{S}) = n \times 1 \qquad (62)$$

$$\dim(\hat{I}) = c \times 1 \qquad (63)$$

where $n = \sum_{k=1}^p s(k)$. From Eq. 57, performing noise reconstruction thus reduces to solving the matrix inverse problem $\hat{S} = \hat{F}^{-1}\hat{I}$. Depending on the particular set of controls and noise sources, and on the dimensions $c$, $p$, and $n$, the exact matrix inverse $\hat{F}^{-1}$ may not exist. Generally, the system may be underdetermined or overdetermined, and the matrix $\hat{F}$ may be singular. Finding solutions to this form of problem is discussed next.

We have developed two distinct machine-learning techniques used to solve this inversion problem which trade accuracy in the presence of complex noise spectra for computational efficiency. Importantly, both approaches go beyond published techniques by accepting arbitrary measurement records and accounting for the full form of the filter function (including harmonics and hardware-induced imperfections), rather than using simplifying approximations. The first method is based on an efficient implementation of pseudo-inverse by singular value decomposition (SVD). The second, based on convex optimiztion (CO), addresses numerical instabilities of the SVD method when noise spectra exhibit narrowly defined features or "spurs". Both techniques enable parameter-free estimation needed to perform reconstructions without *a priori* knowledge of the underlying structure of the noise, and are easily generalized beyond single qubits based on the multi-dimensional filter-function formalism Sec. III B. An experimental demonstration for interacting superconducting qubits is presented in Sec. IV D.

To facilitate efficient numerical solutions to the spectral estimation problem we provide a pseudoinverse technique based on a singular-value decomposition (SVD) method. This approach is numerically efficient and works well in circumstances where noise spectra are expected to vary smoothly as a function of frequency. The general approach may be used to obtain a pseudo-inverse if the problem is undetermined, to perform regression if it is over-determined, or to calculate the exact inverse if it is in fact determined. *Usefully, in all cases, the singular value decomposition of $\hat{F}$ takes the same general form:*

$$\hat{F} = U D V^\dagger, \qquad (64)$$

Here, $U$ is a $(c \times c)$ unitary matrix, $V^\dagger$ is a $(n \times n)$ unitary matrix, and $D$ is a $(c \times n)$ rectangular diagonal matrix with non-negative real numbers on the diagonal. The columns of $U$ and $V$ are the left- and right-singular vectors of $\hat{F}$, and the diagonal elements of $D$, denoted $s_i$, are known as the singular values. The final representation is then given by

$$\hat{S} = V D^+ U^\dagger \hat{I}, \tag{65}$$

where $D^+$ is a diagonal matrix with entries $1/s_i$ for all non-zero singular values, and zero otherwise.

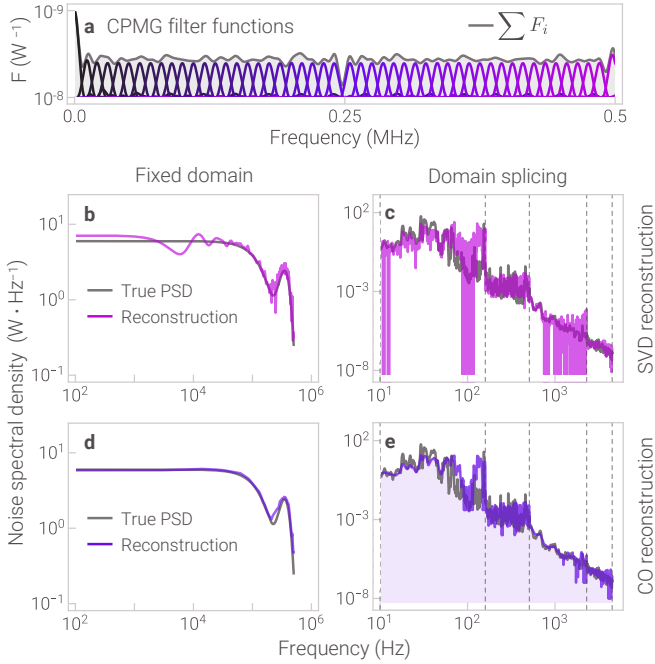

FIG. 7. Spectral reconstruction of simulated dephasing noise using 51 CPMG sequences [115, 116], all with total duration 3 μs. (a) Filter functions $\{F_i\}$ corresponding to CPMG orders $i \in \{0, ..., 50\}$, plotted in 51 colors interpolated between [black, magenta]→ [0, 50]. Filter functions $[F_0, F_{50}]$ have spectral peaks centred at $[0, 0.5]$ MHz. Spectral peaks for intermediate-order filter functions are separated by increments of 10 kHz. Sensitivity to dephasing over the frequency domain is captured by $\sum F_i$ (grey fill). (b,d) Artificially smooth dephasing PSD (grey) reconstructed using SVD (magenta, top) and CO (violet, bottom). Both reconstructions were performed using frequencies uniformly sampled on the domain $[0, 0.5]$ MHz. (c,e) Reconstruction of a complex dephasing PSD with $1/f$ background trend, discrete spurs and finer structure mimicking real hardware noise. Again, the true PSD is plotted in grey, overlaid with SVD (magenta, top) and CO (violet, bottom) reconstructions. In this case the frequency domain was partitioned into 4 intervals. Reconstructed PSDs were spliced together from independent reconstructions on each sub-domain for improved frequency resolution. In all cases (b-e), SVD and CO methods provide reasonable quantitative agreement with the data sets. However the SVD technique exhibits oscillations due to numerical instabilities that are evident for more complex spectra (c). These are absent in the CO reconstruction (e). For all these results, simulated measurement records were generated numerically and reconstructions performed as describe in Algo. 4 allowing performance comparisons focused on the efficacy of the underlying linear-algebraic method.

As mentioned above this technique functions well for smoothly varying noise spectra, but can suffer from instabilities especially in circumstances where the noise exhibits "spurs" that are spectrally narrow relative to the probe filter function. In this case, the lack of a strict positivity criterion for the pseudoinverse can cause oscillations in the reconstruction which result in unphysical estimates of the spectrum (Fig. 7b,c). In order to accommodate these more complex cases we next introduce a second method for spectral estimation which ensures positivity of all estimates at the expense of increased computational complexity.

A convex-optimization technique enables better estimates of a noise power spectrum in the presence of complex spectral characteristics. Returning to the discretized form of Eq. 57, we see that in general there are in fact infinitely many solutions to this equation. The key task is then to find a set of solutions that will be considered reasonable based on some *prior* information about the system; here we add a condition of strict positivity for all inverted representations of $\hat{S}$.

In our CO implementation, instead of directly finding a solution to Eq. 57, we reformulate the problem with the prior information as an objective function. By minimizing that objective function, the optimal solution should represent our best understanding of the problem. This approach can be formalized as follows:

$$\min_{\hat{S}} (\|F\hat{S} - I\|_2^2 + \lambda R(\hat{S})) \quad \text{s.t.} \quad \hat{S} \geq 0 \tag{66}$$

where $\| \cdot \|_2$ denotes the Euclidean norm.

The second term in the objective function is known as the regularization term, and $\lambda > 0$ is a hyper-parameter. In the language of machine learning, a suitable choice of the hyper-parameter and regularization will prevent over-fitting. The hyper-parameter $\lambda$ does not have any physical meaning, but it can be taken as a weight reflecting how our prior information would impact the solution. In our algorithmic approach we follow a conventional "L-curve criterion" for finding the optimal hyper-parameter $\lambda$ [117]. Similarly, there are many approaches to choose the regularization term $R(\hat{S})$, and our algorithmic implementation supports any user-defined form. We elect to incorporate two distinct approaches to regularization which may be used in combination to accommodate the possibility of a smoothly varying $\hat{S}$ incorporating a sparse set of discrete features, without violating positivity. First, we express $R(\hat{S}) = \|D\hat{S}\|_2^2$ using the so-called Tikhonov matrix $D$, whose form depends on the prior information [118]. We select a form for $D$ which corresponds to the first-derivative operator

$$D = \begin{bmatrix} -1 & 1 & & \\ & \ddots & \ddots & \\ & & -1 & 1 \end{bmatrix}_{(n-1) \times n} \tag{67}$$

---

**Algorithm 4** Noise reconstruction

---

$\{N_k(t)\} \leftarrow$ distinct noise operators, $k \in \{1, ..., p\}$
$\{H^{(m)}_{\text{ctrl}}(t)\} \leftarrow$ distinct controls, $m \in \{1, ..., c\}$
**procedure** CONTROLS/MEASUREMENTS
    **for** $m \in \{1, ..., c\}$ **do**
        $\hat{I}^m \leftarrow$ avg. infidelity msmst. ▷ Implementing $H^{(m)}_{\text{ctrl}}$
        **for** $k \in \{1, ..., p\}$ **do**
            $\{f^{(k)}_\nu\} \leftarrow$ frequency domain, $\nu \in \{1, .., s(k)\}$
            $\{F^m_{k,\nu}\} \leftarrow \text{FF}(\{f^{(k)}_\nu\}; H^{(m)}_{\text{ctrl}}, N_k)$ ▷ Algo. 1
        **end for**
    **end for**
    $\hat{\boldsymbol{I}} \leftarrow [\hat{I}^m]$ ▷ $c \times 1$ matrix
    $\hat{\boldsymbol{F}}_k \leftarrow [F^m_{k,\nu}]$ ▷ $c \times s(k)$ matrix
    $\hat{\boldsymbol{F}} \leftarrow [\hat{\boldsymbol{F}}_1 \ldots \hat{\boldsymbol{F}}_p]$ ▷ $c \times n$ matrix, $n = \sum_{k=1}^p s(k)$
**end procedure**

**function** RECONSTRUCTSVD($\hat{\boldsymbol{F}}, \hat{\boldsymbol{I}}$)
    $\hat{\boldsymbol{U}}, \hat{\boldsymbol{D}}, \hat{\boldsymbol{V}} \leftarrow \text{SVD}(\hat{\boldsymbol{F}})$
    Returns $\hat{\boldsymbol{S}}$ ▷ Eq. 65
**end function**

**function** RECONSTRUCTCO($\hat{\boldsymbol{F}}, \hat{\boldsymbol{I}}$)
    $R \leftarrow$ regularization function, default $\|D\hat{\boldsymbol{S}}\|^2_2$ ▷ Eq. 67
    $\lambda \leftarrow$ FINDHYPERPARAMETER($\hat{\boldsymbol{F}}, \hat{\boldsymbol{I}}, R$)
    $\text{cost}(\hat{\boldsymbol{S}}) \leftarrow \|\hat{\boldsymbol{F}}\hat{\boldsymbol{S}} - \hat{\boldsymbol{I}}\|^2 + \lambda R(\hat{\boldsymbol{S}})$ ▷ Eq. 66
    $\hat{\boldsymbol{S}}_{\text{optimal}} \leftarrow$ OPTIMIZE($\text{cost}(\hat{\boldsymbol{S}})$) ▷ Algo. 2
    Returns $\hat{\boldsymbol{S}}_{\text{optimal}}$
**end function**

**function** SVD($\hat{\boldsymbol{F}}$)
    Returns $\hat{\boldsymbol{U}}, \hat{\boldsymbol{D}}, \hat{\boldsymbol{V}}$ as SVD of $\hat{\boldsymbol{F}}$ ▷ Eq. 64
**end function**

**function** FINDHYPERPARAMETER($\hat{\boldsymbol{F}}, \hat{\boldsymbol{I}}, R$)
    Returns the hyperparameter $\lambda$
**end function**

---

such that minimizing $\|D\hat{\boldsymbol{S}}\|^2_2$ minimizes the difference among the elements of $\hat{\boldsymbol{S}}$, indicating an expectation that $\hat{\boldsymbol{S}}$ varies *smoothly* in the parameter space.

Alternatively, the regularization term may be chosen as $\lambda \|\hat{\boldsymbol{S}}\|^2_1$. The $L_1$ norm will enforce the *sparsity* of the optimal solution. This is a reasonable assumption if we expect $\hat{\boldsymbol{S}}$ to be composed of a few non-zero features across a broad range of frequencies. This particular type of $L_1$ optimization problem is well-known for its application in compressed sensing for sparse signal processing [119]. In our protocols we generally combine these two regularization procedures in order to treat a broader range of conditions for the noise spectrum $\hat{\boldsymbol{S}}$.

With this formulation, and employing either regularization condition both the objective function in Eq. 66 and constraints are convex, enabling efficient numeric convex optimization. In our algorithmic implementation, this optimization is handled using the toolkit described in Sec. III C, as per the pseudocode presented in Algo. 4.

The advantages of the CO reconstruction are displayed in Fig. 7d,e. Here, both smoothly varying functions and complex mixed spectra exhibiting narrow features on a smoothly varying background are accurately reconstructed without suffering from numerical instabilities. Both approaches are employed in an experimental setting for a multiqubit gate on a cloud quantum computer in Sec. IV D.

### 2. Hamiltonian parameter estimation

Hamiltonian parameter estimation is based on an efficient model-reduction technique, allowing a system with complex observables to be represented through a finite set of proxy parameters. In such a circumstance, instead of performing an effectively unbounded set of characterization measurements, we may restrict ourselves to identifying this much smaller set of parameters, at some cost in the accuracy of the model achieved. The problem of identifying a system by characterizing its dynamics can be formulated as an optimization problem where we find system parameters using a set of measurement results as input points. If we know how these parameters affect the dynamics of the system, we can establish a cost function that represents how unlikely it is that the input points could have been generated by a given choice of system parameters. With such a cost function, the same set of functions that are used to optimize control operations as in Sec. III C can then be adapted to characterize a system.

More formally, suppose we want to determine $n$ system parameters $\theta_1, \theta_2, \ldots, \theta_n$. To achieve this, we subject the system to $k$ experimental setups that are differently affected by the values of these parameters. Such experiments could consist of different kinds of pulses applied to the qubits, or different interrogation times, for example. The averaged results for each experiment then form a set of $k$ input points $y_1, y_2, \ldots, y_k$, each of them with associated standard deviations $\Delta y_1, \Delta y_2, \ldots, \Delta y_k$. This is the data that will be provided to the optimizer.

To perform the estimation of the parameters of the system, these inputs will be combined inside the optimizer with knowledge about the dynamics of the system. The dynamics will be encapsulated in functions $Y_m(\boldsymbol{\theta})$, which represent the ideal average value of the $m$th experiment if the system evolved according to a vector of parameters $\boldsymbol{\theta} = (\theta_1, \theta_2, \ldots, \theta_n)$. Assuming independent probability distributions for each of the averaged measurement results, the likelihood that a certain choice of values of the parameters $\boldsymbol{\theta}$ was responsible for the vector of input points $\mathbf{y} = (y_1, y_2, \ldots, y_k)$ is given by the product of the individual probabilities for each input point, yielding

$$p(\mathbf{y}|\boldsymbol{\theta}) = \prod_{m=1}^k p(y_m|\boldsymbol{\theta}). \tag{68}$$

Further assuming Gaussian probability distributions for the averaged measurement results, we have

$$p(y_m|\boldsymbol{\theta}) = \frac{1}{\sqrt{2\pi(\Delta y_m)^2}} \exp\left\{-\frac{[Y_m(\boldsymbol{\theta}) - y_m]^2}{2(\Delta y_m)^2}\right\}. \tag{69}$$

This likelihood can be maximized by minimizing its negative logarithm. Removing the constant terms, an appropriate cost function for the optimizer to minimize is

$$C = \sum_{m=1}^{k} \frac{[Y_m(\boldsymbol{\theta}) - y_m]^2}{2(\Delta y_m)^2} \propto -\log\left[p(\mathbf{y}|\boldsymbol{\theta})\right]. \quad (70)$$

This choice of cost function also allows us to assess the precision of the parameter estimates. As $C$ only differs from the negative log likelihood by constant terms, its Hessian (the matrix of second partial derivatives with respect to the parameters $\boldsymbol{\theta}$) can be identified with the Fisher information matrix $\mathcal{I}_{\text{Fisher}}$, where

$$\mathcal{I}_{\text{Fisher}} \equiv \left(\frac{\partial^2 C}{\partial \theta_i \partial \theta_j}\right) = \left(-\frac{\partial^2}{\partial \theta_i \partial \theta_j}\log\left[p(\mathbf{y}|\boldsymbol{\theta})\right]\right). \quad (71)$$

Using the Cramér–Rao bound, the minimum value of the covariances of the parameter estimates is limited by the inverse of the Fisher information matrix:

$$\text{cov}(\boldsymbol{\theta}) \geq \mathcal{I}_{\text{Fisher}}^{-1}, \quad (72)$$

where $\text{cov}(\boldsymbol{\theta})$ is the covariance matrix for the parameters $\boldsymbol{\theta}$.

The way this optimization procedure can be programmed is shown in Algo. 5. The system dynamics encapsulated in the maps $\{Y_m(\boldsymbol{\theta})\}$ can be represented using the same built-in functions from Algo. 2. For example, the same description of a piecewise constant Hamiltonian used there for pulse optimization can be used here to represent the effect of an input PWC pulse in a system whose control Hamiltonian contains parameters $\boldsymbol{\theta}$ that we wish to determine. Likewise, by representing the objects that are part of the cost function from Eq. 70 in the same manner used for pulse optimization, we can re-use the same optimization function used in Algo. 2 to estimate system parameters here.

Once we are in possession of the parameter estimates, we can use them together with Eq. 71 to find the lower bounds of the elements of the covariance matrix of $\boldsymbol{\theta}$. The diagonal elements of this matrix represent the variances of the estimated variables. Assuming a normal distribution, two times the square root of these variances will estimate the errors of the parameters with 95% confidence.

A simple example of system identification using this method consists in characterizing a constant single-qubit Hamiltonian. Excluding terms proportional to the identity, which do not affect the state evolution, a constant single-qubit Hamiltonian is characterized by three coefficients that multiply the Pauli matrices:

$$H = \frac{1}{2}\left(\Omega_x \sigma_x + \Omega_y \sigma_y + \Omega_z \sigma_z\right). \quad (73)$$

---

**Algorithm 5** Sample system identification

---

$k \leftarrow$ number of distinct experiment setups
$\{y_m\} \leftarrow$ distinct input points, $m \in \{1, \ldots, k\}$
$\{\Delta y_m\} \leftarrow$ input standard deviations, $m \in \{1, \ldots, k\}$

**procedure** SYSTEMIDENTIFICATION
  $C(\boldsymbol{\theta}) \leftarrow$ SAMPLECOSTFUNCTION(k, $\{y_m\}$, $\{\Delta y_m\}$)
  $\boldsymbol{\theta} \leftarrow$ OPTIMIZE($C(\boldsymbol{\theta})$)             ▷ Algo. 2
  $V \leftarrow$ COVARIANCEMATRIX($C(\boldsymbol{\theta})$, $\boldsymbol{\theta}$)
  $\boldsymbol{\sigma}^2 \leftarrow$ diagonal elements of $V$
  $\Delta \boldsymbol{\theta} \leftarrow 2\sqrt{\boldsymbol{\sigma}^2}$      ▷ errors estimated as $2\sigma$
**end procedure**

**function** SAMPLECOSTFUNCTION(k, $\{y_m\}$, $\{\Delta y_m\}$)
  $Q(\boldsymbol{\theta}) \leftarrow$ operator as a function of $\boldsymbol{\theta}$   ▷ map to matrix
  **for** $m \in \{1, \ldots, k\}$ **do**
    $\tau_m \leftarrow$ duration of the $m$th experiment
    $O_m \leftarrow$ observable measured in the $m$th experiment
    $|\psi_m\rangle \leftarrow$ initial state for the $m$th experiment
    $\mathbf{v}_m \leftarrow$ pulse segment values for $m$th experiment
    $\alpha_m \leftarrow$ PWCSCALAR($\tau_m$, $\mathbf{v}_m$)      ▷ Algo. 2
    $H_m(t, \boldsymbol{\theta}) \leftarrow$ PWCOPERATOR($\alpha_m(t)$, $Q(\boldsymbol{\theta})$) ▷ Algo. 2
    $U_m(\boldsymbol{\theta}) \leftarrow$ UNITARY($\tau_m$, $H_m(t, \boldsymbol{\theta})$)
    $Y_m(\boldsymbol{\theta}) \leftarrow \langle\psi_m| U_m^\dagger(\boldsymbol{\theta}) O_m U_m(\boldsymbol{\theta}) |\psi_m\rangle$
  **end for**
  $C(\boldsymbol{\theta}) \leftarrow \sum_m [Y_m(\boldsymbol{\theta}) - y_m]^2 / \left[2(\Delta y_m)^2\right]$   ▷ Eq. 70
  Returns $C(\boldsymbol{\theta})$
**end function**

**function** COVARIANCEMATRIX($C(\boldsymbol{\theta})$, $\boldsymbol{\theta}$)
  $\mathcal{I}_{\text{Fisher}} \leftarrow$ Hessian of $C(\boldsymbol{\theta})$ with respect to $\boldsymbol{\theta}$   ▷ Eq. 71
  $V = \mathcal{I}_{\text{Fisher}}^{-1}$          ▷ inverse of a matrix
  Returns $V$
**end function**

· · · · · · · · · · · · · · · · · · · · · · · · · · · · · · · · · · · · · · · · · · · · ·

**Built-in method for building cost functions:**

**function** UNITARY($\tau$, $H(t)$)   ▷ solves the Schrödinger eq.
  $U(\tau) \leftarrow \mathcal{T}\exp\{-i\int_0^\tau dt' H(t')\}$
  Returns $U(\tau)$
**end function**

---

These three parameters $\Omega_x$, $\Omega_y$, and $\Omega_z$ can be identified by performing experiments in which we prepare the qubit in three different initial states, and then measure it after different wait times.

If the qubit measurements were performed in the same eigenbasis in which the qubit was prepared, information about the direction in which the qubit is rotating could be lost, as both clockwise and counterclockwise rotations would decrease the population of the initial state. To avoid this problem, we prepare the qubit in three initial states that are eigenstates of $\sigma_x$, $\sigma_y$, and $\sigma_z$, and measure it in a different eigenbasis ($\sigma_z$, $\sigma_x$, and $\sigma_y$, respectively). Whether the measured observable increases or decreases after the initial time gives information about the direction of the rotation.

To simulate how the Hamiltonian estimation could be performed for a system of this kind, we select a set of true values for $\Omega_x$, $\Omega_y$, $\Omega_z$, and use them to calculate the expectation values in different experimental setups. We generate 20 input points for each of the three initial states. Each of these points will correspond to a different wait time between state preparation and measurement. We allow the measured populations to have errors that are distributed according to a normal distribution with standard deviation 0.01, corresponding to the $\Delta y_m$ in Eq. 70.

In this example, we attempted to identify a system with the following set of parameters:

$$\Omega_x = 0.5 \cdot 2\pi \text{ MHz}, \tag{74}$$

$$\Omega_y = 1.5 \cdot 2\pi \text{ MHz}, \tag{75}$$

$$\Omega_z = 1.8 \cdot 2\pi \text{ MHz}. \tag{76}$$

We ran 30 optimizations following the procedure described in Algo. 5. Each optimization started with different random initial values for the parameters, limited by the Nyquist frequency set by time step between experiments. Out of these 30 runs, the one with lowest cost provided the following estimates for the parameters:

$$\hat{\Omega}_x = (0.494 \pm 0.016) \cdot 2\pi \text{ MHz}, \tag{77}$$

$$\hat{\Omega}_y = (1.499 \pm 0.022) \cdot 2\pi \text{ MHz}, \tag{78}$$

$$\hat{\Omega}_z = (1.808 \pm 0.018) \cdot 2\pi \text{ MHz}. \tag{79}$$

This simple example highlights how the formulation in Algo. 5 may be used with high fidelity to efficiently perform critical parameter estimation tasks.

## IV. QUANTUM CONTROL CASE STUDIES

### A. Open-loop control benefits demonstrated in trapped-ion QCs

In the sections above we described the role of quantum control in combating hardware error, and introduced new technical capabilities for the characterization and optimization of quantum hardware performance. In this section we provide case studies to demonstrate the application of these capabilities in contemporary quantum computing hardware. First, we provide experimental demonstrations of performance of open-loop control solutions in trapped-ion hardware, demonstrating error-robustness as well as error-rate homogenization in space and time. Second, we apply the numerical optimization package described in Sec. III C to generate single and multi-qutrit gates optimized for robustness against leakage and dephasing errors in a coupled-transmon system. Third, we apply the SVD and CO spectral reconstruction techniques outlined in Sec. III E to the IBM Q cloud-based quantum processor to characterize noise affecting two-qubit cross-resonance gates. Finally, we present an example of optimizing the structure of a quantum circuit, producing a logically-equivalent compiled circuit exhibiting suppression of cross-talk errors arising from a constant $ZZ$ interaction. We emphasize that these examples are not exhaustive representations of product capability, and that additional demonstrations for *e.g.* optimizing parallel Mølmer-Sørensen gate implementation, or characterizing simultaneous noise processes in spin qubits will be presented in separate manuscripts.

Trapped-ion quantum computers already exist at medium scales and provide an ideal platform for studies of quantum control efficiency due to long intrinsic lifetimes, high-fidelity operations, and access to multiqubit devices. We have used a trapped-ion quantum computer composed of individual $^{171}\text{Yb}^+$ ions in order to explore the efficacy of quantum control and quantum control optimization in real hardware.

We begin with demonstrations of error-robustness using open-loop control solutions available in the *OpenControls* package of driven single-qubit operations. In Fig. 8 we probe the robustness of various composite control operations implementing an effective $X_\pi$ gate (equivalently a $\pi$ pulse) to quasi-static errors in the pulse amplitude and detuning (Fig. 8b,c). Protocols designed to provide robustness to the associated error channel reveal little deviation from the baseline error rate achieved in the center of the graph (zero induced error) while the measured infidelity ($\mathcal{I}$) increases rapidly in the presence of systematic errors for non-robust controls. This difference is a key signature of error-robust control solutions.

Similarly, using a so-called 'system-identification' technique to probe control robustness to a time-varying perturbation [58] we demonstrate that appropriately crafted controls suppress noise at frequencies slow compared with the control rate (Fig. 8d,e). Experimental measurements compare well with calculation of the control filter function (solid lines), which is available through both BLACK OPAL and BOULDER OPAL. In particular, these experiments demonstrate that it is possible to construct single-qubit logic operations robust to noise in both the control amplitude and qubit-frequency detuning.

Moving beyond physical benefits we can also probe the manner in which these control operations intersect with higher levels of the quantum computing stack. For instance, in Fig. 8f we demonstrate homogenization of Parallel Randomized Benchmarking (RB) error rates across a 10-qubit quantum computer using error-suppressing open-loop gate constructions validated in Fig. 8b-e. Here the dominant error source is a spatial gradient in the coupling of the qubit drive field to the individual ions, due to reflections and interference of the 12.6 GHz microwaves inside the ion-trap vacuum enclosure. We therefore select an control-noise suppressing solution and replace all gates in the randomized benchmarking procedure with their logically equivalent error-robust constructions [123].

In this experiment the best-performing qubit does not exhibit a net improvement in the measured RB error rate, $p_{RB}$—a proxy measure for gate error—beyond measurement uncertainty due to other limiting error sources such as laser-light leakage. However, all other qubits exhibit RB error rates that now approximately match the best reported values, with the standard deviation of RB error rates across the 10-qubit array reduced $10.2\times$ using the appropriate open-loop control solution.

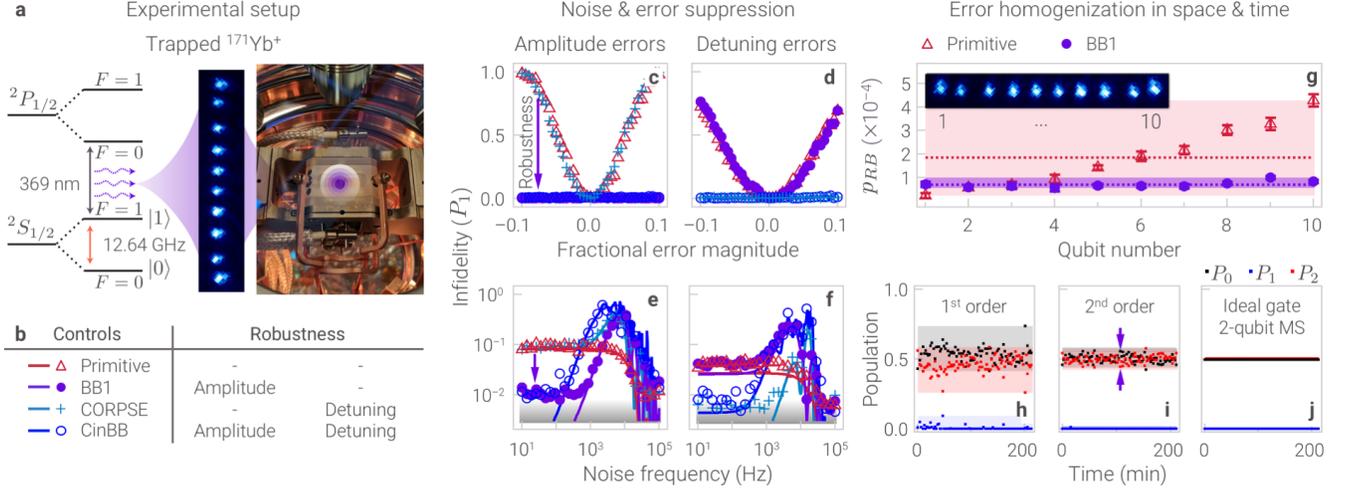

FIG. 8. Experimental validation of dynamic error suppression techniques. (a) Schematic of experimental setup (details in App. E). Qubits defined by Zeeman-split hyperfine ground states of trapped $^{171}$Yb$^+$ ions in a linear Paul trap; qubits resonantly driven by a 12.64 GHz microwave pulses; state detection via 369 nm fluorescence. (b) Four $X_\pi$-pulse constructions to be tested, with varying error-robustness properties. Primitive (red, *no robustness*); BB1 [120] (purple, *amplitude robustness*); CORPSE [121] (cyan, *detuning robustness*); CinBB [122] (blue, *amplitude + detuning robustness*). (c,d) Experimental demonstration of robustness to quasi-static errors. BB1 is most robust to pulse amplitude errors, as indicated by a low measured infidelity as a function of applied over-rotation error. CORPSE shows similar performance in the presence of detuning errors. CinBB shows comparable performance in the presence of both forms of error. The primitive implementation is susceptible to both types of error. (e,f) Demonstration of robustness to time-varying noise. Error-suppressing gates show robustness at low noise frequencies (left of the graph), resulting in lower measured infidelities (markers). Measurements agree with filter function predictions (solid lines). All gates show an onset of error susceptibility at high frequencies near the inverse gate time. (g) Error homogenization across spatially-distributed qubits measured by randomized benchmarking. A global microwave drive simultaneously implements a $X_\pi$-pulse on 10 qubits spatially distributed in an ion chain. Different qubits experience different effective Rabi rates due to the imperfect spatial profile of the microwave amplitude. Under the primitive implementation (red) qubits manifest divergent error rates; the BB1 implementation (purple) is robust to these amplitude-errors, and therefore suppresses and homogenizes them. Shading indicates the range of experimental outcomes while the mean error across the device is indicated by lines. Using the error-suppressing pulse, the mean error is reduced $\sim 5\times$ while the range (measured either by the standard deviation or the difference between minimum and maximum values) is reduced $\sim 10\times$. (h-j) Enhanced stability of 2-qubit Mølmer-Sørensen gates over time, comparing primitive and robust controls. The ideal outcome for both gates is described by $P_0 = P_2 = 0.5$ and $P_1 = 0$, where $P_n$ is the probability of finding $n$-of-2 ions in the $|1\rangle$ state. Gates are repeated over 3.5 hours and final populations logged. Populations $P_0$ (black), $P_1$ (blue), and $P_2$ (red) are estimated via a maximum-likelihood procedure [73]. Panel (H) shows the outcome for the primitive gate with no robustness (1st-order sensitivity to errors). Panel (i) shows the outcome for a phase-optimized gate robust to detuning errors (2nd-order sensitivity to errors). For comparison, panel (j) shows the ideal gate populations assuming no errors. Colored shading represents the range of the associated data set over the measurement window. The range of measurements is dramatically narrower for the robust gate and closer to the ideal case.

Moving beyond the application of control solutions for single-qubit gates, we examine the stability of two-qubit gates realized via the Mølmer-Sørensen interaction on a pair of trapped ions as the system experiences drifts in time. In this experiment we are targeting the creation of a Bell state $(|00\rangle - i\,|11\rangle)/\sqrt{2}$; ideally in this experiment there is an equal probability of measuring two ions in $|0\rangle$ or $|1\rangle$, and one should never observe any experiments with one ion each in these two states. Therefore our key proxy measure for gate performance is the measured population of zero, one, and two ions in state $|1\rangle$. The expected performance of these metrics is shown in Fig. 8i.

We compare two different gate constructions, one being relatively susceptible to drifts and the other designed to reduce sensitivity via a modulation protocol available in BOULDER OPAL, experimentally demonstrated first in [72, 73], and discussed in detail in [74]. Repeatedly performing the same gate shows variations in the measured ion-state populations, corresponding to reductions in gate fidelity. However, by using the error-suppressing gate construction we observe a $\sim 3-4\times$ reduced susceptibility to system drifts, indicated by arrows in Fig. 8h, showing the reduced range of outcomes.

Finally, we demonstrate the efficacy of numerically optimized single-qubit gates against various noise processes in trapped-ion qubits in Fig. 9. Specifically, we have focused on the use of the optimization toolkit described in Sec. III C to produce gates that are simultaneously robust against control noise and dephasing. Typically this requires a concatenated analytic construction which dramatically extends the control duration by up to 24× relative to the primitive gate. For an $X_\pi$ gate the optimizer returns solutions showing simultaneous robustness to error with a gate duration reduced ∼ 5× relative to this analytic approach. In the presence of quasi-static errors the numerically optimized solutions provide show robustness to error in the presence of up to 10% miscalibrations in qubit frequency and Rabi rate, similar to the results of Fig. 8b,c. Further details on the execution of these experiments is included in App. E.

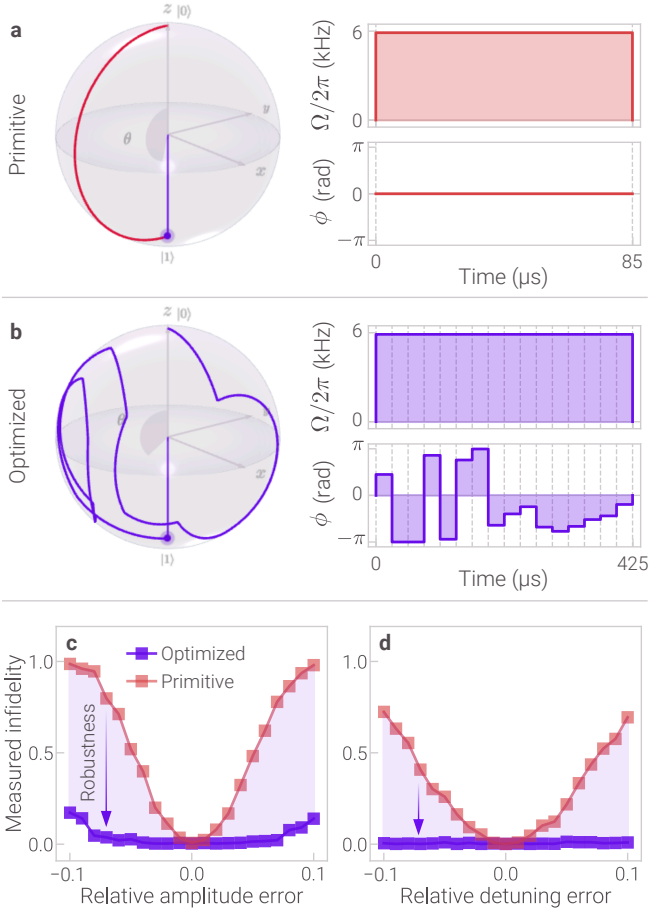

FIG. 9. Experimental demonstration of optimized controls in a trapped-ion quantum computer. (a, c) Bloch-sphere representation of primitive and Q-CTRL optimized robust controls, respectively. Both nominally implement a $X_\pi$ gate. (b,d) Corresponding waveforms plotted in Polar coordinates. The Q-CTRL waveform was optimized to provide dual suppression of both detuning and amplitude errors, and constrained to ensure a fixed pulse amplitude (phase-modulation only). (e,d) Experimental demonstration of robustness against quasi-static control amplitude and detuning errors for each pulse. Gates were implemented on an ion-trap experiment (Fig. 8a). Infidelities were measured while scanning over engineered amplitude and detuning offsets. These are plotted on the $x$-axis in fractional units relative to the Rabi rate or qubit frequency respectively. Shading represents the net improvement in error robustness afforded by the optimized solution. Further details of the experimental implementation are provided in App. E.

Similar results have been obtained using superconducting circuits on IBM Q, and full code for generating optimized controls and experimentally demonstrating them on hardware is available via Ref. [124]. The particular solution employed for the superconducting circuit was implemented using IBM's OpenPulse format [21] through Qiskit, and was filtered in order to comply with bandlimits in transmission lines (such constraints are generally not germane in trapped-ion systems due to the relatively long pulse durations and use of microwave antennae rather than transmission lines). We have also identified that pulses smoothed with a sinc filter and discretized in time using the flexible optimization engine described in Sec. III C generally perform better than slew-rate-limited pulses on IBM Q hardware, as the latter can occasionally include spectral components matched to hardware resonances (e.g. two-level systems). Overall these results demonstrate that the flexibility of the control-optimization approach described here allows for high-fidelity error-robust gates to be implemented on diverse hardware systems.

### B. Simultaneous leakage and noise-robust controls for superconducting circuits

For superconducting qubits, a two-level system is typically singled out from the many levels of an anharmonic oscillator. When driven by naive single-qubit controls, the system is subject to off-resonant coupling to leakage levels outside the qubit manifold, resulting in substantial *leakage errors*. In addition, these qubits face the common challenges of decoherence from ambient dephasing, control-phase and control-amplitude noise.

Suppression of leakage errors has been the focus of considerable research in the superconducting community and has been demonstrated to improve gate performance. The standard approach at present employs an analytic optimal control technique to implement target quantum operations via so-called DRAG pulses [103], or variants thereof. For example, Half-DRAG is designed to suppress leakage out of the qubit subspace via dual-quadrature control, typically involving a Gaussian pulse on $\sigma_x$, and its time derivative simultaneously applied on $\sigma_y$.

Unfortunately this technique does not combine successfully with other open-loop error-suppression strategies needed to combat decoherence from additional noise channels, e.g. NMR-inspired composite pulses [125]. For example, concatenation of pulse segments defined by DRAG into an overall CORPSE structure, known to suppress detuning noise (see Sec. IV A), fails as the dual-axis DRAG control breaks the geometric construction required to provide noise robustness.

Advancing on previous work, we present optimal and robust controls that simultaneously reduce sensitivity to both leakage and dephasing errors by orders of magnitude, using the optimization tools described in Sec. III C. Our starting point is the Hamiltonian for an anharmonic three-level qutrit subject to dephasing noise and leakage to the lowest-lying excited state in the system (Fig. 10a):

$$H(t) = (\gamma(t)a + \text{H.C.}) + \frac{\eta}{2}(a^\dagger)^2(a^2) + \beta_z(t)\sigma_z \quad (80)$$

where $a = |0\rangle\langle 1| + \sqrt{2}|1\rangle\langle 2|$. We encode this anharmonic oscillator using a *drift* control with operator $\frac{\eta}{2}(a^\dagger)^2 a^2$; a microwave *drive* control with operator $a$ and complex pulse envelope $\gamma(t) = \Omega(t)e^{-i\phi(t)}$; and an additive noise operator with Pauli operator $\sigma_z$ (see App. B for further details on this representation).

We perform two robust-control optimizations subject to different constraints (Table II). First, we implement a concurrent optimization allowing dual-quadrature controls similar in required controls to Half-DRAG (*e.g.* IQ modulation). Next we perform a fixed-waveform optimization that holds the amplitude of the control pulse associated with the microwave drive fixed while allowing its phase to vary, as in phase-modulation. The resulting waveforms are displayed in Fig. 10.

We compare performance in three distinct ways which illustrate the simultaneous robustness to both leakage and dephasing errors in a pulse whose duration is the same as the Half-DRAG solution. First, we represent the dephasing-noise operator associated filter function. We see that the filter functions for the two optimized controls show a low-frequency-noise suppressing character similar to that illustrated in Fig. 3, while all other controls indicate broadband noise admittance. Next, we use the numerical simulation tools described in Sec. III D to determine control robustness to quasi-static dephasing errors. In this circumstance the two optimized solutions demonstrate a broad *plateau* of fixed detunings over which the infidelity remains low, again following the experimental results of Fig. 8b,c. Finally, we simulate the full evolution of the three states of the qutrit under application of the net $X_\pi$ rotation and applied noise. Here we see that despite the complex dynamics at times less than the gate time, at the conclusion of the gate the optimized solutions show the appropriate net state transfer. Other optimization approaches such as RC-filtered and slew-rate-bounded controls have been used to demonstrate similar performance. Overall these solutions represent new controls that – for the first time – allow simultaneous leakage-error and dephasing-noise suppression in a single optimized construction.

## C. Robust control for parametrically-driven superconducting entangling gates

Parametric activation of entangling gates presents a paradigm enabling tunable, high-fidelity two-qubit gates [92, 93]. This overcomes the scaling penalty imposed by frequency crowding in conventionally coupled transmons [126], though it suffers from decoherence channels arising from control noise in the parametric drive. We perform first-principles analyses of dominant error channels and introduce novel gate structures incorporating numeric optimization via tools described in Sec. III C in order to suppress the influence of these control-induced errors.

Two-qubit parametrically-driven gates may be implemented between one fixed- and one tunable-frequency transmon. A control flux drive $\Phi(t)$, with frequency $\omega_p$ and phase offset $\theta_p$, is applied to the tunable-frequency transmon resulting in a modulated transition frequency of the form

$$\omega_T(t) = \bar{\omega}_T + \tilde{\omega}_T \cos(2\omega_p t + 2\theta_p) \quad (81)$$

where $\bar{\omega}_T$ is the average shift in qubit frequency and $\tilde{\omega}_T$ is the amplitude of the modulation caused by the applied flux drive. The Hamiltonian for the system under this modulation, transforming to an interaction picture, takes the form

$$\begin{aligned}
H_{\text{int}}(t) = g(t) &\sum_{n=-\infty}^{\infty} J_n\left(\frac{\tilde{\omega}_T}{2\omega_p}\right)e^{+i(2\omega_p t + 2\theta_p)n} \\
&\times \left\{ e^{-it\Delta}|10\rangle\langle 01| \right. \\
&\quad + \sqrt{2}e^{-i(\Delta+\eta_F)t}|20\rangle\langle 11| \\
&\quad + \sqrt{2}e^{-i(\Delta-\eta_T)t}|11\rangle\langle 02| \\
&\quad \left. + 2e^{-i(\Delta+\eta_F-\eta_T)t}|21\rangle\langle 12| + \text{H.C.} \right\}
\end{aligned} \quad (82)$$

Here $g(t)$ describes the capacitive coupling between the transmon qubits; $\eta_T(\eta_F)$ are the positively-defined anharmonicities for the tunable-frequency (fixed-frequency) transmons; $\Delta = \bar{\omega}_T - \omega_F$ is the detuning between the average transition frequency of the tunable-frequency qubit and the fixed transition frequency of the fixed-frequency qubit; and $J_n(x)$ is the $n$th-order Bessel function of the first kind. A detailed description of the underlying physical system and the derivation of the associated Hamiltonians can be found in [91–93, 127].

Target entangling gates are activated by matching the modulation frequency $\omega_p$ to the detuning between relevant energy levels of the capacitively-coupled transmons. For typical experimental parameters, the time-dependent phase factors on the Hamiltonian operators above lead to rapidly-oscillating terms in the system evolution that effectively suppress the coupling rate to the associated transitions. Activation of a target transition is achieved by resonantly tuning the drive frequency to cancel the associated phase factor. In particular:

$$\begin{aligned}
&i\text{SWAP}: &&|10\rangle \leftrightarrow |01\rangle &&2n\omega_p = \Delta &&(83) \\
&\text{CZ}_{20}: &&|11\rangle \leftrightarrow |20\rangle &&2n\omega_p = \Delta + \eta_F &&(84) \\
&\text{CZ}_{02}: &&|11\rangle \leftrightarrow |02\rangle &&2n\omega_p = \Delta - \eta_T &&(85)
\end{aligned}$$

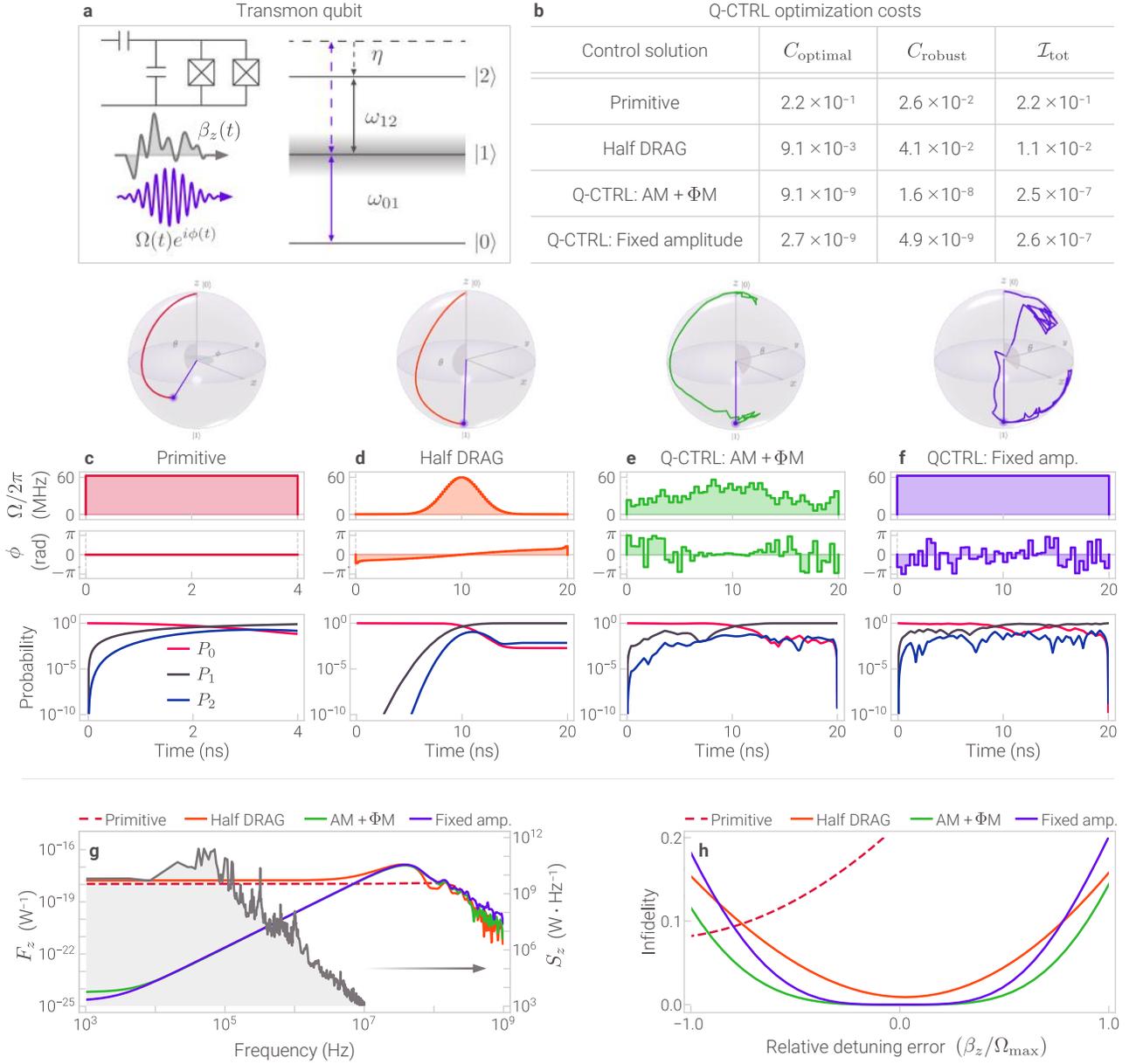

FIG. 10. Q-CTRL pulses optimized to suppress leakage and dephasing in a transmon qubit. (a) Energy level diagram and schematic of transmon qubit, modelled as a 3-level system with anharmonicity $\eta$. Dephasing is modelled as a time-varying shift $\beta_z(t)$ in the qubit energy splitting, resulting in an effective detuning of the drive $\Omega(t)e^{i\phi(t)}$ form resonance. (b) Performance metrics for control solutions presented in (c-f). Cost functions $C_{\text{optimal}}$ and $C_{\text{robust}}$ are defined in Table I. The total infidelity is computed as $\mathcal{I}_{\text{tot}} = \mathcal{I}_{\text{optimal}} + \mathcal{I}_{\text{robust}}$ with $\mathcal{I}_{\text{robust}}$ given by the integral in Eq. 31, evaluated using the PSD plotted on the right axis in panel (g). (c-f) primitive and Half-DRAG compared to optimized Q-CTRL pulses. Optimization constraints (Table II): (green) none (purple) fixed drive amplitude. All pulses are designed to implement a $X_{\pi}$ gate, with ideal population transfer $[P_0, P_1, P_2] : [1, 0, 0] \rightarrow [0, 1, 0]$. In each panel: (top) state evolution visulized on the Bloch sphere (App. F); (middle) waveforms plotted in polar coordinates (Eq. B17); (bottom) evolution of state populations simulated during pulse application, including leakage and dephasing (Algo. 3). Q-CTRL solutions suppress errors in final populations $P_0$ and $P_2$ (leakage channel) by orders of magnitude compared to both primitive and Half-DRAG in the presence of noise. (g) Robustness in the frequency domain: (left axis) dephasing filter functions computed for all pulses (Algo. 1); (right axis) PSD for dephasing field $\beta_z(t)$. Filter functions for Q-CTRL pulses are small at low frequencies, indicating superior dephasing suppression. (h) Robustness to quasi-static detuning errors for each pulse. Gate infidelities are computed while scanning over a constant value of $\beta_z$. Flatter response of Q-CTRL pulses indicates superior robustness.

The use of a parametric drive with a user-defined amplitude, phase, and frequency introduces a new control-induced channel for errors in the gate. To account for these control errors we assume the flux bias, and consequently the parametric drive in Eq. 81, experience three distinct error processes

$$\text{modulation offset error:} \quad \bar{\omega}_T \rightarrow \bar{\omega}_T + \bar{\epsilon}_T \quad (86)$$
$$\text{modulation amplitude error:} \quad \tilde{\omega}_T \rightarrow \tilde{\omega}_T + \tilde{\epsilon}_T \quad (87)$$
$$\text{modulation frequency error:} \quad \omega_p \rightarrow \omega_p + \epsilon_p \quad (88)$$

where the $\epsilon$ are assumed to be small errors. These generate additional Hamiltonian terms which, performing a Taylor expansion in the small offset parameters and moving to the interaction picture, result in the noise Hamiltonian $H_{\text{noise}}(t) = \beta(t)N$, where $\beta(t)$ captures the effective noise strength. We introduce a noise-operator of the form

$$N = \mathbb{I}_F \otimes (\Pi_1 + 2\Pi_2). \quad (89)$$

Here $\Pi_i = |i\rangle\langle i|$ defines the projection operator onto the $i$th eigenstate of the tunable-frequency transmon, and $\mathbb{I}_F$ is the identity on the fixed-frequency transmon.

The $i$SWAP interaction is activated by resonantly driving the $|10\rangle\langle01|$ term, for example by setting the 1st-order ($n = 1$) resonance condition $\omega_p = \Delta/2$. Assuming this configuration we may therefore restrict attention to the relevant ($4 \times 4$) $i$SWAP subspace, spanned by the eigenstates

$$|00\rangle, \quad |10\rangle, \quad |01\rangle, \quad |11\rangle. \quad (90)$$

In this case the remaining rapidly-oscillating terms may be ignored, and Eq. 82 reduces to

$$H_{i\text{SWAP}}(t) \approx \frac{1}{2}\Lambda(t)e^{+i\xi(t)}|10\rangle\langle01| + \text{H.C.} \quad (91)$$

where the parametric coupling rate $\Lambda = 2g(t)J_1\left(\frac{\tilde{\omega}_T}{2\omega_p}\right)$ and the parametric drive phase $\xi = 2\theta_p$.

On the $i$SWAP subspace the noise operator defined in Eq. 89 reduces to

$$N = \frac{1}{2}\mathbb{I}_F \otimes \sigma_z = \frac{1}{2}\begin{bmatrix} -1 & 0 & 0 & 0 \\ 0 & 1 & 0 & 0 \\ 0 & 0 & -1 & 0 \\ 0 & 0 & 0 & 1 \end{bmatrix} \quad (92)$$

resembling dephasing on the subspace of the tunable-frequency qubit. This new analysis thus shows that the three different channels for the introduction of noise via the parametric drive are manifested at the Hamiltonian level as an effective dephasing process.

Our objective is now to craft control solutions which are able to suppress this effective dephasing channel using the control available to us. Incorporating the fixed-frequency transmon term

$$H_{\text{qubit-F}}(t) = \left(\frac{1}{2}\Omega(t)e^{+i\phi(t)}|0\rangle\langle1| + \text{H.C.}\right) \otimes \mathbb{I}_T \quad (93)$$

the full control Hamiltonian is then written

$$H_{\text{ctrl}}(t) = H_{i\text{SWAP}}(t) + H_{\text{qubit-F}}(t). \quad (94)$$

This Hamiltonian is then parameterized according to the prescription in App. B, $H_{\text{ctrl}}(t) = \boldsymbol{\gamma}(t)\boldsymbol{C} + \text{H.C.}$ in terms of the drive pulses and operators

$$\boldsymbol{\gamma}(t) = \left[\Lambda(t)e^{+i\xi(t)}, \ \Omega(t)e^{+i\phi(t)}\right],$$

$$\boldsymbol{C} = \begin{bmatrix} \frac{1}{2}|10\rangle\langle01| \\ \frac{1}{2}|0\rangle\langle1| \otimes \mathbb{I} \end{bmatrix}.$$

We introduce three novel solutions providing robustness against control errors in the parametrically activated gate. All combine an $i$SWAP coupling drive $\Lambda(t)$ with single-qubit rotations which yields full control over the relevant subspace. This may be examined by observing that sandwiching an $i$SWAP operation between a pair of single-qubit $X_\pi$ gates permits the realization of the (Hermitian) operator

$$A_{\text{eff}} \propto \begin{bmatrix} 0 & 0 & 0 & 1 \\ 0 & 0 & 0 & 0 \\ 0 & 0 & 0 & 0 \\ 1 & 0 & 0 & 0 \end{bmatrix}, \quad (95)$$

similar in structure to a general NOT gate. Solutions need not employ this particular gate, but leverage the full controllability afforded by the *combination* of single-qubit and $i$SWAP control modulation.

With this formulation of the control problem we are able to directly deploy numeric optimization in order to find control solutions combing modulation of the parametric drive and single-qubit control. In Fig. 11c,d we present two representative numerically optimized solutions subject to constraints outlined in Table II. First, we produce a fixed-amplitude solution which combines a phase-modulated coupling drive with an *interleaved* single-qubit control, compatible with situations in which both controls cannot be applied simultaneously. The single-qubit operations are enacted with fixed amplitude, but variable durations (hence variable rotation angles) and variable phase. Similarly we present a *band-limited, concurrent* solution incorporating these two drives. Here we have enforced, as an example, an RC-filter on the controls (see Table II) to match potential band limits as may be experienced in a system with finite transmission-line bandwidth. In these cases we observe enhanced robustness to quasi-static errors, validated by time-domain simulation, as well as time-varying noise captured through filter functions (Fig. 11e,f).

We can compare these solutions to an approach defined analytically (Fig. 11b), recognizing that dephasing noise can be mitigated by the action of the spin echo and Walsh-modulation on a driven operation. We combine single-qubit $X_\pi$ operations with an $i$SWAP control envelope defined using a superposition of Walsh functions. This approach has previously been used to craft dephasing-robust single-qubit driven operations [34, 58] and provide a simple means to realize error robustness. This solution provides similar overall performance, but requires $\sim 18.5\%$ more time to execute than gate resulting from a fixed-amplitude constrained optimization.

Overall these examples demonstrate how high-performance solutions may be achieved under a wide range of constraints for entangling gates in superconducting circuits. We have achieved similar results for the CZ gate and non-parametric cross-resonance gates.

### D. Experimental noise characterization of multiqubit circuits on an IBM cloud QC

Microscopic noise characterization is widely employed in the development of optimized control solutions for a range of devices including superconducting qubits [128]. However, existing protocols have focused on the characterization of global fields measured at the single-qubit level [107, 110, 129]. The combination of multi-dimensional filter functions (Sec. III B) and flexible noise reconstruction algorithms (Sec. III E) permit new insights to be gleaned from real experimental quantum computing hardware.

We have focused on the characterization of previously unidentified microscopic noise sources present in entangling gates executed on cloud-based superconducting quantum computers [23]. At present access to such systems is highly restricted, making the arbitrary application of complex modulated controls on subspaces within the machines impossible. Accordingly, we have developed and deployed a simplified probe protocol consisting of sequences of entangling gates for two-qubit noise characterization. This probing sequence is readily implementable on the current IBM quantum computing platform, with filter functions for the individual sequences calculable using the software functionality introduced in Sec. III E.

Our probing sequence is designed to characterize two-qubit dephasing noise defined by the noise operator

$$N = \frac{1}{2}(Z_a - Z_b) \qquad (96)$$

where $Z_a$ and $Z_b$ are the Pauli $Z$ operator on qubits $a$ and $b$, respectively. This probing sequence consists of a fixed number $M$ of single-qubit and two-qubit quantum gates, in which each quantum gate has a fixed gate duration of $T_g$. Fixing $M$ - and in particular the number of two-qubit gates used - ensures that signatures arising from imperfect execution of the entangling gates do not vary between sequences and swamp the noise signals to be measured. In general, $M$ can be chosen suitably based on the hardware specification and $T_g$; here, we choose $M = 66$ to ensure the total duration of the experiment is within the coherence time of the IBM NISQ computers and $T_g = 110$ ns.

We first prepare the Bell state $(|01\rangle + |10\rangle)/\sqrt{2}$ by ap-

plying a sequence of 3 quantum gates defined by

$$U_E = X_b \text{CNOT}_{ab} H_a, \qquad (97)$$

where $\text{CNOT}_{ab}$ is the controlled-not gate on control qubit $a$ and target qubit $b$, $H_a$ is the Hadamard operator on qubit $a$ and $X_b$ is a NOT gate on qubit $b$. Then $i$ identity gates are applied to both qubits followed by a swap operation defined as

$$\text{SW} = \text{CNOT}_{ab} H_{ab} \text{CNOT}_{ab} H_{ab} \text{CNOT}_{ab}, \qquad (98)$$

where $H_{ab} = H_a \otimes H_b$. A second SW operation is applied after an additional $j$ identity gates. Then $M - 16 - i - j = 50 - i - j$ identity gates are employed before reverting the quantum state via the $U_E^\dagger$ operation. The schematic of the probing sequence is shown in Fig. 12b. This probing sequence effectively implements a series of $\pi$ pulses in the two-qubit system, analogous to implementing a CPMG (Carr-Purcell-Meiboom-Gill) sequence [115, 116] used in single-qubit dynamic decoupling, or measuring spin relaxation in NMR.

A set of generalized filter functions and the composite spectral sensitivity function for the set of control sequences can be constructed for each sequence as $i, j$ vary (Fig. 12d). Subsequently, when the two-qubit system is subject to noise, a corresponding set of infidelity measurements can be obtained and used to infer the noise power spectral density via the SVD noise reconstruction method described in Sec. III E. The hardware constraints we face on implementation - result in a somewhat inefficiently conditioned set of measurements, each probing overlapping spectral regions. This makes standard matrix inversion approaches to spectrum reconstruction impossible and helps demonstrate the value of flexible spectral estimation techniques.

We implement the new probing sequence on the IBM Q 5 Tenerife (ibmqx4) quantum computer [130]. Fig. 12a depicts the device chip layout and qubit couplings; here, we choose the qubit pair $Q_0$ and $Q_2$ for the experimental results displayed, but all pairs have been characterized in detail yielding qualitatively similar results. For this device, $\text{CNOT}_{20}$ can be natively implemented between these two qubits.

The values of $i$ and $j$ for the probing sequences are varied to construct a set of filter functions with high sensitivity over a broad range of frequencies of interest, up to 250 kHz. For each chosen values of $i, j$, the corresponding quantum circuit was executed over 8192 shots, yielding the average infidelity depicted in Fig. 12c as a colorscale heatmap. For a fixed $i$, as $j$ increases until the second SW operation is placed roughly half-way between the first SW and the $U_E^\dagger$ operations, the average infidelity becomes small, resembling an echo-effect similar to that of CPMG sequences. However there is additional structure present which breaks the symmetry of these graphs, indicating other noise contributions at higher frequencies.

Utilizing both the SVD and CO methods (Sec. III E), a reconstructed noise power spectral density based on these results is shown in Fig. 12e. These result shows that with high confidence, the two-qubit dephasing noise exhibits a low-frequency noise component and a repeatable higher-frequency noise contribution in the range $150 - 200$ kHz.

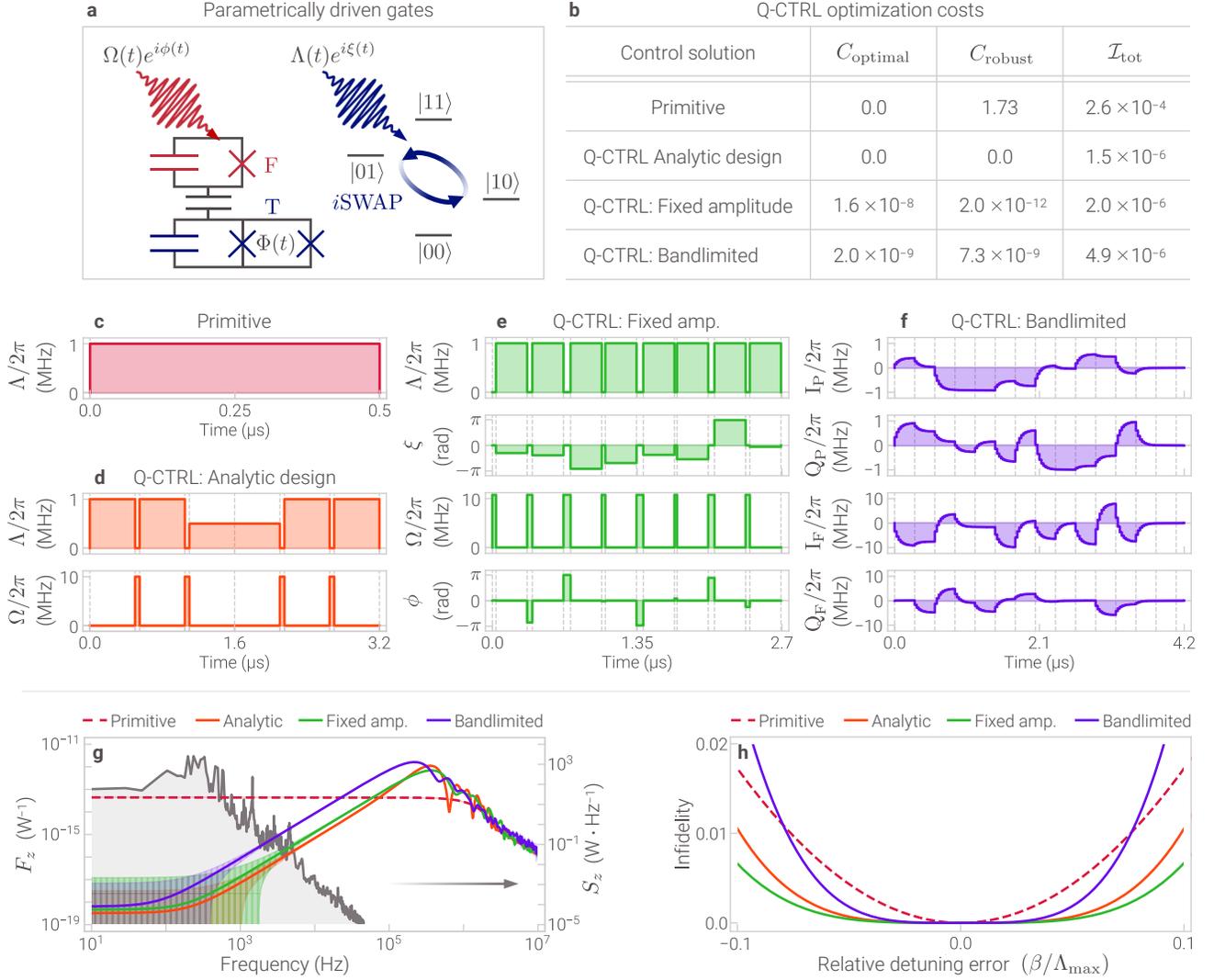

FIG. 11. Q-CTRL pulses optimized to suppress control errors for parametrically driven two-qubit entangling gates in super-conducting circuits. (a) Schematic of capacitively-coupled transmon qubits with fixed frequency ($F$) and tunable frequency ($T$). Single-qubit interactions are driven on $F$ by the pulse $\Omega(t)e^{i\phi(t)}$, while parametric modulation of the flux $\Phi(t)$ generates effective $i$SWAP interactions described by the pulse $\Lambda(t)e^{i\xi(t)}$. Noise enters the system via hardware errors on the flux modulation. (b) Performance metrics for control solutions presented in (c-f). Cost functions $C_{\text{optimal}}$ and $C_{\text{robust}}$ are defined in Table I. The total infidelity is computed as $\mathcal{I}_{\text{tot}} = \mathcal{I}_{\text{optimal}} + \mathcal{I}_{\text{robust}}$ with $\mathcal{I}_{\text{robust}}$ given by the integral in Eq. 31, evaluated using the PSD plotted on the right axis in panel (g). (c) Primitive coupling gate. (d) Analytically designed Walsh-modulated gate. (e) Q-CTRL optimized gate: interleaved single- and two-qubit controls with fixed-amplitude constraint (phase modulation only). (f) Q-CTRL optimized gate: simultaneous single- and two-qubit controls with band-limited 5 MHz RC filter constraint (Table II). (g) Robustness in the frequency domain: (left axis) filter functions computed for all pulses (Algo. 1); (right axis) PSD for noise field $\beta(t)$ associated effective detuning errors generated by the noise operator $N$ (Eq. 92). Filter functions for Q-CTRL pulses are small at low frequencies, indicating superior dephasing suppression. (h) Robustness to quasi-static detuning errors for each pulse. Gate infidelities are computed while scanning over the relative error defined by $\beta/\Lambda_{\text{max}}$, where $\beta$ scales the magnitude of the noise Hamiltonian as $H_{\text{noise}} = \beta N$, and $\Lambda_{\text{max}}$ denotes the maximum permissible value for the parametric drive amplitude $\Lambda(t)$ ($2\pi \times 1$ MHz for these simulations). Flatter response of Q-CTRL pulses indicates superior robustness.

The breadth of spectral features observed is due to the Fourier limits of the individual measurements employed in the reconstruction routine, as confirmed by numerical simulations.

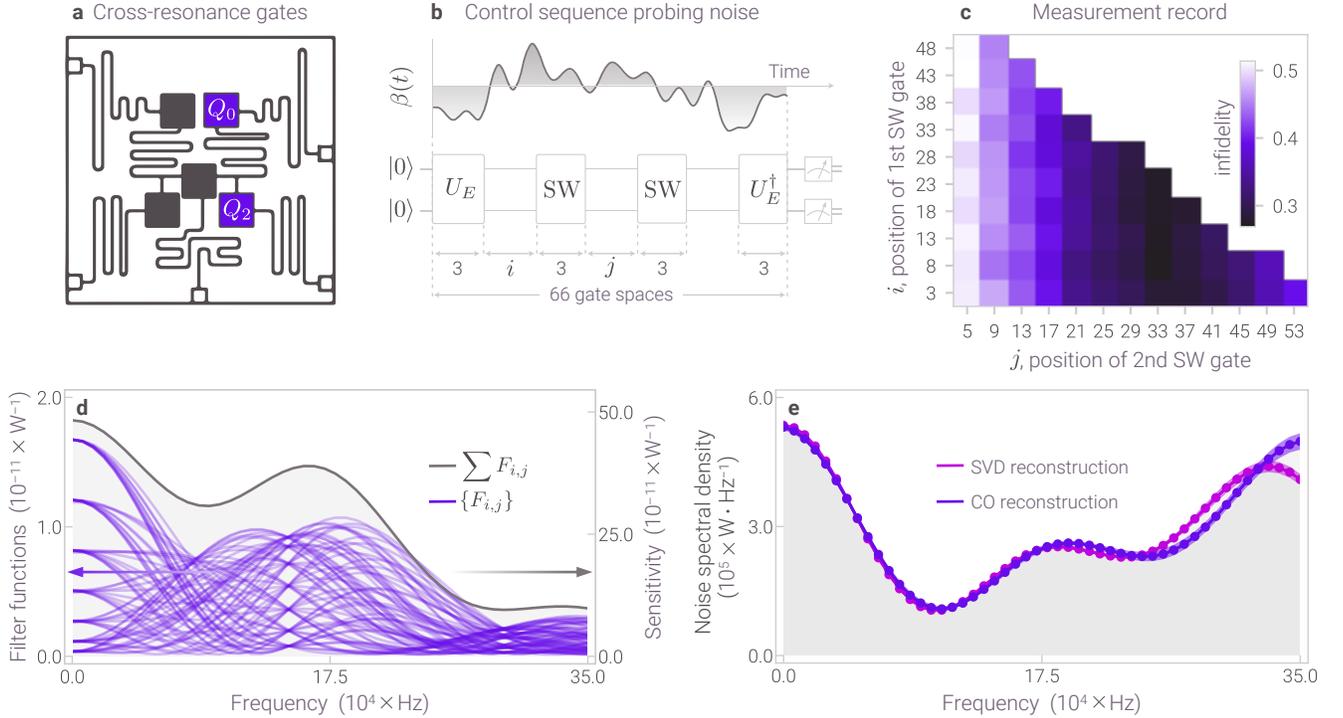

FIG. 12. Noise characterization in the IBM Q 5 Tenerife (ibmqx4) device. (a) Chip layout is shown identifying the two qubits for which data are presented. (b) Schematic of measurement procedure: (top) time-varying two-qubit dephasing noise associated with the noise operator in Eq. 96; the control sequence probing the noise, parameterized by the timing positions $(i, j)$ of the two SW gates. (c) Measured infidelity from all probe sequences $(i, j)$. (d) Multi-dimensional filter functions $\{F_{i,j}\}$ computed as in Algo. 1 for the noise channel in Eq. 96, for all probe sequences $(i, j)$. Sensitivity to the target noise process over the frequency domain $[0, 350]$ kHz is captured by $\sum F_i$ (grey fill). (e) Reconstructed two-qubit dephasing PSD using SVD (magenta) and CO (violet) methods described in Algo. 4.

We have observed similar performance across multiple qubit pairs, with variations in the strength of the quasi-static component. Numerical simulations have been used to demonstrate that similar results to those shown in Fig. 12e arise for a simple spectrum composed of a quasi-static noise component and a single fixed-frequency spur at higher frequencies. Similarly we have confirmed by engineering numerically synthesized data used in the reconstruction that the presence of the feature in the range $150 - 200$ kHz does not appear to be an artefact of either the measurement routine or the reconstruction method.

These experimental results represent an early demonstration of microscopic noise characterization within two-qubit gates using a new software toolkit permitting greater flexibility in control-based quantum noise spectroscopy. The identification of a high-frequency spectral component in a range commonly associated with electronic noise provides guidance on system improvement and noise suppression strategies for these machines.

### E. Crosstalk-resistant circuit compilation

In this subsection we demonstrate the use of optimal control for algorithmic design and "hardware-aware com-pilation". We consider a complex circuit composed of multiple interacting transmons and subject to unwanted cross-coupling. We use numerical optimization in order to implement a target circuit, subject to constraints on available controls and circuit duration, and optimized to combat always-on cross-talk errors through the structure of the circuit itself (no gate-level optimization). Our objective is to demonstrate the utility of the optimization software tools introduced in Sec. III C for a new class of problem in a high-dimensional Hilbert space.

Due to the low anharmonicity of transmons, quantum computations can be designed to exploit the three lowest-energy levels. The relative detunings between the various energy levels in an ensemble of transmons gives rise to an always-on effective ZZ-type coupling Hamiltonian that can be exploited for generating entangling operations. However, this coupling also leads to residual cross-talk errors that degrade algorithmic performance. Below we describe an example physical system and create an optimized circuit construction which suppresses these residual couplings. In spirit, this approach is similar to the low-level compilation of collections of logical subcircuits using analog control waveforms - rather than a universal gate set - as relayed in [131].

We consider a linear arrangement of 5 qutrits, labelled

$q \in \{1, ...5\}$. For a given qutrit pair $p = (q, q+1)$, characteristic detunings between respective energy levels generate relative phases on states $|11\rangle$, $|12\rangle$, $|21\rangle$, and $|22\rangle$. In this case the total coupling Hamiltonian is written

$$\begin{aligned}
H_{zz} = \ &H_{zz}^{(1,2)} \otimes \mathbb{I} \otimes \mathbb{I} \otimes \mathbb{I} \\
&+ \mathbb{I} \otimes H_{zz}^{(2,3)} \otimes \mathbb{I} \otimes \mathbb{I} \\
&+ \mathbb{I} \otimes \mathbb{I} \otimes H_{zz}^{(3,4)} \otimes \mathbb{I} \\
&+ \mathbb{I} \otimes \mathbb{I} \otimes \mathbb{I} \otimes H_{zz}^{(4,5)}
\end{aligned} \tag{99}$$

where nearest-neighbour interactions between pair $p$ are described by

$$\begin{aligned}
H_{zz}^p = \ &\alpha_{11}^p \, |11\rangle\langle 11| + \alpha_{12}^p \, |12\rangle\langle 12| \\
&+ \alpha_{21}^p \, |21\rangle\langle 21| + \alpha_{22}^p \, |22\rangle\langle 22|
\end{aligned} \tag{100}$$

and the $\alpha_{ij}^p$ are effective coupling strengths, tabulated in Table III for all pairs. Here $\mathbb{I}$ is the identity on a 3-dimensional single-qutrit Hilbert space, $H_{zz}^p$ operates on a $3^2$-dimensional Hilbert space associated with the $p$th qutrit pair, and $H_{zz}$ operates on the $3^5$-dimensional Hilbert space associated total 5-qutrit system.

As an example algorithm, we consider a circuit on this 5-qutrit system which seeks to simultaneously execute controlled-sum (CSUM) gates on qutrit pairs $(1, 2)$ and $(3, 4)$, while leaving the 5th qutrit unaffected. This may be expressed formally as

$$U_{\text{target}} = U_{C\phi} \otimes U_{C\phi} \otimes \mathbb{I} \tag{101}$$

where $U_{C\phi}$ is a 2-qutrit phase gate, locally equivalent to a CSUM, defined as

$$U_{C\phi} = \begin{array}{c} \\ \\ \\ \begin{matrix}{\scriptstyle |00\rangle} \\ {\scriptstyle |01\rangle} \\ {\scriptstyle |02\rangle} \\ {\scriptstyle |10\rangle} \\ {\scriptstyle |11\rangle} \\ {\scriptstyle |12\rangle} \\ {\scriptstyle |20\rangle} \\ {\scriptstyle |21\rangle} \\ {\scriptstyle |22\rangle} \end{matrix} \end{array} \overset{\begin{matrix}{\scriptstyle |00\rangle} \ {\scriptstyle |01\rangle} \ {\scriptstyle |02\rangle} \ {\scriptstyle |10\rangle} \ {\scriptstyle |11\rangle} \ {\scriptstyle |12\rangle} \ {\scriptstyle |20\rangle} \ {\scriptstyle |21\rangle} \ {\scriptstyle |22\rangle}\end{matrix}}{\begin{bmatrix} 1 & & & & & & & & \\ & 1 & & & & & & & \\ & & 1 & & & & & & \\ & & & 1 & & & & & \\ & & & & \omega^* & & & & \\ & & & & & \omega & & & \\ & & & & & & 1 & & \\ & & & & & & & \omega & \\ & & & & & & & & \omega^* \end{bmatrix}}$$

and $\omega \equiv e^{2\pi i/3}$.

Assuming excitations between qutrit states $|0\rangle \leftrightarrow |2\rangle$ cannot be controlled, we consider a restricted control basis spanning only single-qutrit operations coupling states $|0\rangle \leftrightarrow |1\rangle$ and $|1\rangle \leftrightarrow |2\rangle$. We further assume these may be implemented instantaneously and in parallel across all qutrits within the circuit.

The total control Hamiltonian may be written

$$H_{\text{ctrl}}(\boldsymbol{\Omega}, \boldsymbol{\phi}) = \sum_{q=1}^{5} \sum_{\nu \in \{01, 12\}} H_\nu^q(\Omega_\nu^q, \phi_\nu^q) \tag{102}$$

where

$$\boldsymbol{\Omega} = \left(\Omega_{01}^1, \ \dots \ \Omega_{01}^5, \Omega_{12}^1, \ \dots \ \Omega_{12}^5\right), \tag{103}$$

$$\boldsymbol{\phi} = \left(\phi_{01}^1, \ \dots \ \phi_{01}^5, \phi_{12}^1, \ \dots \ \phi_{12}^5\right). \tag{104}$$

The assumption of instantaneous single-qutrit operations allows us to absorb the duration $\Delta t$ over which the corresponding unitary is implemented, yielding a total evolution operator

$$U_{\text{ctrl}}(\boldsymbol{\theta}, \boldsymbol{\phi}) = \exp\left[-iL(\boldsymbol{\theta}, \boldsymbol{\phi})\right], \qquad \boldsymbol{\theta} = \Delta t \, \boldsymbol{\Omega} \tag{105}$$

where

$$L(\boldsymbol{\theta}, \boldsymbol{\phi}) = \sum_{q=1}^{5} \sum_{\nu \in \{01, 12\}} \theta_\nu^q \exp\left[+i\phi_\nu^q\right] C_\nu^q + \text{H.C.} \tag{106}$$

Here we refer to driven operations on the $q$th qutrit and $\nu$th transition, with Rabi rate $\Omega_\nu^q$ and phase $\phi_\nu^q$.

Within this formulation we have defined single-qutrit drive operators

$$C_{10} = \frac{1}{2} \, |1\rangle\langle 0| = \frac{1}{2} \begin{bmatrix} 0 & 0 & 0 \\ 1 & 0 & 0 \\ 0 & 0 & 0 \end{bmatrix} \tag{107}$$

$$C_{21} = \frac{1}{2} \, |2\rangle\langle 1| = \frac{1}{2} \begin{bmatrix} 0 & 0 & 0 \\ 0 & 0 & 0 \\ 0 & 1 & 0 \end{bmatrix} \tag{108}$$

which are generalized for the $q$th qutrit within the multi-qutrit system

$$C_\nu^q \equiv \mathbb{I}^{\otimes(q-1)} \otimes C_\nu \otimes \mathbb{I}^{\otimes(n-q-1)}, \quad \nu \in \{01, 12\}. \tag{109}$$

Our approach to circuit-level optimization for residual-cross-talk suppression is through the integration of deterministic dynamic decoupling. We partition the circuit implementing Eq. 101 into $m$ periods of free evolution under the always-on coupling Hamiltonian Eq. 99, interleaved with $m + 1$ gates of the form

$$P_j = \prod_{\ell=1}^{k} U_{\text{ctrl}}(\boldsymbol{\theta}_{j,\ell}, \boldsymbol{\phi}_{j,\ell}), \quad j \in \{0, ...m\} \tag{110}$$

each composed as products of $k$ distinct control unitaries of the form Eq. 105. In this expression the $j$th period, of duration $\tau_j$, starts and ends with instantaneous unitaries $P_{j-1}$ and $P_j$ respectively. This structure corresponds to a generalized dynamic decoupling sequence and augments the controllability of the 5-qutrit system due to the non-commuting terms in the operator products.

| qutrit pairs $p = (q, q+1)$ | $\alpha_{11}^p$ $(2\pi \cdot \text{MHz})$ | $\alpha_{12}^p$ $(2\pi \cdot \text{MHz})$ | $\alpha_{21}^p$ $(2\pi \cdot \text{MHz})$ | $\alpha_{22}^p$ $(2\pi \cdot \text{MHz})$ |
|---|---|---|---|---|
| $(1, 2)$ | $-0.27935$ | $0.1599$ | $-0.52793$ | $-0.74297$ |
| $(2, 3)$ | $-0.1382$ | $0.15827$ | $-0.33507$ | $-0.3418$ |
| $(3, 4)$ | $-0.276$ | $-0.6313$ | $0.24327$ | $-0.74777$ |
| $(4, 5)$ | $-0.26175$ | $-0.49503$ | $0.14497$ | $-0.70843$ |

TABLE III. Example ZZ-type coupling strengths between nearest-neighbor qutrit pairs within a circuit.

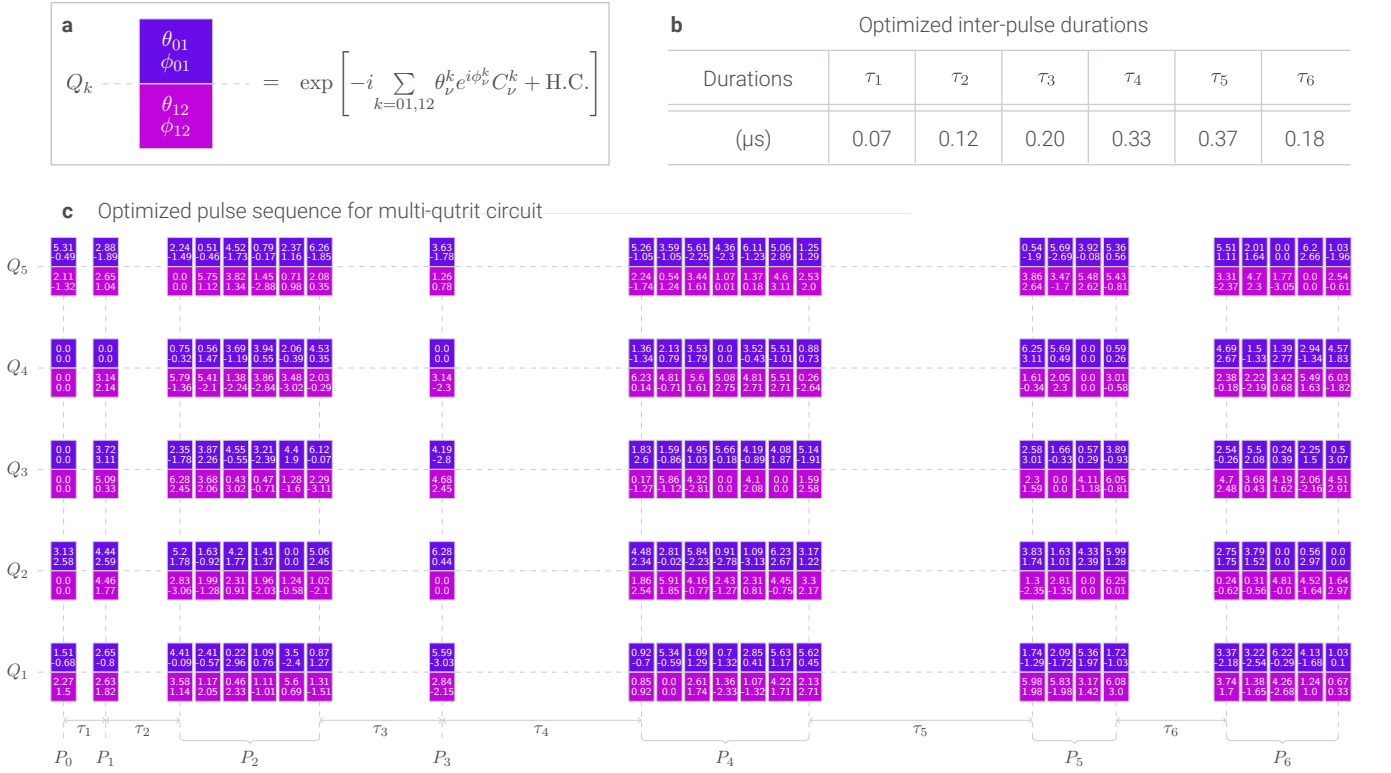

FIG. 13. Optimized 5-qutrit circuit. Each qutrit is controlled via $|0\rangle \leftrightarrow |1\rangle$ and $|1\rangle \leftrightarrow |2\rangle$ transitions. Control on each transition is captured via the phasor $\gamma_\nu = \theta_\nu e^{i\phi_\nu}$, comprising rotation, $\theta_\nu$, and phase angles, $\phi_\nu$, where $\nu \in 01, 12$. As shown in the inset, each single-qutrit operation is indicated via the block of angles $[\theta_{01}, \phi_{01} | \theta_{12}, \phi_{12}]$, with corresponding colours. Consecutive sequences of block arrays indicate product operations of the form $P_j$ defined in Eq. 110. Blocks of unitary operations are separated by durations $\tau_j$. All selected phasor components on the control operations and values of $\tau_j$ are returned via the circuit optimization procedure. The optimal cost for this circuit compilation is $\mathcal{I}_{\text{optimal}} = 5.7 \times 10^{-3}$.

This generalized dynamic decoupling sequence structure must now be numerically optimized in order to return the target circuit functionality expressed in Eq. 101, which necessarily entails cancellation of the $ZZ$ cross-coupling. Optimization of the control structure is performed on the search space spanned by the timing variables $\boldsymbol{\tau} = (\tau_1, ..., \tau_m)$, rotation variables $\boldsymbol{\theta}_{\ell,j}$, and phase variables $\boldsymbol{\phi}_{\ell,j}$ for $\ell \in \{1, ..., k\}$ and $j \in \{1, ..., m+1\}$. It is the combination of appropriately timed free-evolution periods with specific unitary operations (rather than simple bit flips as in standard dynamic decoupling), that allows the decoupling of the unwanted $ZZ$ interaction while implementing a non-identity operation.

Following the procedure described in Sec. III C 1, the timing, rotation, and phase variables form the basis for defining the array of generalized controls $\boldsymbol{v}$, appropriately normalized for efficient TensorFlow optimization. We also introduce an experimentally motivated constraint in that the circuit must not exceed the qutrit coherence time, set by $T_1$. We therefore compose the cost function

in Eq. 35 as

$$C(\boldsymbol{v}) = \mathcal{I}_{\text{optimal}}(\boldsymbol{v}) + C_{\text{duration}}(\boldsymbol{\tau}) \qquad (111)$$

where $\mathcal{I}_{\text{optimal}}(\boldsymbol{v})$ is defined in Sec. III A and $C_{\text{duration}}(\boldsymbol{\tau})$ imposes a penalty for exceeding the upper limit chosen for the circuit duration. We set a threshold of $\sim 1.5$ µs, chosen assuming a qutrit coherence time of $\sim 20$ µs.

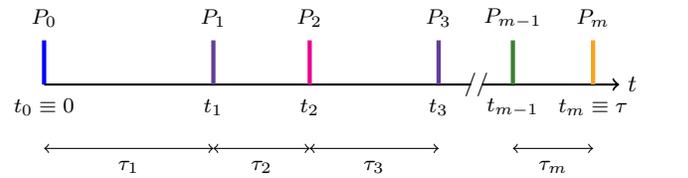

FIG. 14. Dynamic decoupling sequence composed of unitaries Eq. 105, with arbitrary inter-pulse separation times $\tau_i = t_i - t_{i-1}$, for $i \in \{1, ..., m\}$, resulting in a total duration $\tau = \sum_{i=1}^{m} \tau_i$. Here the different colours are used to indicate non-uniformity of the unitaries $P_i$, applied at times $t_i$.

The optimizer returns variations on the circuit structure composed of compound rotations on different qutrit levels with variable timing between these operations. The combination of driven rotations and their timing in the sequence is essential in performing the target net unitary with low infidelity; compactifying the circuit structure in order to reduce nominal dead time changes the cross-talk-suppressing nature of the circuit. In the example optimization realized in Fig. 13 we are able to improve the cross-talk limited fidelity $(1 − \mathcal{I}_{\text{optimal}})$ in the target unitary from ∼ 2.2% (under the simplest compilation, without any form of dynamic cross-talk suppression) to 99.4%.

This demonstration validates the premise of circuit-level optimization in order to realize deterministic error robustness. Treating hardware-aware circuit compilation as a challenge in optimal control - for either deterministic error suppression or decomposition of a complex circuit into a constrained set of physical-layer controls - is a feature set incorporated in the forthcoming FIRE OPAL package.

# V. CONCLUSION AND OUTLOOK

In this manuscript we have provided an overview of a new toolset built to allow users to incorporate quantum control into their research and application development. The software architecture combines locally installed packages coupled with a cloud-compute engine in order to deliver computational benefits for complex computations such as control optimizations. The specific products we introduced range from intuitive web interfaces with interactive visualizations through to advanced python toolkits for integration into professional programming and hardware, targeting a range of users from consultants and students through to hardware R&D teams.

As background, we have provided a thorough mathematical treatment of key tasks and approaches in quantum control - including the introduction of new techniques developed by our team - and describe how they are implemented algorithmically in software. We contextualized these capabilities through a series of theoretical case studies demonstrating the utility of these software capabilities in solving challenging problems in quantum control. In addition, we provided experimental evidence derived from real quantum computing hardware demonstrating quantum-control benefits such as suppression of noise susceptibility, error homogenization in multiqubit devices, gate-fidelity stabilization in time, and noise-spectroscopy in multiqubit gates.

Future development will expand functionality to integrate novel machine-learning tools for data analysis and hardware characterization at scale. For instance, we are investigating a number of time-series analysis techniques which enable the identification and extraction of system dynamics from discretized measurement records. It's common practice in quantum computing experiments to simply average together large data sets in order to obtain probabilistic information about *e.g.* quantum-state populations in algorithms. This procedure, however, is confounded by the presence of large-scale temporal drifts in hardware that can introduce dynamics in measurements that are not captured by simple averaging. The tools we are building include features for Gaussian Process Regression and Autoregressive Kalman Filtering [35], targeting both data fitting for the removal of background dynamics [132] and also predictive estimation for feedforward control stabilization of qubits and clocks [81]. These time-domain analytic frameworks are also useful for data fusion incorporating multiple measurement streams from sensors or measured qubits.

Similarly, we will be implementing novel automated and adaptive strategies for the tuneup, calibration, and optimization of mesoscale systems, moving beyond the brute-force strategy of independent calibration of all devices. In this space, advanced machine learning, reinforcement learning, and robotic control concepts provide new opportunities to facilitate rapid, autonomous bring up of devices in a way that will grow in importance as system sizes increase. We already have considerable effort in this area, taking inspiration from autonomous robotic control to facilitate adaptive measurement and data inference on large qubit arrays [36, 133], and will be investing heavily in this area in the future.

By combining novel advances in quantum control engineering with high-efficiency algorithmic development, we hope these tools will prove a valuable resource for a wide range of users. We believe that the integration of highly maintainable, professionally engineered software solutions targeting specialized tasks in the quantum computing stack will ultimately provide major benefits to the research and business communities, much like the introduction of specialist cloud security software has accelerated many aspects of cloud-service businesses.

# ACKNOWLEDGEMENTS

Q-CTRL efforts supported by Data Collective, Horizons Ventures, Main Sequence Ventures, Sequoia Capital (China), Sierra Ventures, and SquarePeg Capital. Development of multi-dimensional filter functions and SVD spectrum inversion technique by Q-CTRL supported by the US Army Research Office under Contract W911NF-12-R-0012. Q-CTRL is grateful to I. Siddiqi and D. Santiago for provision of device data which inspired circuit optimization results. Experimental work using trapped ions at USYD partially supported by the ARC Centre of Excellence for Engineered Quantum Systems CE170100009,the Intelligence Advanced Research Projects Activity (IARPA) through the US Army Research Office Grant No. W911NF-16-1-0070, and a private grant from H. & A. Harley.

The authors are grateful to all other colleagues at Q-CTRL whose technical and design work has supported the results presented in this paper. Backend: Kevin Nguyen, Ryan Barker, Stefano Tabacco and Luigi Cristofolini. Frontend: Rob Harkness and Yashar Zolmajdi. Design: Damien Metcalf and Christina Maresca. Quan-tum engineering: Viktor Perunicic for assistance demon-strating and describing the visualizer.

## Appendix A: Technical definitions

### 1. Frobenius inner product and Frobenius norm

For matrices $A, B \in \mathbb{C}^{m \times n}$, the Frobenius inner product is defined as

$$\langle A, B \rangle_F = \sum_{i,j} A_{ij}^* B_{ij} = \mathrm{Tr}\left(A^\dagger B\right) \tag{A1}$$

The inner product in Eq. A1 induces a matrix norm. For a matrix $A \in \mathbb{C}^{m \times n}$, the Frobenius norm is defined by

$$\|A\|_F = \sqrt{\langle A, A \rangle_F} = \sqrt{\sum_{i,j} |A_{ij}|^2} = \sqrt{\mathrm{Tr}\left(A^\dagger A\right)} \tag{A2}$$

### 2. Fourier transform

In this paper we exclusively use the non-unitary angular-frequency convention for Fourier transform pairs, defining

$$Q(\omega) \equiv \int_{-\infty}^{\infty} dt\, e^{-i\omega t} Q(t) \tag{A3}$$

$$Q(t) \equiv \frac{1}{2\pi} \int_{-\infty}^{\infty} d\omega\, e^{i\omega t} Q(\omega) \tag{A4}$$

where $Q(t)$ denotes any scalar-, matrix- or operator-valued function of time, and $Q(\omega)$ is its Fourier transform, implemented element-wise for matrices. For ease of notation we reuse the same symbol and simply change the argument to distinguish time- or frequency-domain transforms. To avoid confusion we also write $\mathscr{F}\left\{Q(t)\right\}(\omega) \equiv Q(\omega)$ and $\mathscr{F}^{-1}\left\{Q(\omega)\right\}(t) \equiv Q(t)$.

### 3. Power spectral density

Here we develop the relationship between noise processes in the time-domain and their frequency-domain representations. Let $\beta_k(t)$ for $k \in \{1, ..., n\}$ denote a set of scalar-valued noise fields. Using the definition for the Fourier transform set out in App. A 2, we establish the following relationships between time- and frequency-domain variables

$$\beta_k(t) = \frac{1}{2\pi} \int_{-\infty}^{\infty} d\omega\, e^{i\omega t} \beta_k(\omega), \tag{A5}$$

$$\beta_k(\omega) = \int_{-\infty}^{\infty} dt\, e^{-i\omega t} \beta_k(t). \tag{A6}$$

We assume the noise fields are independent[134], zero-mean random variables. The cross-correlation functions consequently vanish, namely

$$\langle \beta_j(t_1)\beta_k^*(t_2) \rangle = 0, \qquad j \neq k \in \{1, ..., n\} \tag{A7}$$

where the angle brackets denote an ensemble average over the stochastic variables. The frequency-domain variables inherit the equivalent property, namely

$$\langle \beta_j(\omega_1)\beta_k^*(\omega_2) \rangle = 0, \qquad j \neq k \in \{1, ..., n\}, \tag{A8}$$

which may be shown by substituting in Eq. A6, and invoking Eq. A7. We further assume the noise processes are *wide sense stationary*, implying the autocorrelation functions, defined as

$$C_k(t_2 - t_1) \equiv \langle \beta_k(t_1)\beta_k^*(t_2) \rangle, \qquad i \in \{1, ..., n\}, \tag{A9}$$

depend only on the time *difference* $\tau = t_2 - t_1$. Under these conditions the autocorrelation function for each noise field may be related to its power spectral density $S_i(\omega)$ using the Wiener-Khinchin Theorem [135]. Specifically,

$$C_k(t_2 - t_1) = \frac{1}{2\pi} \int_{-\infty}^{\infty} S_k(\omega) e^{i\omega(t_2 - t_1)} d\omega, \tag{A10}$$

which is consistent with *defining* the power spectral density as

$$S_k(\omega) \equiv \frac{1}{2\pi} \left\langle |\beta_k(\omega)|^2 \right\rangle. \tag{A11}$$

To show this observe

$$\langle \beta_k(\omega_1) \beta_k^*(\omega_2) \rangle = \left\langle \left( \int_{-\infty}^{\infty} dt_1 e^{-i\omega_1 t_1} \beta_k(t_1) \right) \left( \int_{-\infty}^{\infty} dt_2 e^{-i\omega_2 t_2} \beta_k(t_2) \right)^* \right\rangle \tag{A12}$$

$$= \int_{-\infty}^{\infty} dt_1 \int_{-\infty}^{\infty} dt_2 \, \langle \beta_k(t_1) \beta_k^*(t_2) \rangle \, e^{i\omega_2 t_2} e^{-i\omega_1 t_1} \tag{A13}$$

$$= \int_{-\infty}^{\infty} dt_1 \int_{-\infty}^{\infty} dt_2 \left( \frac{1}{2\pi} \int_{-\infty}^{\infty} S_k(\omega) e^{i\omega(t_2 - t_1)} d\omega \right) e^{i\omega_2 t_2} e^{-i\omega_1 t_1} \tag{A14}$$

$$= \frac{1}{2\pi} \int_{-\infty}^{\infty} d\omega S_k(\omega) \int_{-\infty}^{\infty} dt_1 e^{-i(\omega + \omega_1) t_1} \int_{-\infty}^{\infty} dt_2 e^{i(\omega + \omega_2) t_2} \tag{A15}$$

$$= \frac{1}{2\pi} \int_{-\infty}^{\infty} d\omega S_k(\omega) \big( 2\pi \cdot \delta(\omega + \omega_1) \big) \big( 2\pi \cdot \delta(-\omega + \omega_2) \big) \tag{A16}$$

$$= \begin{cases} 0 & \omega_1 \neq \omega_2 \\ 2\pi S_k(\omega_1) & \omega_1 = \omega_2 \end{cases} \tag{A17}$$

## Appendix B: Formal definition of the control Hamiltonian

As outlined in the main text, the central objectives of quantum control is to enhance the performance of a quantum system by leveraging the available controls against the influence of relevant noise sources. Delivering this for arbitrary quantum systems (qubits, qutrits, multi-qubit ensembles, etc.) requires a generalized formalism for describing the control Hamiltonian. In this appendix we introduce this formalism.

### 1. Generalized formalism

Let $\mathcal{H}$ be a $d$-dimensional Hilbert space for the controlled quantum system. The control Hamiltonian is written

$$H_{\text{ctrl}}(t) = \left( \sum_{j=1}^{d} \gamma_j(t) C_j + \text{H.C.} \right) + \sum_{l=1}^{s} \alpha_l(t) A_l + D \tag{B1}$$

in terms of the control operators $A_l, C_j, D \in \mathcal{H}$ and control pulses (waveforms) $\alpha_l(t) \in \mathbb{R}$ and $\gamma_j(t) \in \mathbb{C}$. To unpack this notation we introduce the nomenclature of *drives*, *shifts* and *drifts*, useful for mapping generalized quantum control concepts to common physical control variables. These are detailed in Table IV. In this framework, system evolution under $H_{\text{ctrl}}$ may be viewed as a combination of generalized rotations, driven by control pulses (real or complex) about effective *control axes*, defined by the associated operators.

| control term | operator | | pulse | |
|---|---|---|---|---|
| drive | $C_j$ | non-Hermitian | $\gamma_j(t)$ | $\mathbb{C}$: complex |
| shift | $A_l$ | Hermitian | $\alpha_l(t)$ | $\mathbb{R}$: real |
| drift | $D$ | Hermitian | - | - |
| | *symbol* | *type* | *symbol* | *type* |

TABLE IV. Decomposition of control Hamiltonian into generalized drive, shift and drift terms. Drive terms are defined by non-Hermitian operators, $C_j \neq C_j^\dagger$, and complex-valued control pulses $\gamma_j(t)$. Shift terms are defined by Hermitian operators, $A_l = A_l^\dagger$, and real-valued control pulses $\alpha_l(t)$. The operator $D$ is a time-independent Hermitian operator, which we refer to as the drift Hamiltonian.

### 2. Control solutions

Assuming the operator basis defined above, the control Hamiltonian may be expressed more compactly as

$$H_{\text{ctrl}}(t) = \left( \boldsymbol{\gamma}(t)\boldsymbol{C} + \text{H.C.} \right) + \boldsymbol{\alpha}(t)\boldsymbol{A} + D \tag{B2}$$

in terms of the vectorized objects defined by drive terms: $\qquad \boldsymbol{\gamma}(t) = \begin{bmatrix} \gamma_1(t), & \gamma_2(t), & \dots & \gamma_d(t) \end{bmatrix}, \qquad \boldsymbol{C} = \begin{bmatrix} C_1 \\ C_2 \\ \vdots \\ C_d \end{bmatrix}, \qquad t \in [0, \tau] \tag{B3}$ shift terms: $\qquad \boldsymbol{\alpha}(t) = \begin{bmatrix} \alpha_1(t), & \alpha_2(t), & \dots & \alpha_s(t) \end{bmatrix}, \qquad \boldsymbol{A} = \begin{bmatrix} A_1 \\ A_2 \\ \vdots \\ A_s \end{bmatrix}, \qquad t \in [0, \tau] \tag{B4}$ with drive and shift pulses listed as complex and real row vectors respectively, and corresponding drive and shift operators listed as column vectors. Given the operator-basis defined by $\boldsymbol{A}$ and $\boldsymbol{C}$, the most general description of control is therefore specified by the set of functions $\boldsymbol{\alpha}(t)$ and $\boldsymbol{\gamma}(t)$, defined on the time interval $t \in [0, \tau]$, defining the duration over which the control is applied. We refer to this structure as a *control solution*.

### 3. Control segments

It is often more useful to specify the form of the control Hamiltonian, or functional form of the control pulses $\boldsymbol{\gamma}(t)$ and $\boldsymbol{\alpha}(t)$, locally in time. In this case the time-domain $t \in [0, \tau]$ is partitioned into a series of intervals, or segments. The functional form of the control pulses are then defined on each segment. This is illustrated below for a shift pulse $\alpha_j(t)$. The time domain $t \in [0, \tau]$ has been formally partitioned into $m$ subintervals

$$[t_{i-1}, t_i], \qquad i \in \{1, ..., m\}, \qquad t_0 \equiv 0, \qquad t_m \equiv \tau \tag{B5}$$

where $t_{i-1}$ and $t_i$ are respectively the start and end times of the $i$th segment, and

$$\tau_i = t_i - t_{i-1} \tag{B6}$$

is its duration. The shift pulse $\alpha_j(t)$ is piecewise-constant, defined to take the constant value $\alpha_{i,j}$ on the $i$th segment, $t \in [t_{i-1}, t_i]$.

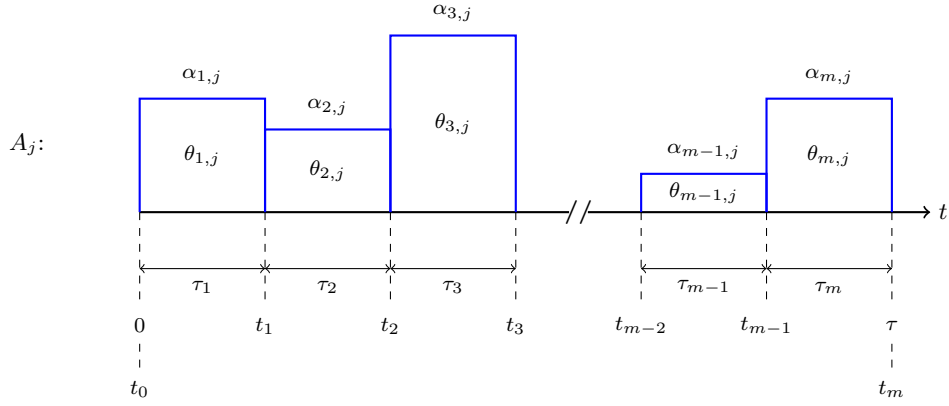

FIG. 15. Segmentation of control amplitude $\alpha_j(t)$ for control axis $A_j$ into $m$ segments. The area under the $i$th segment is given by the variable $\theta_{i,j} \equiv \alpha_{i,j} \tau_j$.

### 4. Generic shaped control segments

It may not always be desirable to treat each segment as constant-valued, as illustrated in Fig. 15. To facilitate a more general description of pulse shaping on a given segmentation in time, we introduce some further notation. Let the time domain be partitioned into $m$ segments, and define the window function on the $i$th segment by

$$\Theta_i(t) = \begin{cases} 1 & t \in [t_{i-1}, t_i] \\ 0 & \text{else} \end{cases}. \tag{B7}$$

Let $f(t)$ be a scalar or vector function of time, locally defined on the $i$th segment by

$$f_i(t) = \Theta_i(t) f(t) \qquad i \in \{1, ..., m\} \tag{B8}$$

such that

$$f(t) = \sum_{i=1}^{m} f_i(t). \tag{B9}$$

This partitioning define the map

$$S: \quad f(t) \quad \mapsto \quad \begin{bmatrix} f_1(t) \\ f_2(t) \\ \vdots \\ f_m(t) \end{bmatrix} \tag{B10}$$

which we refer to as the segments of $f(t)$. Using this notation we define the drive control segments

$$S(\boldsymbol{\gamma}(t)) = \begin{bmatrix} \gamma_{1,1}(t) & \gamma_{1,2}(t) & \dots & \gamma_{1,d}(t) \\ \gamma_{2,1}(t) & \gamma_{2,2}(t) & \dots & \gamma_{2,d}(t) \\ \vdots & \vdots & \ddots & \vdots \\ \gamma_{m,1}(t) & \gamma_{m,2}(t) & \dots & \gamma_{m,d}(t) \end{bmatrix}, \qquad \gamma_{i,j}(t) \equiv \Theta_i(t)\gamma_j(t) \tag{B11}$$

$$S(\boldsymbol{\alpha}(t)) = \begin{bmatrix} \alpha_{1,1}(t) & \alpha_{1,2}(t) & \dots & \alpha_{1,s}(t) \\ \alpha_{2,1}(t) & \alpha_{2,2}(t) & \dots & \alpha_{2,s}(t) \\ \vdots & \vdots & \ddots & \vdots \\ \alpha_{m,1}(t) & \alpha_{m,2}(t) & \dots & \alpha_{m,s}(t) \end{bmatrix}, \qquad \alpha_{i,l}(t) \equiv \Theta_i(t)\alpha_l(t) \tag{B12}$$

where the doubly-subscripted functions $\gamma_{i,j}(t)$ and $\alpha_{i,l}(t)$ define the *time-dependent* modulation envelopes for the $j$th phasor and $l$th amplitude on the $i$th segment. Each column maps to a distinct control, while each row maps a distinct segment. The segments of the control Hamiltonian are given by

$$S(H_{\text{ctrl}}(t)) = \begin{bmatrix} H_{\text{ctrl},1}(t) \\ H_{\text{ctrl},2}(t) \\ \vdots \\ H_{\text{ctrl},m}(t) \end{bmatrix} \tag{B13}$$

where the control Hamiltonian on the $i$th segment is given by

$$H_{\text{ctrl},i}(t) = \left( \Big[ S(\boldsymbol{\gamma}(t)) \Big]_i \boldsymbol{C} + \text{H.C.} \right) + \Big[ S(\boldsymbol{\alpha}(t)) \Big]_i \boldsymbol{A} + D \tag{B14}$$

$$= \left( \sum_{j=1}^{d} \gamma_{i,j}(t) C_j \;+\; \text{H.C.} \right) \;+\; \sum_{l=1}^{s} \alpha_{i,l}(t) A_l \;+\; D. \tag{B15}$$

Control is therefore completely specified by the $m \times (d+s)$ functions tabulated in the time-dependent control-space array as

$$\left[ \begin{array}{c|ccc|ccc} - & \gamma_1(t) & \cdots & \gamma_d(t) & \alpha_1(t) & \cdots & \alpha_s(t) \\ \hline \tau_1 & \gamma_{1,1}(t) & \dots & \gamma_{1,d}(t) & \alpha_{1,1}(t) & \dots & \alpha_{1,s}(t) \\ \vdots & \vdots & \ddots & \vdots & \vdots & \ddots & \vdots \\ \tau_m & \gamma_{m,1}(t) & \dots & \gamma_{m,d}(t) & \alpha_{m,1}(t) & \dots & \alpha_{m,s}(t) \end{array} \right] \tag{B16}$$

### 5. Control coordinates

Here we introduce notation conventions followed by Q-CTRL to define drive pulses, their decomposition, and their relationship to Hermitian and non-Hermitian control operators. Since the following structure applies to every pulse-operator pair $(\gamma_j(t), C_j)$, we drop the subscript $j$ for simplicity. The complex-valued pulse $\gamma(t) \in \mathbb{C}$ may be written in polar or Cartesian form. Namely,

| | | |
|---|---|---|
| Polar form: | $\gamma(t) = \Omega(t)e^{+i\phi(t)}$ | (B17) |
| Cartesian form: | $\gamma(t) = I(t) + iQ(t)$ | (B18) |

allowing us establish the following control forms

| | | | | | |
|---|---|---|---|---|---|
| modulus: | $\Omega(t)$ | $=$ | $\lvert \gamma(t) \rvert$ | | (B19) |
| phase: | $\phi(t)$ | $=$ | $\text{Arg}\,(\gamma(t))$ | | (B20) |
| in-phase: | $I(t)$ | $=$ | $\text{Re}\,(\gamma(t))$ | $=$ | $\Omega(t)\cos(\phi(t))$ | (B21) |
| in-quadrature: | $Q(t)$ | $=$ | $\text{Im}\,(\gamma(t))$ | $=$ | $\Omega(t)\sin(\phi(t))$ | (B22) |

where the drive phase $\phi(t) = +\text{Arg}(\gamma(t))$ is defined as the *positive* argument. The drive term in the control Hamiltonian is therefore expressed

$$\gamma(t)C + \text{H.C.} = \gamma(t)C + \gamma^*(t)C^\dagger \tag{B23}$$

$$= \Big(I(t) + iQ(t)\Big)C + \Big(I(t) - iQ(t)\Big)C^\dagger \tag{B24}$$

$$= I(t)\left(C + C^\dagger\right) + Q(t)\left(iC - iC^\dagger\right) \tag{B25}$$

$$= I(t)A_I + Q(t)A_Q \tag{B26}$$

where we have defined the *Hermitian* operators

$$A_I = C + C^\dagger, \qquad A_Q = i(C - C^\dagger). \tag{B27}$$

Each drive term therefore decomposes into a pair of shift-terms $(I(t), A_I)$ and $(Q(t), A_Q)$ in the familiar form of quadrature controls. These are related to the non-Hermitian operator as

$$C = \frac{1}{2}\left(A_I - iA_Q\right) \tag{B28}$$

The *modulus* $\Omega(t)$ sets the rotation rate, while the *phase* $\phi(t)$ sets the direction of rotation, oriented between the control axes defined by $(A_I, A_Q)$.

## Appendix C: Derivation of multidimensional filter functions

### 1. Magnus expansion

The first few Magnus terms are computed as [89, 90]

$$\Phi_1(\tau) = \int_0^\tau dt \tilde{H}_{\text{noise}}(t) \tag{C1}$$

$$\Phi_2(\tau) = -\frac{i}{2} \int_0^\tau dt_1 \int_0^{t_1} dt_2 \left[ \tilde{H}_{\text{noise}}(t_1), \tilde{H}_{\text{noise}}(t_2) \right] \tag{C2}$$

$$\Phi_3(\tau) = \frac{1}{6} \int_0^\tau dt_1 \int_0^{t_1} dt_2 \int_0^{t_2} dt_3 \left[ \tilde{H}_{\text{noise}}(t_1), \left[ \tilde{H}_{\text{noise}}(t_2), \tilde{H}_{\text{noise}}(t_3) \right] \right] + \left[ \tilde{H}_{\text{noise}}(t_3), \left[ \tilde{H}_{\text{noise}}(t_2), \tilde{H}_{\text{noise}}(t_1) \right] \right] \tag{C3}$$

$$\vdots$$

The Magnus series establishes a framework to define error *cancellation order*. Implementing a control with fidelity defined in Eq. 16 up to order $\alpha$ means choosing controls such that $\Phi_k(\tau) \approx 0$ for all $k \leq \alpha$. We now describe a useful framework for computing the $\Phi_\alpha(\tau)$ in the Fourier domain using filter functions, as expressed in Eq. 28. Using the definition for the Fourier transform set out in App. A 3, we have

$$\beta_k(t) = \frac{1}{2\pi} \int_{-\infty}^\infty d\omega e^{i\omega t} \beta_k(\omega), \tag{C4}$$

$$N_k'(t) = \frac{1}{2\pi} \int_{-\infty}^\infty d\omega e^{i\omega t} N_k'(\omega). \tag{C5}$$

Substituting into Eq. 27 we therefore obtain

$$\Phi_1(\tau) = \sum_{k=1}^p \int_{-\infty}^\infty dt \left( \frac{1}{2\pi} \int_{-\infty}^\infty d\omega_1 e^{i\omega_1 t} N_k'(\omega_1) \right) \left( \frac{1}{2\pi} \int_{-\infty}^\infty d\omega_2 e^{i\omega_2 t} \beta_k(\omega_2) \right) \tag{C6}$$

$$= \left( \frac{1}{2\pi} \right)^2 \sum_{k=1}^p \int_{-\infty}^\infty d\omega_1 \int_{-\infty}^\infty d\omega_2 N_k'(\omega_1) \beta_k(\omega_2) \int_{-\infty}^\infty dt e^{i\omega_1 t} e^{i\omega_2 t} \tag{C7}$$

$$= \left( \frac{1}{2\pi} \right)^2 \sum_{k=1}^p \int_{-\infty}^\infty d\omega_1 \int_{-\infty}^\infty d\omega_2 N_k'(\omega_1) \beta_k(\omega_2) \cdot 2\pi \delta(-\omega_2 - \omega_1) \tag{C8}$$

where $\delta(x)$ is the Dirac delta function. Consequently

$$\Phi_1(\tau) = \frac{1}{2\pi} \sum_{k=1}^p \int_{-\infty}^\infty d\omega_2 N_k'(-\omega_2) \beta_k(\omega_2) \tag{C9}$$

$$= \frac{1}{2\pi} \sum_{k=1}^p \int_{-\infty}^\infty d\omega G_k(\omega) \beta_k(\omega) \tag{C10}$$

where we have defined

$$G_k(\omega) \equiv N_k'(-\omega) \equiv \int_{-\infty}^\infty dt e^{i\omega t} N_k'(t). \tag{C11}$$

### 2. Leading order robustness infidelity in terms of filter functions

The leading-order error action operator Eq. 25 may be Taylor expanded as

$$\tilde{U}_{\text{noise}}(\tau) = \mathbb{I} - i\Phi_1 - \frac{1}{2}\Phi_1^2 + \dots \tag{C12}$$

$$= \mathbb{I} - i\Phi_1 - \frac{1}{2}\Phi_1\Phi_1^\dagger + \dots \tag{C13}$$

where we have used the property that the Magnus terms are Hermitian. Substituting into Eq. 16 the leading order contribution to the robustness fidelity takes the form

$$\mathcal{F}_{\text{robust}}(\tau) = \left\langle \left| \frac{1}{\text{Tr}\,(P)} \text{Tr}\left(P\left\{\mathbb{I} - i\Phi_1 - \frac{1}{2}\Phi_1\Phi_1^\dagger + ...\right\}\right)\right|^2 \right\rangle \tag{C14}$$

$$\approx \left\langle \left| \frac{1}{\text{Tr}\,(P)}\left\{\text{Tr}\,(P) - i\text{Tr}\,(P\Phi_1) - \text{Tr}\left(\frac{1}{2}P\Phi_1\Phi_1^\dagger\right)\right\}\right|^2 \right\rangle. \tag{C15}$$

Due to our choice of gauge transformation in Eq. 24, we additionally use the property that $\text{Tr}\,(P\Phi_1) = 0$, yielding

$$\mathcal{F}_{\text{robust}}(\tau) = \left\langle \left|1 - \frac{1}{\text{Tr}\,(P)}\text{Tr}\left(\frac{1}{2}P\Phi_1\Phi_1^\dagger\right)\right|^2 \right\rangle \tag{C16}$$

$$= \left\langle \left[1 - \frac{1}{\text{Tr}\,(P)}\text{Tr}\left(\frac{1}{2}P\Phi_1\Phi_1^\dagger\right)\right]^* \left[1 - \frac{1}{\text{Tr}\,(P)}\text{Tr}\left(\frac{1}{2}P\Phi_1\Phi_1^\dagger\right)\right] \right\rangle \tag{C17}$$

$$= \left\langle 1 - \frac{2}{\text{Tr}\,(P)}\text{Tr}\left(\frac{1}{2}P\Phi_1\Phi_1^\dagger\right) + \mathcal{O}\left(|\Phi_1|^4\right) \right\rangle \tag{C18}$$

where the last line uses the result that $\text{Tr}\left(P\Phi_1\Phi_1^\dagger\right)$ is real-valued, following from the Hermiticity of $\Phi_1$. Ignoring terms beyond $\mathcal{O}\left(|\Phi_1|^2\right)$, and observing the ensemble average over noise-realizations only affects terms dependent on $\Phi_1$, we therefore obtain

$$\mathcal{I}_{\text{robust}}(\tau) = 1 - \mathcal{F}_{\text{robust}}(\tau) \tag{C19}$$

$$\approx \frac{2}{2\text{Tr}\,(P)}\left\langle\text{Tr}\left(P\Phi_1\Phi_1^\dagger\right)\right\rangle \tag{C20}$$

$$= \frac{1}{\text{Tr}\,(P)}\text{Tr}\left(P\left\langle\Phi_1\Phi_1^\dagger\right\rangle\right). \tag{C21}$$

We may now explicitly calculate this term, substituting Eq. 28 into Eq. 26 we obtain

$$\left\langle\Phi_1(\tau)\Phi_1^\dagger(\tau)\right\rangle = \sum_{i=1}^{p}\sum_{j=1}^{p}\left(\frac{1}{2\pi}\right)^2\int_{-\infty}^{\infty}d\omega_1\int_{-\infty}^{\infty}d\omega_2 G_i(\omega_1)G_j^\dagger(\omega_2)\left\langle\beta_i(\omega_1)\beta_j^*(\omega_2)\right\rangle \tag{C22}$$

$$= \sum_{k=1}^{p}\left(\frac{1}{2\pi}\right)^2\int_{-\infty}^{\infty}d\omega_1\int_{-\infty}^{\infty}d\omega_2 G_k(\omega_1)G_k^\dagger(\omega_2)\left\langle\beta_k(\omega_1)\beta_k^*(\omega_2)\right\rangle \tag{C23}$$

$$= \sum_{k=1}^{p}\frac{1}{2\pi}\int_{-\infty}^{\infty}d\omega G_k(\omega)G_k^\dagger(\omega)S_k(\omega) \tag{C24}$$

where in the second line we invoke the independence property of the frequency-domain variables $\beta_{i,j}(\omega)$ defined by Eq. A8, and in the third line we use Eq. A17. Substituting into Eq. 26 we therefore obtain

$$\mathcal{I}_{\text{robust}}(\tau) \approx \sum_{k=1}^{p}\int_{-\infty}^{\infty}\frac{d\omega}{2\pi}\left\{\frac{1}{\text{Tr}\,(P)}\text{Tr}\left(PG_k(\omega)G_k^\dagger(\omega)\right)\right\}S_k(\omega) \tag{C25}$$

## Appendix D: Optimization benchmarking

In this appendix we provide additional details regarding the performance benchmarking of the Q-CTRL optimization engine.

### 1. Optimization tools

Four different optimization tools were compared:

**NumPy:** A simple in-house implementation of gradient-based pulse optimization, using mostly vectorized NumPy functions for calculating the system time evolution, the operational infidelity and the gradient of the infidelity, for given piecewise-constant control segments. The SciPy [136] implementation of the L-BFGS-B algorithm (via the `scipy.optimize.minimize` function with default arguments) was used to perform the optimization updates.

**QuTiP:** The QuTiP [99, 100] implementation of the gradient ascent pulse engineering algorithm. The `create_pulse_optimizer` function was called with the physical parameters defining the system (controls, drift, duration, target, segment count, and pulse bounds), with the `dyn_type` parameter set to 'UNIT', and convergence conditions consistent with the other tools used in the comparison (`max_iter=100000`, `max_wall_time=1800`, `fid_err_targ=0`, and `min_grad=1e-5`). All controls were scaled to have pulse bounds of $[-1, 1]$ prior to being passed to the function, to ensure the default `pulse_scaling` value of 1.0 was suitable. The resulting `Optimizer` object was then used to perform all necessary optimization runs (for the given system configuration).

**Q-CTRL (local):** The Q-CTRL implementation of gradient-based pulse optimization, running on the same hardware as the NumPy and QuTiP tools. Note that the optimization updates were performed using the same SciPy L-BFGS-B algorithm as the NumPy optimizer described above (again with default arguments).

**Q-CTRL (cloud):** Same as Q-CTRL (local), but running on cloud hardware with custom backend resource management.

### 2. Software and hardware versions

The following software versions were used:

**Base Docker image:** `jupyter/tensorflow-notebook:2c0af4ab516b` [137]

**libopenblas:** 0.3.7 (installed from Anaconda [138])

**NumPy:** 1.18.4 (installed from Anaconda)

**SciPy:** 1.4.1 (installed from Anaconda)

**Cython:** 0.29.19 (installed from Anaconda)

**TensorFlow:** 2.2.0 (installed from PyPI [139])

**QuTiP:** 4.4.1 (installed from PyPI)

Optimizations using local-instance code—NumPy, QuTiP and Q-CTRL (local)—were all run on a single dedicated machine with a 2.30 GHz 4-core Intel® Xeon® CPU and 16 GB RAM (note that this machine was running in the cloud and accessed remotely). The Q-CTRL (cloud) optimizations were run on a cluster of 20 machines, each with a 2.30 GHz 4-core Intel® Xeon® CPU, an NVIDIA® T4 GPU, and 16 GB RAM.

### 3. Physical systems

In all cases, for each of the two physical system configurations below, 20 optimization runs were performed with randomly-generated initial seed solutions. The numbers of objective function evaluations and the final obtained infidelities were compared qualitatively to ensure approximate consistency.

### a. Single controllable qubit in four-qubit space

For the comparison of optimizer performance against number of control segments, we used a system consisting of four qubits, with full three-axis control of a single qubit (note that while this system is separable, and thus could be solved efficiently by optimizing the controllable qubit in isolation, none of the tools were configured to take advantage of this fact). Specifically, we used the Hamiltonian:

$$H(t) = \frac{I(t)}{2}\sigma_x \otimes \mathbb{I}^{\otimes 3} + \frac{Q(t)}{2}\sigma_y \otimes \mathbb{I}^{\otimes 3} + \frac{\alpha(t) + \nu}{2}\sigma_z \otimes \mathbb{I}^{\otimes 3},$$

where $\sigma_{\{x,y,z\}}$ are the Pauli operators, $I(t), Q(t), \alpha(t)$ are optimizable controls, and $\nu$ is a constant dephasing drift.

The optimizations were performed with the following parameters:

**Target gate:** $H \otimes \mathbb{I}^{\otimes 3}$ (Hadamard on first qubit)

**Gate duration:** $0.5\,\text{s}$

**Pulse bounds:** $|\alpha(t)| \leq 2\pi \times 2\,\text{Hz}$ and $|I(t) + iQ(t)| \leq 2\pi \times 2\,\text{Hz}$ (note that QuTiP does not support this type of complex constraint, so the relaxed constraint $|I(t)|, |Q(t)| \leq 2\pi \times 2\,\text{Hz}$ was used instead)

**Dephasing:** $\nu = 2\pi \times 1\,\text{Hz}$

In this case, system complexity was tuned by varying the number of segments used for the piecewise-constant pulses $I(t), Q(t), \alpha(t)$, between 10 and 500. This segment count maps directly to the dimensionality of the optimization search space (there are three controls, so for $m$ segments per control there are $3m$ optimizable parameters), and to the computational complexity of calculating the system dynamics.

### b. Linear array of Rydberg atoms

For the comparison of optimizer performance against number of qubits, we used a system consisting of a linear array of Rydberg atoms, with controllable global coupling and detuning. The Hamiltonian for the system, assuming an array of $N$ atoms, is[140]:

$$H(t) = \frac{\Omega(t)}{2}\sum_{i=1}^{N}\sigma_x^{(i)} - \Delta(t)\sum_{i=1}^{n}n^{(i)} - \sum_{i=1}^{N}\delta_i n^{(i)} + \sum_{i<j}\frac{V}{|i-j|^6}n^{(i)}n^{(j)},$$

where $\sigma_x$ is the Pauli X operator, $n = |1\rangle\langle 1|$ is the number operator, the superscripts indicate a single-qubit operator embedded in the full space (e.g. $\sigma_x^{(i)} = \mathbb{I}^{\otimes i-1} \otimes \sigma_x \otimes \mathbb{I}^{\otimes n-i}$), $\Omega(t)$ and $\Delta(t)$ are optimizable controls, $\delta_i$ are fixed per-atom detunings, and $V$ is the interaction strength.

The optimization was performed with the following parameters (note that the timescales are arbitrary and do not affect the optimization procedure, so here we have chosen them to be physically realistic):

**Target gate:** $|\text{GHZ}\rangle\langle 000\cdots|$, where $|\text{GHZ}\rangle = \frac{1}{\sqrt{2}}(|0101\cdots\rangle + |1010\cdots\rangle)$ (preparation of a Greenberger-Horne-Zeilinger state)

**Gate duration:** $1.1\,\mu\text{s}$

**Pulse bounds:** $|\Omega(t)| \leq 2\pi \times 5\,\text{MHz}$ and $|\Delta(t)| \leq 2\pi \times 20\,\text{MHz}$

**Interaction strength:** $V = 2\pi \times 24\,\text{MHz}$

**Detunings per-atom:** $\delta_i = \begin{cases} -2\pi \times 4.5\,\text{MHz} & \text{for } i = 1, N \\ 0 & \text{for } 2 \leq i \leq N-1 \end{cases}$

**Control segments:** 40 segments each for piecewise-constant controls $\Omega(t)$ and $\Delta(t)$

In this case, system complexity was tuned by varying the number of atoms $N$ from 2 to 8 (although results were not taken if the 20 optimizations were projected to take any more than roughly one hour, which was true for $N = 7, 8$ with the NumPy optimizer and for $N = 8$ with the QuTiP and Q-CTRL (local) optimizers).

## Appendix E: Methods for experimental demonstrations of quantum-control benefits

### 1. Quasi-static error robustness

In Fig. 8b,c we implement a net $X_\pi$ gate using four different pulse constructions, with varying error-robustness properties:

**Primitive:** no robustness (red).

**BB1 [120]:** robust to pulse amplitude errors (purple).

**CORPSE [121]:** robust to pulse detuning errors (cyan).

**CinBB [122]:** robust to both pulse amplitude and detuning errors (blue).

We compare the performance of all four against quasi-static errors in both the rotation angle (amplitude) and qubit frequency (detuning). A single trapped ion qubit is prepared in $|0\rangle$ and a sequence of $X_\pi$ pulses is applied to amplify the error. The probability of finding the qubit in the $|1\rangle$ state, $P_1$, is then measured providing a proxy for the sequence infidelity $\mathcal{I}$ (zero for error-free rotations).

In Fig. 8b, an over-rotation error is engineered by scanning the pulse length either side of the the $\pi$-time (the ideal value). For each error strength we measure $P_1$ after implementing a sequence of 10 $X_\pi$ pulse. This sequence amplifies the effect of over-rotation errors. In Fig. 8c, an engineered detuning error is created by driving the qubit off-resonantly. The absolute frequency detuning is normalized by the Rabi rate to quantify a relative error that is varied between ±10%. To amplify the effect of detuning errors we use an alternating sequence of 10 ±$X_\pi$ pulses.

### 2. Suppression of time-varying noise

The experimental filter function reconstructions shown in Fig. 8d,e in the main text were performed through the application of a single frequency disturbance at $\omega_{\rm sid}$ added to either the control or the dephasing quadrature. We denote these time-dependent noise fields with $\beta_k(t)$ with $k \in \{\Omega, z\}$ and our corresponding Hamiltonian reads

$$H(t) = H_{\rm ctrl}(t)(1 + \beta_\Omega(t)) + \beta_z(t)\sigma_z, \tag{E1}$$

where $H_{\rm ctrl}(t)$ is the control Hamiltonian that represents the driven evolution through the microwave field, $\beta_\Omega(t)$ is the amplitude noise and $\beta_z(t)$ is the dephasing noise. Typically, we write $H_{\rm ctrl}(t)$ in a rotating frame with respect to the qubit splitting such that it takes the form of a time-dependent $X$- or $Y$-rotation with

$$H_{\rm ctrl}(t) = \frac{\Omega(t)}{2}(\cos\phi(t)\sigma_x + \sin\phi(t)\sigma_y), \tag{E2}$$

where $\Omega(t)$ is the time-dependent Rabi rate and $\phi(t)$ is the control phase (see [110] for further details on the experimental system and the control synthesis). The noise fields take the explicit form of

$$\beta_k(t) = \alpha_k \cos(\omega_{\rm sid}t + \varphi) \quad \text{for} \quad k \in \{\Omega, z\}, \tag{E3}$$

where $\alpha_k$ is a constant factor to set the modulation depth. Through averaging over phase parameter $\varphi \in \{0, 2\pi\}$, this form of modulation produces a $\delta$-function like noise spectrum $S_k(\omega) \approx \delta(\omega - \omega_{\rm sid})$ which, using the relationship $\chi \propto \int d\omega S(\omega)F(\omega)$ allows us to directly extract the value of the filter function at the frequency $\omega_{\rm sid}$ [58, 110]. For the experiments here, we used $\alpha_\Omega = 0.25$ and $\alpha_z = 0.7$ and the points were averaged over 5 values of $\varphi$ spaced linearly between 0 and $2\pi$.

Experimentally, this is achieved via amplitude and frequency modulation of the microwave field that drives the qubit transition. The amplitude noise is added digitally to the I/Q control waveforms before upload to the microwave signal generator (Keysight E8267D), such that we obtain a noisy drive waveform with

$$\Omega(t) \to \Omega(t)(1 + \beta_\Omega(t)). \tag{E4}$$

The dephasing noise is engineered through an additive term in the phase of the control Hamiltonian (Eq. E2). The total phase offset is formally split into a control and noise term: $\phi(t) \to \phi_{\rm ctrl}(t) + \phi_{\rm noise}(t)$. In the interaction picture defined by the frame transformation

$$H_{\rm int}(t) = U_\phi(t)^\dagger H_{\rm ctrl}(t)U_\phi(t) - \dot{U}_\phi(t)U_\phi^\dagger(t), \qquad\qquad U_\phi(t) = \exp\left(-i\frac{\phi_{\rm noise}(t)}{2}\sigma_z\right) \tag{E5}$$

the total Hamiltonian takes the form

$$H_{\text{int}}(t) = -\frac{\Omega(t)(1 + \beta_\Omega(t))}{2}(\cos\phi(t)\sigma_x + \sin\phi(t)\sigma_y) - \frac{\dot{\phi}_{\text{noise}}(t)}{2}\sigma_z. \qquad (E6)$$

Using our notation from Eq. E1, we can identify the dephasing noise term $\beta_z(t) = \dot{\phi}_{\text{noise}}(t)/2$. The corresponding dephasing noise waveforms are generated using an external arbitrary waveform generator (Keysight 33600A), whose output is fed into the analog FM port of the microwave generator, which produces the desired dephasing noise term. For more details, see chapter 2 in [110].

### 3. Error homogenization characterized via 10-qubit parallel randomized benchmarking

In Fig. 8f, we measure a spatially varying average error rate along a string of 10 trapped ion qubits (shown in Fig. 8f inset). In this system, ions are simultaneously addressed by a global microwave control field to drive global single-qubit rotations. However, due to a gradient in the strength of the microwave field, the qubits rotate with a spatially varying Rabi rate meaning that the control cannot be synchronously calibrated for all 10 qubits. Qubit 1 in the figure is used for calibration in this experiment, yielding the lowest error rate.

Error is measured using parallel randomized benchmarking in which all ions are illuminated with microwaves simultaneously and experience the same RB sequence. The general approach to RB sequence construction and experimental implementation is described in detail in the supplemental material of Ref. [123]. Sequences here are composed of up to 500 operations selected from the Clifford set. Measurement is conducted using a spatially-resolved EMCCD (electron-multiplying CCD) camera in order to extract average error rates for each individual qubit. Given the relatively low quantum efficiency of this detection method, state-preparation and measurement (SPAM) errors are in the range of $\sim 3-5\%$, approximately an order of magnitude higher than achieved using single-qubit RB, as measured via an avalanche photodetector.

### 4. Mølmer-Sørensen drift measurements

In Fig. 8g-i, we compare two different constructions of phase-modulated two-qubit Mølmer-Sørensen gates, both designed to produce the entangled Bell state $(|00\rangle - i\,|11\rangle)/\sqrt{2}$. The gates are implemented by illuminating two ions with a pair of orthogonal beams from a pulsed laser near 355 nm, driving stimulated Raman transitions as described in [73]. The geometry of the beams enables coupling to the radial motional modes of the ions, which have approximate frequencies $\omega_k/2\pi = (1.579, 1.498, 1.485, 1.398)$ MHz and are denoted from highest to lowest frequency by $k = 1$ to $k = 4$. One of the Raman beams is controlled by an acousto-optic modulator driven by a two-tone radio-frequency signal produced by an arbitrary waveform generator. This results in a bichromatic light field that off-resonantly drives the red and blue sideband transitions, creating the state-dependent force used in the gate. We modulate the phase of the driving force $\phi(t)$ by adjusting the phase difference between the red and blue frequency components, $\phi(t) = [\phi_b(t) - \phi_r(t)]/2$.

After an initial calibration of the mode frequencies and gate Rabi frequency, we repeatedly perform the entangling operations over a period of several hours without further calibration, alternating between the two different phase-modulated gate constructions (panel g vs i) in order to mitigate any systematic differences between measurements. The ions are optically pumped to $|00\rangle$ and the selected gate is performed by applying the Raman beams for a duration of 500 µs. For both constructions, the gate detuning set to $-2$ kHz from the $k = 2$ mode. In this configuration, only the $k = 2$ and $k = 3$ modes are significantly excited during the operation. Each gate is repeated 500 times and the ion fluorescence measured after each repetition. We use a maximum likelihood procedure described in [73] to extract the state populations $P_n$, the probability of measuring $n$ ions bright, for each set of repetitions.

The first phase-modulated gate construction consists of four phase segments and is calculated to ensure modes $k = 2$ and $k = 3$ are de-excited at the conclusion of the operation. The second is calculated to provide additional robustness to low frequency noise affecting the closure of mode $k = 2$, which necessitates doubling the number of phase segments as per the analytic procedure outlined in [73].

The required gate Rabi frequencies are $\Omega = 2\pi \times 18.3$ kHz and $\Omega = 2\pi \times 22.9$ kHz, for the first and second gates, respectively. The laser amplitude required to produce the desired $\Omega$ for a particular gate construction is calibrated by fixing the amplitude of the single-tone Raman beam and varying the amplitude of the bichromatic beam. As the amplitude of the beam is increased, the populations $P_0$ and $P_2$ will converge to the point at which $P_0 = P_2 \approx 0.5$, indicating the creation of the Bell state and the correct laser amplitude.

## Appendix F: Visualizations of noise and control in quantum circuits

Q-CTRL provides an advanced visual interface that enables users to compose quantum circuits (Fig. 16a and Fig. 17a) and interactively track state evolution subject using Bloch spheres and a visual representation of entanglement based on *correlation tetrahedra*. These tools offer unique, interactive, 3-dimensional visualizations, assisting users to build intuition for key logical operations performed in quantum circuits.

In our tools, an arbitrary sequence of single and multiqubit operations may be graphically composed or sequenced in python. The associated state evolution is calculated and displayed using interactive three.js objects. These may then be rendered in Q-CTRL products, or directly embedded in Jupyter notebooks or websites. Current visualization packages are limited to two-qubit subspaces, with forthcoming development expanding to larger circuits.

As an example consider the time-domain evolution of the state of a single qubit subject to control. A series of gates is described by an ideal target Hamiltonian:

$$H_{\text{tot}} = \frac{1}{2}\Omega(\cos(\phi)\sigma_x + \sin(\phi)\sigma_y) + \frac{1}{2}\Delta\sigma_z, \tag{F1}$$

where, $\Omega$ is the Rabi rate, $\Delta$ the detuning and $\sigma_{x,y,z}$ are the standard Pauli matrices. Depicted in Fig. 16b is a snapshot of the state vector evolution (purple pointer, Fig. 16b) along its present trajectory (solid red line, Fig. 16b) given by the last Hadamard gate (highlighted through a green indicator). The visualization is dynamic and evolves in time; a user can interact with a time-indication slider in order to move through the time sequence. Moreover the view of the Bloch sphere and its color palette may be adjusted by the user, allowing for a user to gain insights that may be challenging in a simple 2D representation. In this specific example, for instance, it becomes immediately obvious that a Hadamard gate has the action not only of transforming a state $|+z\rangle \rightarrow |+x\rangle$, but that it does so as a rotation about an axis tilted out of the equatorial plane of the Bloch sphere.

The visualizer module also provides a means by which one may intuitively explore the effect of noise on unitary operations performed within quantum circuits. Two noise channels are available: control-amplitude noise $\Omega \rightarrow \Omega(1 + \beta)$; and ambient dephasing noise $\Delta \rightarrow \Delta + \eta$, where $\beta$ is the fractional fluctuation away from the target driving rate $\Omega$ and $\eta$ is a fluctuation away form the target detuning $\Delta$. In this circumstance the ideal state evolution is perturbed by the presence of noise, as illustrated by a displaced trajectory on the Bloch sphere. Ultimately, the discrepancy between the final location of the state at the end of the circuit and the ideal transformation, illustrated by a dotted line, provides a clear visual representation of how noise reduces the fidelity of a unitary transformation.

Producing an intuitive visual representation of entanglement poses a significant challenge due to the presence of non-classical correlations between quantum systems. The Q-CTRL visualizer provides an exact representation of a two-qubit system exhibiting entanglement and subject to unitary controls utilizing two Bloch spheres and three correlation tetrahedra. The Bloch spheres depict the standard Pauli observables corresponding to each of the qubits (Fig. 17b), given for qubits one and two respectively by:

$$X_1 = \langle\sigma_x \otimes \mathbb{I}\rangle, \quad X_2 = \langle\mathbb{I} \otimes \sigma_x\rangle,$$
$$Y_1 = \langle\sigma_y \otimes \mathbb{I}\rangle, \quad Y_2 = \langle\mathbb{I} \otimes \sigma_y\rangle,$$
$$Z_1 = \langle\sigma_z \otimes \mathbb{I}\rangle, \quad Z_2 = \langle\mathbb{I} \otimes \sigma_z\rangle. \tag{F2}$$

The surfaces of the Bloch spheres fully describe the space of separable states; the presence of any entanglement necessitates that the Bloch vectors shrink and the states move off of the Bloch sphere surfaces. For instance, in Fig. 17b, the magnitude of both Bloch vectors shrinks to zero following the first CNOT gate, as the system evolves into a maximally entangled Bell state as a result of this gate.

We visually depict entanglement using a set of observables corresponding to correlations between pairs of Cartesian observables for the two qubits. The correlation value between the observables is given by:

$$V(AB) = \langle\sigma_A \otimes \sigma_B\rangle - \langle\sigma_A \otimes \mathbb{I}\rangle\langle\mathbb{I} \otimes \sigma_B\rangle,$$
$$A, B \in \{X, Y, Z\}. \tag{F3}$$

The nine correlation-pairs are organized into three coordinate systems bounded by a tetrahedral geometry [141] given by the following axial arrangement of the observable pairs: $(XX, YY, ZZ)$, $(XY, YZ, ZX)$ and $(XZ, YX, ZY)$. For separable states all correlation values are zero and the visual indicator is set to the center of the tetrahedra. In the presence of non-zero entanglement, however, these visual indicators emerge from the origin of the correlation coordinate systems and grow towards the extrema of the convex hull for maximally entangled qubit pairs. The degree of entanglement is also visually represented using the concurrence $C(\psi)$ defined as:

$$C(\psi) = 2\left|\langle00|\psi\rangle\langle11|\psi\rangle - \langle01|\psi\rangle\langle10|\psi\rangle\right| \tag{F4}$$

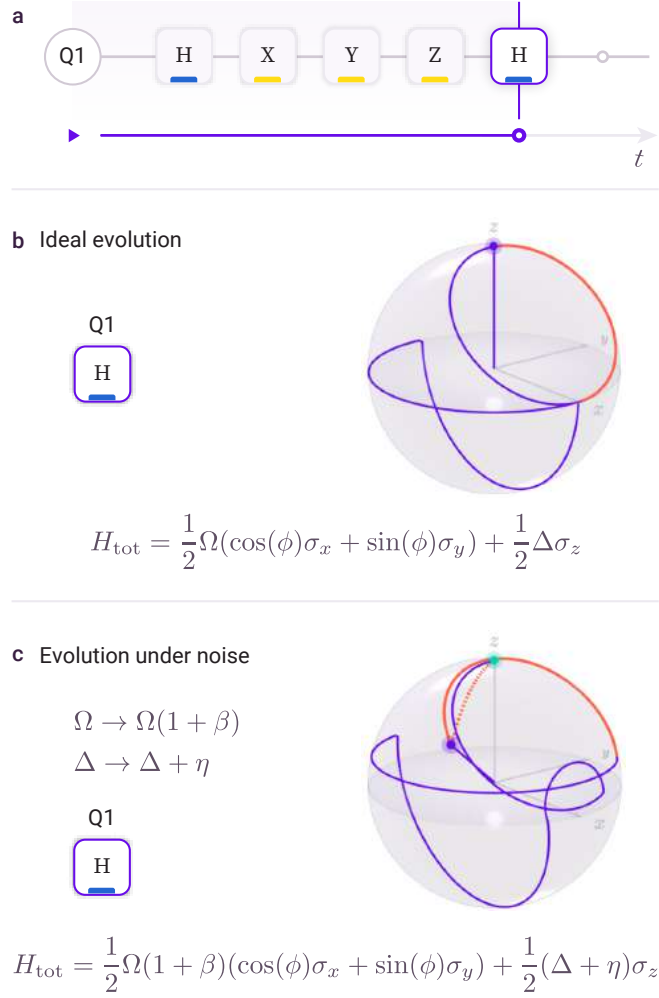

FIG. 16. Circuit evolution under noise. (a) The interactive Q-CTRL quantum circuit interface, paused during the second Hadamard gate (highlighted). (b) Evolution of the state vector (purple pointer) on the Bloch sphere, under ideal, noise-free conditions. The current trajectory on is highlighted in red (H-gate) which sequentially continues on the prior circuit evolution (purple). (c) Evolution of the state in the presence of two noise channels: control amplitude ($\beta$) and ambient dephasing ($\eta$). The cumulative error in the qubit state (dashed red) due to the presence of these noise channels is indicated by the dashed line, relative to the ideal target trajectory (green pointer).

which varies between 0 (separable states) and 1 (maximally entangled) throughout the system evolution, and shown using a horizontal indicator. Maximally entangled states may traverse the correlation tetrahedra when local unitaries are applied to the individual qubits, but the Bloch vectors remain at the centers of the Bloch spheres. Once again, noise processes may be added to the system's evolution in order to represent how the presence of noise perturbs the entangled states.

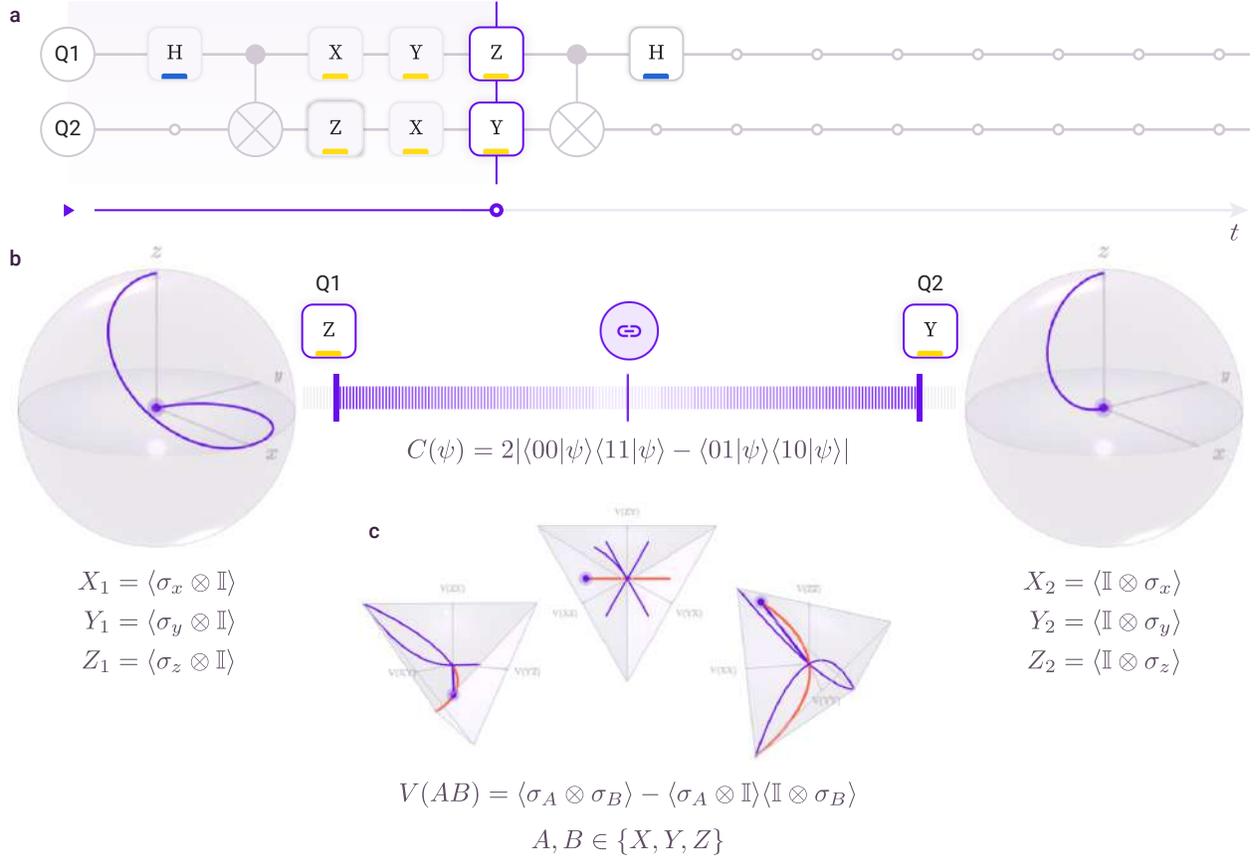

FIG. 17. Visualization of two-qubit entangled states using BLACK OPAL. (a) Evolution of a two-qubit entangling circuit paused during a pair of gates (circled in red). (b) The trajectories of individual qubit observables depicting separable states are represented on an interactive 3D Bloch sphere. The bloch vector goes to zero as qubits become entangled through the action of the CNOT gate. (c) Three interactive entanglement tetrahedra track correlations between the nine pairs of observables enabling visual representation of complete two-qubit state evolution. Concurrence indicates the level of entanglement (red markers) between the qubits throughout the evolution of the circuit. Control and/or dephasing noise may be added to assist in understanding its impact on the evolution of entangled states.